\documentclass[twocolumn]{aastex631}
\usepackage{placeins}

\usepackage{multirow}

\usepackage{hyperref}
\begin{document}

\title{SHORES II:\\  Multi-frequency Characterisation of the Sub-mJy Radio Population in FIR-selected Fields}

\author[0000-0002-6444-8547]{Meriem Behiri}
\email{meriembehiri96@gmail.com}
\affiliation{Scuola Internazionale Superiore di Studi Avanzati, Via Bonomea 265, 34136 Trieste, Italy}
\affiliation{INAF-OAS Bologna, via Gobetti 101, I-40129 Bologna, Italy}
\affiliation{INAF - Istituto di Radioastronomia, Via Gobetti 101, 40129 Bologna, Italy}
\affiliation{Institute for Fundamental Physics of the Universe (IFPU), Via Beirut 2, 34014 Trieste}

\author[0000-0002-0375-8330]{Marcella Massardi}
\affiliation{INAF - Istituto di Radioastronomia - Italian ALMA Regional Centre,
Via Gobetti 101, 40129 Bologna, Italy}
\affiliation{Scuola Internazionale Superiore di Studi Avanzati, Via Bonomea 265, 34136 Trieste, Italy}

\author[0000-0003-1394-7044]{Vincenzo Galluzzi}
\affiliation{INAF - Istituto di Radioastronomia, Via Gobetti 101, 40129 Bologna, Italy}
\affiliation{INAF-Osservatorio Astronomico di Trieste - Italian Astronomical Archives, via Tiepolo 11,34131 Trieste, Italy}

\author[0000-0002-1847-4496]{Marika Giulietti}
\affiliation{INAF - Istituto di Radioastronomia, Via Gobetti 101, 40129 Bologna, Italy}
\author[0000-0002-7472-7697]{Gayathri Gururajan}
\affiliation{Scuola Internazionale Superiore di Studi Avanzati, Via Bonomea 265, 34136 Trieste, Italy}

\affiliation{Institute for Fundamental Physics of the Universe (IFPU), Via Beirut 2, 34014 Trieste}
\author[0000-0001-9680-7092]{Isabella Prandoni}
\affiliation{INAF - Istituto di Radioastronomia, Via Gobetti 101, 40129 Bologna, Italy}

\author[0000-0002-4882-1735]{Andrea Lapi}
\affiliation{Scuola Internazionale Superiore di Studi Avanzati, Via Bonomea 265, 34136 Trieste, Italy}
\affiliation{INAF - Istituto di Radioastronomia, Via Gobetti 101, 40129 Bologna, Italy}
\affiliation{Institute for Fundamental Physics of the Universe (IFPU), Via Beirut 2, 34014 Trieste}
\affiliation{Istituto Nazionale Fisica Nucleare (INFN), Sezione di Trieste, Via Valerio 2, 34127 Trieste, Italy}

\begin{abstract}
We present a new deep multi-frequency radio survey of two extragalactic fields observed with the Australia Telescope Compact Array (ATCA) as part of the SHORES project (Serendipitous H-ATLAS-fields Observations of Radio Extragalactic Sources). The observations, centred at 2.1, 5.5, and 9 GHz, cover the central 0.5 deg$^2$ of two Herschel Astrophysical Terahertz Large Area Survey (H-ATLAS) fields down to rms sensitivities of 9–17 $\mu$Jy/beam at 2.1 GHz, 28–39 $\mu$Jy/beam at 5.5 GHz and 38-61 $\mu$Jy/beam at 9 GHz. This setup allows us to investigate the spectral energy distributions (SEDs) of faint radio sources and probe the nature of the sub-mJy population. We extract and validate a robust catalogue of 489 sources at 2.1 GHz, 101 of which are also detected at 5.5 GHz. We perform a multi-frequency analysis of the radio number counts and derive the spectral indices of sources in the deep fields. The spectral index distribution of our sources peaks around $\alpha \sim -0.7$, consistent with synchrotron emission from the faint radio population. The number counts at 2.1 GHz are consistent with previous deep surveys and theoretical models, and provide a lower limit on the star-forming galaxy population, which is expected to dominate the faint end. The 5.5 GHz data offer new, direct constraints on the sub-mJy radio sky at higher frequencies.
By cross-matching with the H-ATLAS catalogue, we identify a sample of sources with far-infrared (FIR) counterparts and explore the far-infrared–radio correlation (FIRRC). The sources with $q_{\mathrm{FIR}} \geq 1.69$ exhibit radio spectral indices typical of star-forming galaxies. Furthermore, we identify a population of radio-only sources with similar indices that may correspond to high-redshift SFGs, lacking counterparts in the FIR survey{ due to its limited} resolution and sensitivity. 

\end{abstract}


\section{Introduction} \label{sec:intro}

Over the past decades, radio astronomy has become a cornerstone in the study of galaxy formation and evolution, providing insights into active galactic nuclei (AGN) activity across cosmic time. However, technical limitations—particularly in sensitivity and resolution—have historically limited the detection and identification of star-forming galaxies (SFGs) in statistical radio studies. In fact, faint radio sources, including SFGs and radio-quiet AGN, trace the build-up of stellar mass and black hole growth, even in heavily dust-obscured environments where optical and UV diagnostics fail (e.g. \citealt{talia21,Behiri+23,gentile24}). As such, deep radio surveys reaching sub-mJy sensitivities (e.g. \citealt{smith21,best23,whittam22,hale23}) are essential to constrain the cosmic star formation history and to investigate the complex interplay between star formation and AGN feedback.

Radio emission from SFGs arises primarily from two physical processes tied to massive star formation: thermal free-free (bremsstrahlung) emission from ionised gas in HII regions and non-thermal synchrotron radiation from relativistic electrons accelerated in supernova remnants (e.g. \citealt{Condon+92,Murphy+09}). The balance between these two components varies with galaxy age and star formation efficiency, imprinting distinctive spectral energy distributions (SEDs) observable at radio wavelengths. In parallel, dust grains formed in the ejecta of massive stars absorb UV photons from young stellar populations and re-radiate in the far-infrared (FIR), linking FIR and radio emission through the well-known far-infrared–radio correlation (FIRRC; e.g. \citealt{helou85,Condon+92,yun01,bell03,ivison10b,sargent10,Giulietti+22}). This correlation constitutes a powerful diagnostic of star formation, provided that contamination from AGN-related emission is carefully addressed.

AGN, especially in their radio-loud phase, contribute significantly to the radio sky through synchrotron radiation from relativistic jets and lobes. These structures are powered by the extraction of rotational energy from spinning supermassive black holes, mediated by magnetic fields (e.g. \citealt{Blanford}). The spectral properties of AGN-related radio emission depend on several factors, including jet power, age, environment, and viewing angle, and can differ significantly from those of SFGs. At GHz frequencies, compact flat-spectrum cores, steep-spectrum lobes, peaked flaring jetted components, and curved spectra due to synchrotron ageing or self-absorption are commonly observed (e.g. \citealt{massardi11}).

To disentangle these populations and build a comprehensive picture of radio source evolution, a multi-wavelength approach is essential, together with a combination of wide and deep observations. While wide-area shallow surveys at $\sim$GHz frequencies, such as RACS \citep{racs1}, excel at detecting bright, extended radio sources, but they often lack the resolution or depth needed to characterise the faint, compact population of SFGs.  
{Further, the MIGHTEE survey \citep{jarvis16} provides wide-area deep radio imaging, complementary to the intermediate-depth, small-area observations presented here.}
On the other hand, higher-frequency observations —particularly in the 2–10 GHz range— offer higher resolution, probe rest-frame frequencies where thermal processes may become relevant. This spectral window captures the contribution by both AGN and star-formation, making it particularly suitable to study faint galaxies and disentangle the physical processes that fuel the radio emission.

Within this context, we started the SHORES project (Serendipitous H-ATLAS-fields Observations of Radio Extragalactic Sources, \citealt{shores}), designed to exploit the synergy between deep radio observations and the rich far-infrared coverage provided by the Herschel Astrophysical Terahertz Large Area Survey (H-ATLAS; \citealt{eales10}). SHORES includes both shallow and deep fields, both surveyed at 2.1 GHz with the Australia Telescope Compact Array (ATCA), centred on a sample of candidate lensed galaxies in the South Galactic Pole (SGP) selected by \cite{negrello2017} based on their FIR fluxes in H-ATLAS. This selection guarantees comprehensive FIR coverage across the observed fields. The shallow component, described in \citet{shores} (hereafter referred as SHORES-I), focused on the bright end of the 2.1 GHz radio population, dominated by AGN, and reached 95\% completeness above 0.5 mJy over a 26 {deg$^2$} surveyed area. Further, the deep fields have been also followed up at 5.5 and 9 GHz, to improve the radio photometric characterisation. In this paper, we present the analysis of the deep fields, which target the sub-mJy regime and are tailored to study the faint radio sources at 2.1, 5.5 and 9 GHz.

The deep observations reach rms sensitivities of 9–17 $\mu$Jy/beam at 2.1 GHz and 28–39 $\mu$Jy/beam at 5.5-9 GHz, over the two 0.5 {deg$^2$} H-ATLAS fields. This allows robust source extraction, spectral index analysis, and cross-matching with FIR-selected galaxies. Combining radio and FIR data allows us to investigate the FIRRC down to faint flux levels, identify candidate high-redshift SFGs, and assess AGN contamination using spectral diagnostics.

This paper is structured as follows. In Section \ref{sec:observations}, we describe the observations, data reduction, and source extraction. Section \ref{sec:catalogue} presents the multi-frequency catalogue and spectral index analysis. Section \ref{sec:FIRRC} focuses on the FIR cross-matching and the analysis of the FIRRC. In Section \ref{sec:counts}, we derive the radio number counts at 2.1 and 5.5 GHz. Finally, we discuss the implications of our results in Section \ref{sec:conclusions}.

Throughout this paper we define spectral indices $\alpha$ so that flux densities scale as $S_\nu\propto \nu^\alpha$, and we adopt a flat $\Lambda$CDM cosmology (\citealt{planck20}) with round parameter values $h\approx 0.67$, $\Omega_m\approx 0.3$ and $\Omega_{\Lambda}\approx 0.7$.

\begin{figure}[htbp]
    \centering
    
    \includegraphics[width=\linewidth]{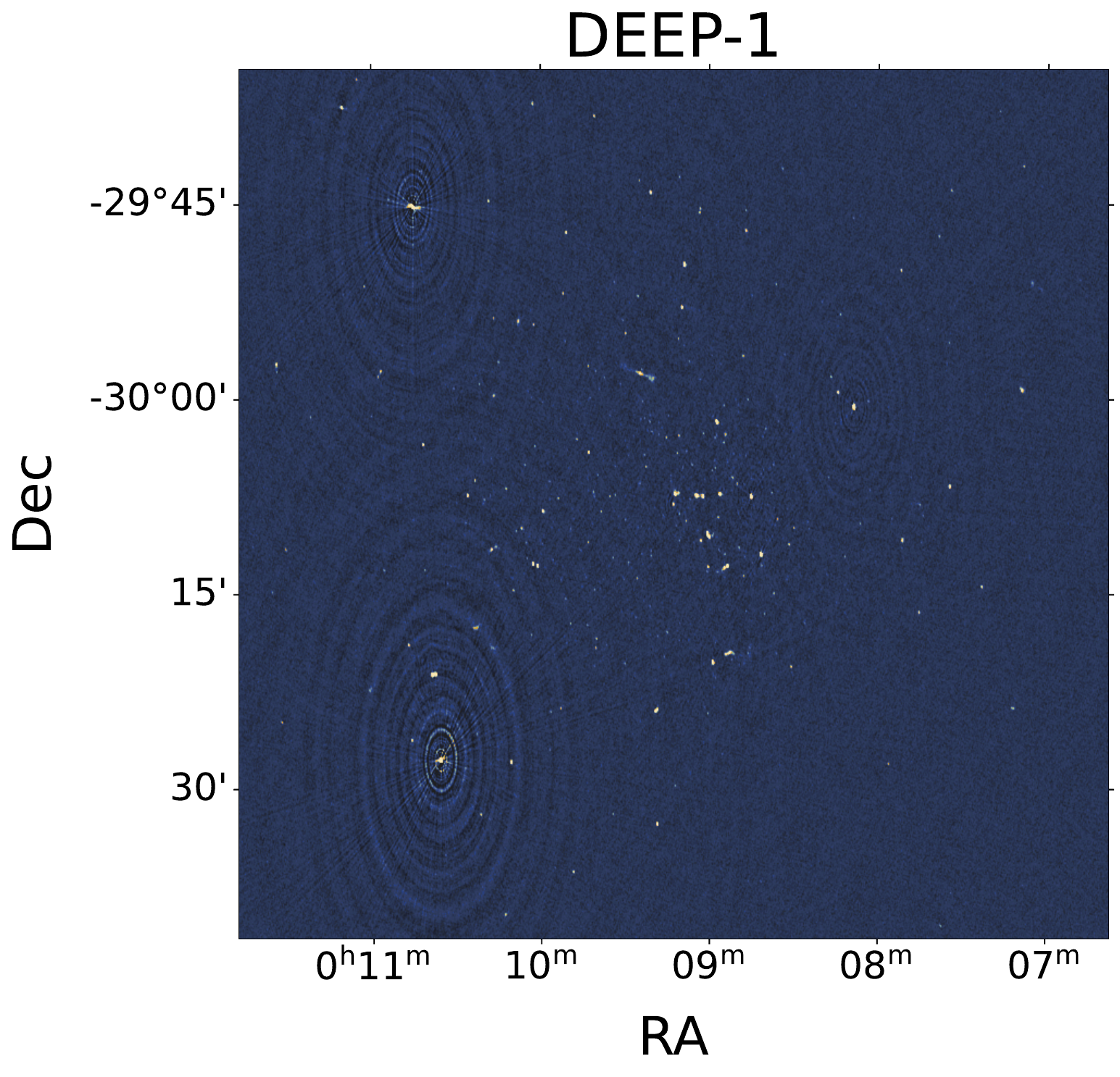}\\[-0.2em]  
    
    \includegraphics[width=\linewidth]{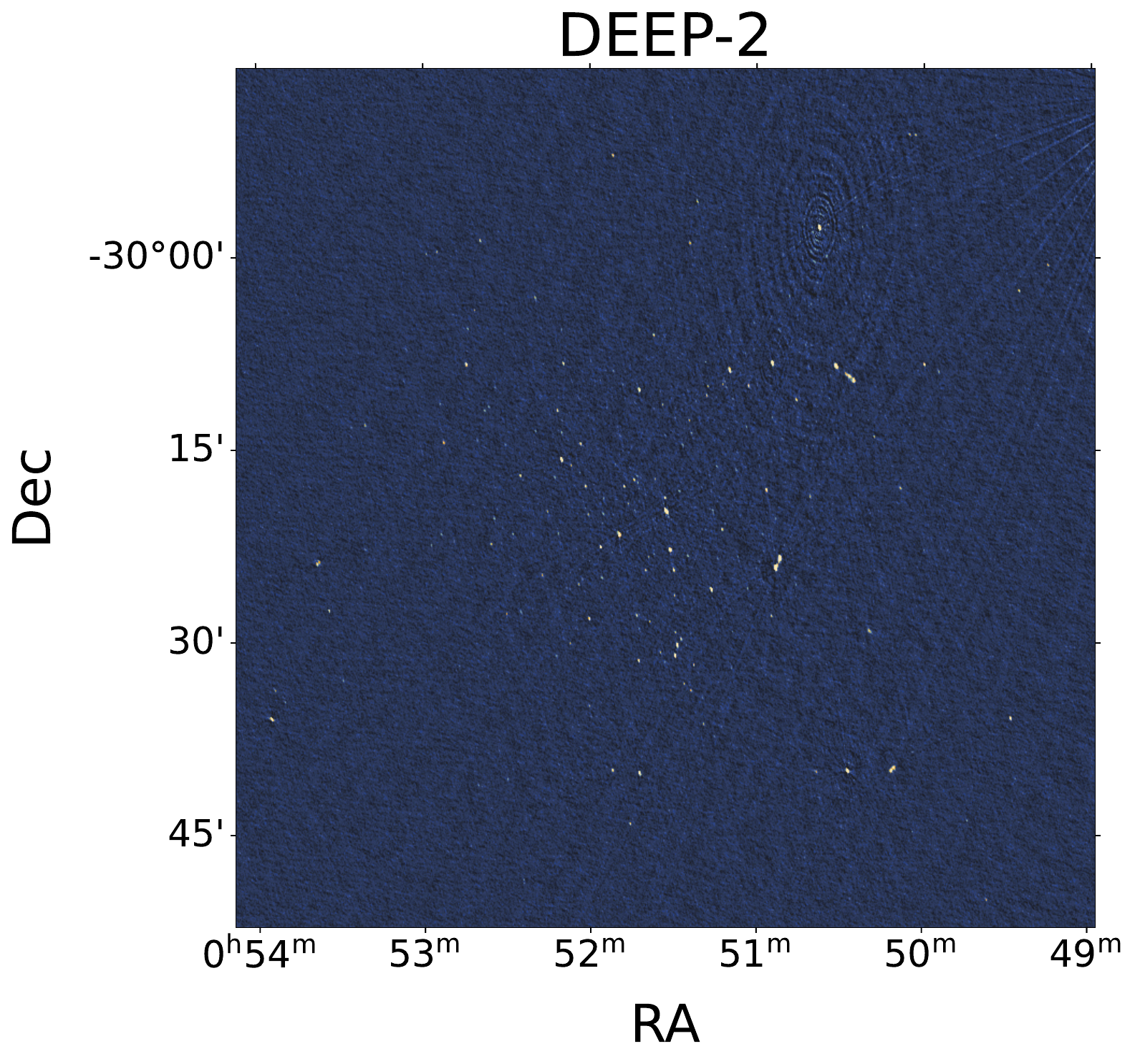}\\[0.8em]   

    \includegraphics[width=\linewidth]{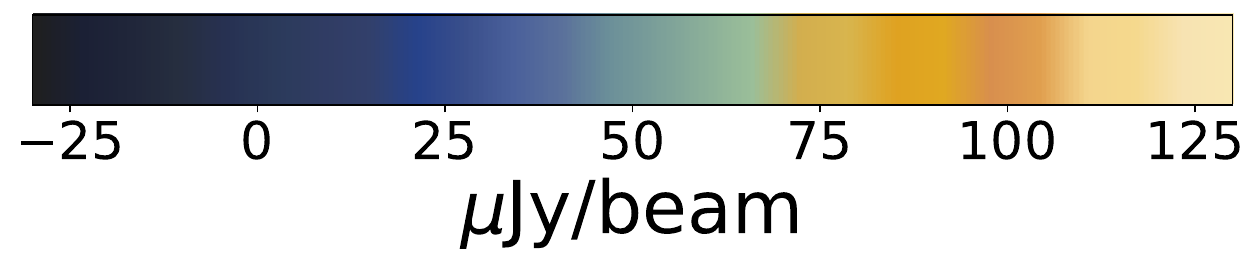}
    
    \caption{2.1 GHz maps of the DEEP~1 (\textit{top}) and DEEP~2 (\textit{bottom}) fields. 
             A single colour scale is shown below.}
    \label{fig:maps2d1}
\end{figure}

\section{Observations and Data Reduction Process for SHORES Deep Fields}  \label{sec:observations}
\subsection{The deep fields selection}
In the H/ATLAS SGP fields targeted by the SHORES survey, \cite{negrello2017} identified 30 potential lensed sources. Among these, we identified the field at HATLAJ000912.7-300807 as notable due to its abundance of ancillary data. This particular field, especially its central lensed galaxy, had been previously observed with various instruments including the Atacama Large Millimeter Array (ALMA), the Hubble Space Telescope (HST), the Wide-field Infrared Survey Explorer (WISE), the GALaxy Evolution eXplorer (GALEX), the Kilo-Degree Survey (KiDS), and Spitzer surveys. Importantly, this field was free from interference by any extremely bright sources in radio surveys such as the NRAO VLA Sky Survey (NVSS) or the Rapid ASKAP Continuum Survey (RACS), which could affect our deeper observations. Hence, we selected it as the optimal candidate for more in-depth observations, naming it "DEEP-1".

The second deep field was selected combining two 2.1 GHz ATCA pointings centred to the lensed sources  HATLASJ005132.8-01848 and HATLASJ005132.0-302011. These two fields happen to be closer than the 22.2 arcmin FWHM of the 2.1 GHz ATCA FoV, thus overlapping in FoV coverage. Therefore, we combined their observations to double the time on source for the area with respect to any other shallow field of the SHORES survey, building a "DEEP-2" field. 

The two fields were observed in different epochs and with different calibrators; thus, they were reduced separately and consequently reached different sensitivity levels: extracted sources were combined in the analysis, accounting for the differences in the field maps where necessary, as described in the following section.  

\begin{table}[ht]
    \centering
        \caption{ ATCA configurations and corresponding observing times for the SHORES deep fields}
    \begin{tabular}{c|c|c|c}
    \hline
        Date & Field & Band & Configuration \\
        \hline
        22OCT09 & DEEP-1 & 4 cm & 6D \\
        22NOV29 &  DEEP-1,2 & 16 cm & 6C  \\
        22NOV30 & DEEP-1 & 16 cm &6C  \\
        22DEC20 & DEEP-2 & 16 cm &6C  \\
        22DEC23 & DEEP-1 &  4 cm & 6C  \\
        22DEC29 & DEEP-2 & 16 cm &6C  \\
        22DEC24 & DEEP-1 &  16cm & 6C \\
        22DEC25 &  DEEP-1 & 16 cm & 6C \\
        22DEC27 & DEEP-1 &  16 cm & 6C \\
        22DEC28 &  DEEP-1 & 4 cm & 6C \\
        22DEC29 & DEEP-1 &  16 cm & 6C \\
        22DEC30 &  DEEP-1 & 16 cm & 6C \\
        23JEN26 &  DEEP-1 & 16 cm & 6C \\
        23DEC25 &  DEEP-1,2 & 16 cm & 6D \\
        24FEB29 &  DEEP-1 & 4 cm & 6A \\   
        24MAR14 &  DEEP-1 & 4 cm & 6A \\   
        24MAR18 &  DEEP-1,2 & 16 cm & 6A \\  
        24APR28 &  DEEP-2  & 4 cm & 6A \\
        24MAY03 &  DEEP-2 & 4 cm & 6A \\  
        24JUN09 &  DEEP-2 & 4 cm & 6D \\
        \hline

\end{tabular}

    \label{tab:observations}
\end{table}

\subsection{Observation, data reduction and imaging}

{In this section, we first describe the 2.1 GHz observations, which constitute the low-frequency component of the survey, and then the 5.5 and 9 GHz high-frequency observations.}
\subsubsection{2.1 GHz observations}
The {low frequency }observations span a 2 GHz bandwidth centered at 2.1 GHz and were run with the six 22-meter antennas of ATCA in the most extended E-W configurations and processed according to the procedures applied to all the other SHORES shallow fields, described in SHORES-I (see Tab. 1 of \citealt{shores} for the complete list of observing epochs, and a summary of it limited to the deep fields in Tab.\ref{tab:observations} below). PKS1934-638 was used as flux density and bandpass calibrator, and a set of bright sources (including PKS0008-421, PKS0008-264, PKS0118-272, PKS0426-380) were used as phase and polarization calibrators. 

We also split the 2.1\,GHz visibilities into four equally spaced sub-bands, centred at 1.3, 1.8, 2.3, and 2.9\,GHz, and applied the same calibration and imaging procedures described in SHORES-I. This allowed us, where sensitivity permitted, to recover a more refined characterisation of the spectral behaviour of the detected sources, as discussed later in this paper.
\subsubsection{5.5 and 9 GHz observations}

The central region ($\sim$0.5\,deg$^2$) of the deep fields was mapped in two Nyquist-sampled mosaics of 72 pointings each with the 4\,cm receivers, configured with two simultaneous basebands centred at 5.5 and 9.0\,GHz, respectively.{ Both the 5.5 and 9 GHz datasets were observed with a 2 GHz bandwidth.} Observations were carried out with the ATCA over several nights between October 2022 and June 2024, as detailed in Table \ref{tab:observations} (Projects C3502-CX542-C3605; PI: M. Behiri).

Calibration was performed according to standard procedures, using phase and polarisation calibrators such as PKS0023$-$263\footnote{Note that PKS0023$-$263 shows extended components at 9.0\,GHz at our sensitivity; we accounted for this and also used additional calibrators.}, PKS2357$-$318, PKS0118$-$272, and PKS2337$-$334.

\begin{figure*}
    \centering
    \includegraphics[width=\linewidth]{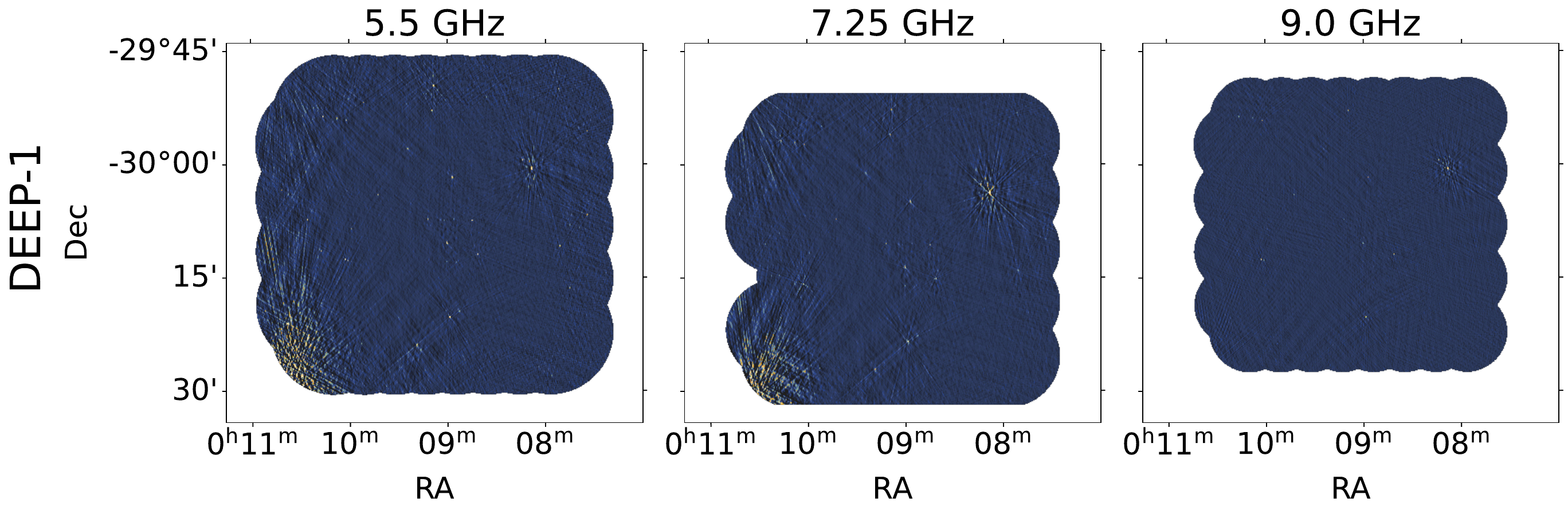}
    \includegraphics[width=\linewidth]{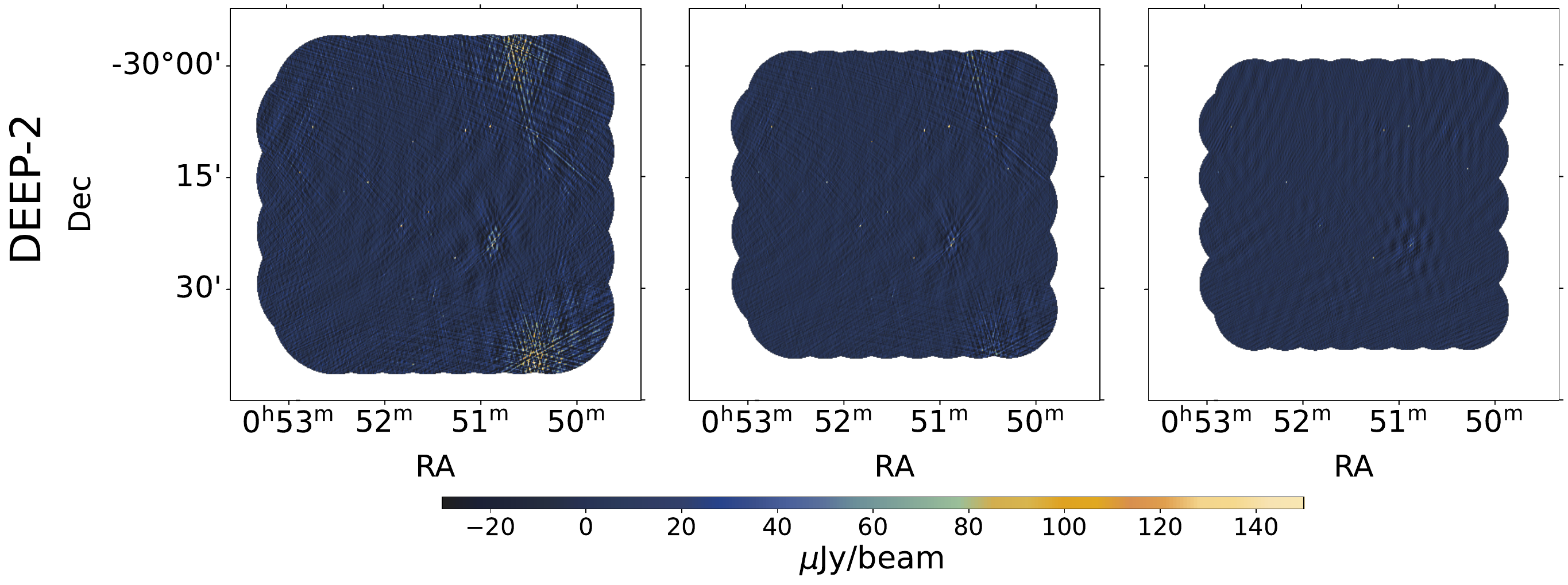}  
    \caption{Maps of the DEEP-1 (\textit{top}) and DEEP-2 (\textit{bottom}) at 5.5 (left), 7.25 (center), and 9.0 (right) GHz. Notice that the 7.25 GHz is the composed by multi-frequency synthesis of the other two frequencies, on which each 2 GHz baseband was centered. }
    \label{fig:maps5p5-9}
\end{figure*}

Maps of the calibrated data for each pointing were generated for each observed frequency and for the combined band (corresponding to a nominal frequency of 7.25 GHz) using the imaging software \textsc{WSClean} \citep{wsclean}, similarly to{ what was done} for the 2.1 GHz observations (SHORES-I).
We adopt a robust (Briggs parameter set to $0.5$) weighting scheme to strike a balance between artefact suppression and the best sensitivity, and we exploit the auto-masking feature to identify all emissions above $5\sigma$ significance level, hence perform a deep cleaning down to the noise level ($1\sigma$, autothreshold set to 1) over those regions. 

We finally combine the various pointings to a mosaic for each field, exploiting the \textsc{Miriad} task \textsc{LINMOS}.

\begin{table}
  \centering
   \caption{Properties of the maps.}
  \begin{tabular}{c|c|c|c|c|c}
    \hline
     &Frequency & Resolution & FOV & Area & Rms  \\ 
     & [GHz] & [arcsec]  &  & [deg$^2]$ &[$\mu$Jy/beam] \\ \hline
    \multirow{3}{*}{\rotatebox{90}{DEEP-1}} & 2.1 &  4.2$\times$9.4& 12.3' & 1.25 & 9 \\ 
     & 5.5 &  0.9$\times$5.87 &6.6'& 0.6 &39.1  \\
      & 7.25&  0.8$\times$7.4 &6.6'& 0.6 &28.5\\ 
     & 9.0 & 0.8$\times$3.8 & 5.3'& 0.43 & 37.9 \\ 
     \hline
    \multirow{3}{*}{\rotatebox{90}{DEEP-2}} & 2.1 &3.2$\times$9.8& 12.3' & 1.25 & 18.2 \\ 
       & 5.5 &  1.34$\times$4.16 &6.6'& 0.6 &55.3  \\ 
      & 7.25&  1.1$\times$3.7 &6.6'& 0.6 &39.5\\ 
     & 9.0 & 0.9$\times$3.9& 5.3'& 0.43 & 60.1 \\ 
    \hline
  \end{tabular}
    \begin{minipage}{0.95\linewidth}
    \footnotesize
    \textit{Notes} — 
    Frequency: central (or effective) observing frequency of each map; 
    $^{\dagger}$ 7.25 GHz is the effective frequency of the multi-frequency synthesis (MFS) image combining the 5.5 and 9.0 GHz basebands. 
    Resolution: synthesised beam FWHM. 
    FOV: primary-beam FWHM of a single pointing at the given frequency. 
    Area: total area of the map at that frequency. 
    Rms: median image rms noise in the mosaic.
  \end{minipage}

  \label{tab:map-properties}
\end{table}

Both fields host sources brighter than expected at the selection stage, whose effect is not negligible at our final sensitivity level. In particular, DEEP-1 contains two sources with a peak flux density of $\sim 2\,$mJy. This generates an appreciable noise pattern in the maps. We exploited the bright source to apply a self-calibration procedure to reduce it. 
Therefore, at $2.1\,$GHz we reached an rms of $\sim 9\mu\,$Jy/beam and of $\sim 18.2\mu\,$Jy/beam respectively for the two fields.

The same bright sources contaminating the 2.1 GHz maps happen to be at the edge of the mosaicked areas at 5.5 and 9.0 GHz (Figure \ref{fig:maps5p5-9}). We identified the closest pointing and derived improved gain solutions to refine the calibration of any other pointings affected nearby. This{ allows us to reach} a rather homogenous noise pattern in mosaicked maps. 

Table \ref{tab:map-properties} summarises the properties of the maps at all the reference frequencies.

{Following the SHORES-I pipeline \citep{shores}, we use PYSE exclusively to generate rms maps, as it provides stable noise estimates across mosaics with spatially varying sensitivity. Source extraction is performed with BLOBCAT, which was shown in \citealt{shores} to be more suitable than PySE or Aegean in ATCA multi-pointing mosaics affected by residual sidelobes, given our purposes.}

\begin{figure}
    \centering
        \includegraphics[width=\linewidth]{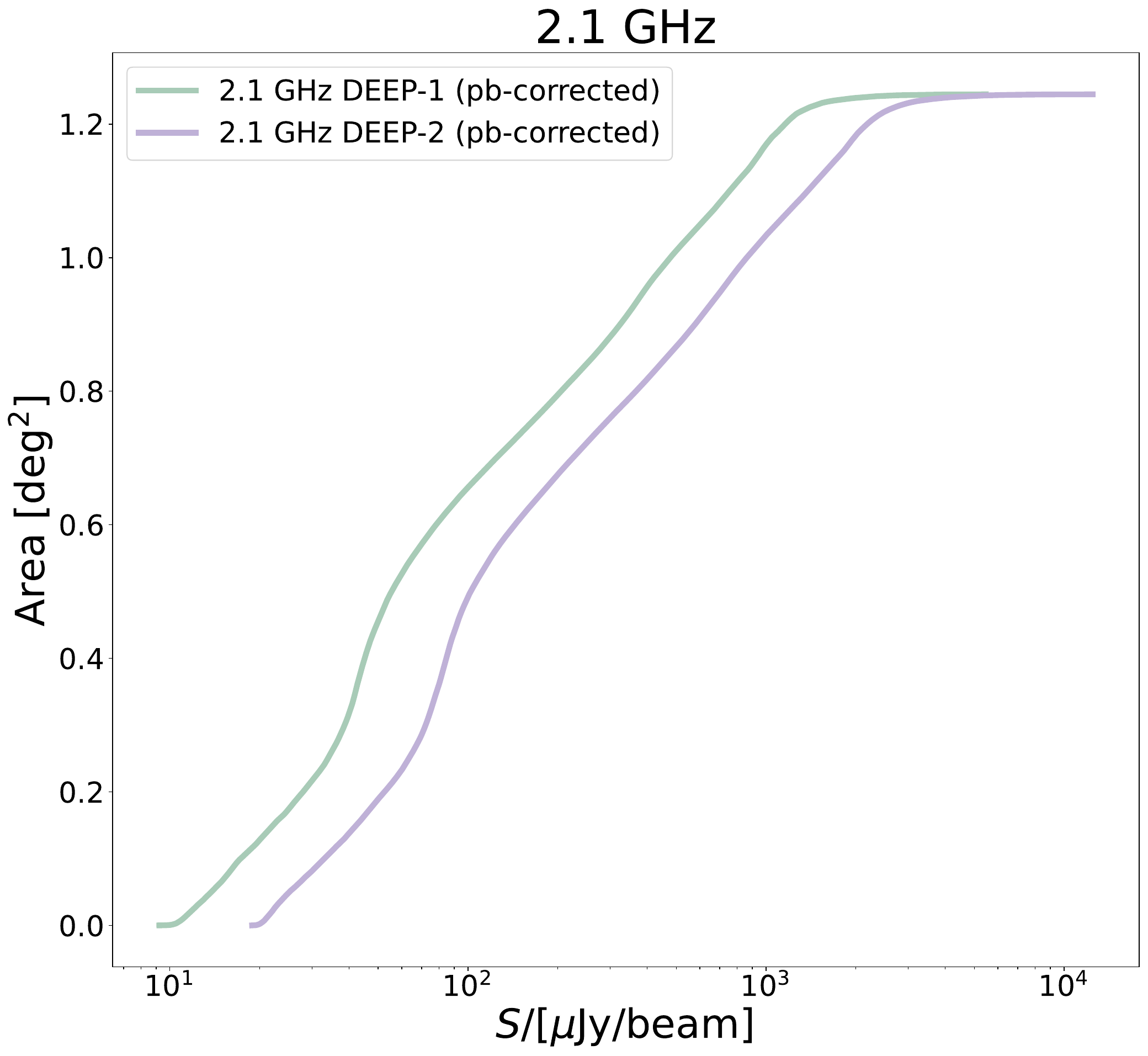} \quad\includegraphics[width=\linewidth]{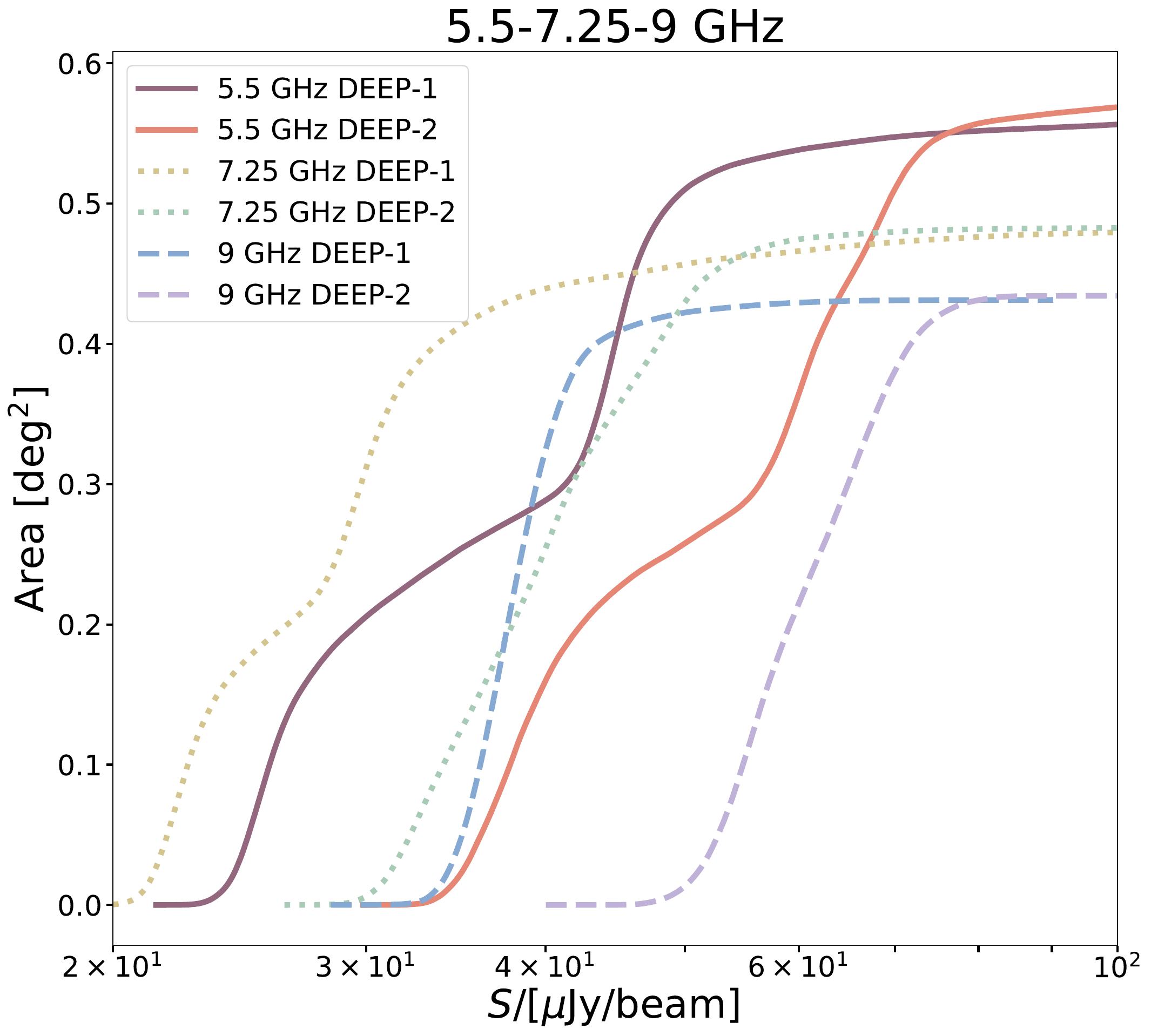}
    \caption{\textit{Upper}: Effective area curves at 2.1 GHz (DEEP-1/DEEP-2), with primary-beam correction. \textit{Lower}: Effective area at 5.5-7.25-9 GHz  for DEEP-1 and DEEP-2.}
    \label{fig:eff_area}
\end{figure}

Finally, Figure \ref{fig:eff_area} illustrates the visibility region as a function of rms, i.e. the cumulative area of the maps where the local noise is below a given rms threshold. A slight shoulder in the PB-corrected curves is visible: in the single 2.1 GHz PB-corrected map the edges are noisier than the centre, so the effective-area curve can show a small bump (see also \citealt{shores}). This effect is also present in the mosaics, where changes in pointing overlap create rings with similar sensitivity, making the bump more noticeable at 5.5 GHz, where the mosaicked area is wider.

\begin{figure}
    \centering
    \includegraphics[width=\linewidth]{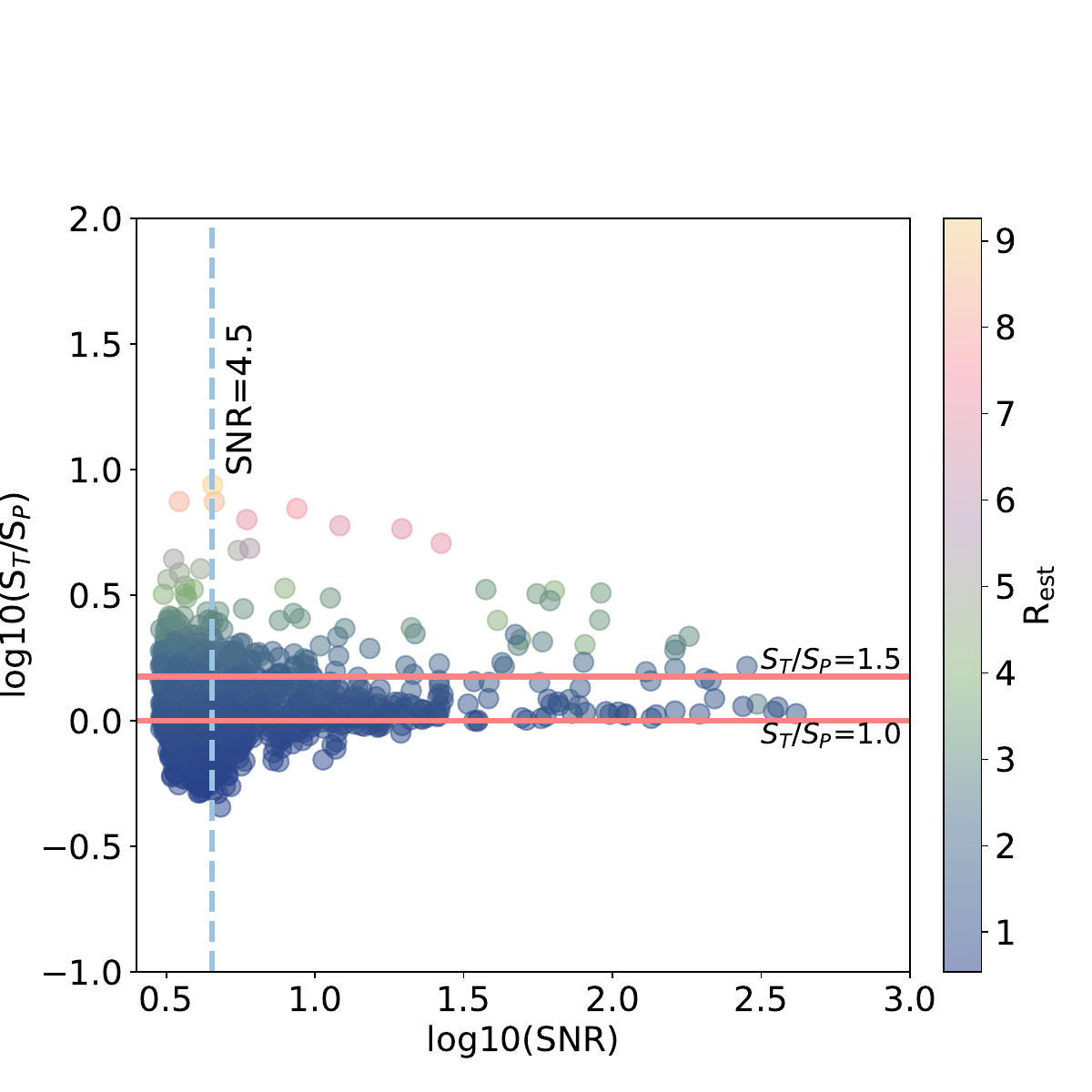}
    \caption{Ratio between total flux and peak flux as a function of the SNR for the SHORES sources.}
    \label{fig:stsp}
\end{figure}

\section{Source extraction and catalogues}\label{sec:catalogue}

\textsc{BLOBCAT} \citep{BLOBCAT} is based on the ``flood-fill'' algorithm and is specifically designed to perform efficient source extraction on radio survey data: we use it to identify candidate sources in the maps, following the procedure already tested in the SHORES shallow fields (SHORES-I). 

At $2.1\,$GHz, we perform the detections on the non-primary beam-corrected maps. The primary beam correction is applied \textit{a posteriori} using the polynomial form that was computed in SHORES-I which characterises the beam response out to the first null region.
The \textsc{MIRIAD} task \textsc{LINMOS} used to combine the pointings into the 5.5-9.0 GHz mosaics applies the primary beam correction (pb-correction). Thus, the detection maps are already pb-corrected.  In all the cases, we provide in input the rms maps produced by \textsc{PySE}.

At 2.1 GHz, \textsc{BLOBCAT} extracts a total of 1480 sources above 3$\sigma$, and 497 above 4.5$\sigma$. 
Consistent with \cite{shores} and the standard practice adopted in radio surveys (e.g. \citealt{franzen15,smolcic2017,hancock18,norris21}), all 4.5–5$\sigma$ candidates were visually inspected to identify artefacts associated with ATCA sidelobes and non-Gaussian noise features. Detections were rejected only when showing clear sidelobe morphology, inconsistent centroiding, or placement in regions of unstable rms. This conservative step ensures a clean low-SNR catalogue while maintaining consistency with the formal false detection rate.

After visual inspection, we confirmed 483 genuine sources, including 16 composite objects, i.e. sources composed of multiple blobs that were initially detected as separate components.

We use the \textsc{BLOBCAT}-generated parameter $R_{ext}$ to differentiate between unresolved and extended sources, classifying as extended those with $R_{ext}>1.4$. This criterion ensures that in point sources the total-to-peak flux density ratio, $S_{T}/S_{P}$, is lower than 1.15 (Figure \ref{fig:stsp}).
As a result, at SNR$>$4.5 we get 429 point-like and 68 extended sources (including the composite ones).

To compute the reliability of the deep fields we changed the sign to the map pixel values to build the negative map of each field and extract the sources using \textsc{BLOBCAT} again. Thus, we estimate the False Detection Rate (FDR), i.e. the ratio between the detections on the negative maps (false detections) and the detections on the positive maps (real detections) at a given flux density level. We define the reliability as $1-\rm{FDR}$. 

\begin{figure}[ht]
    \centering
    \includegraphics[width=\linewidth]{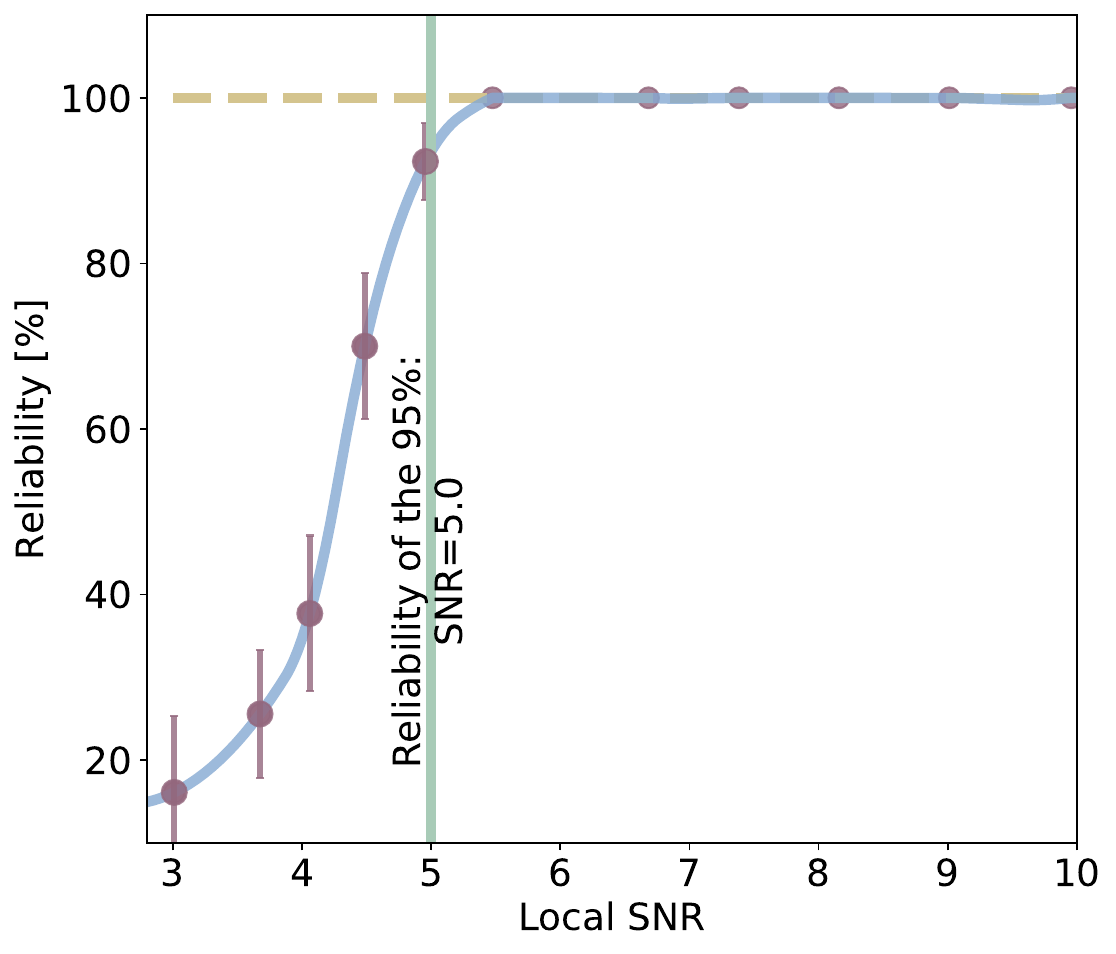}
    \includegraphics[width=\linewidth]{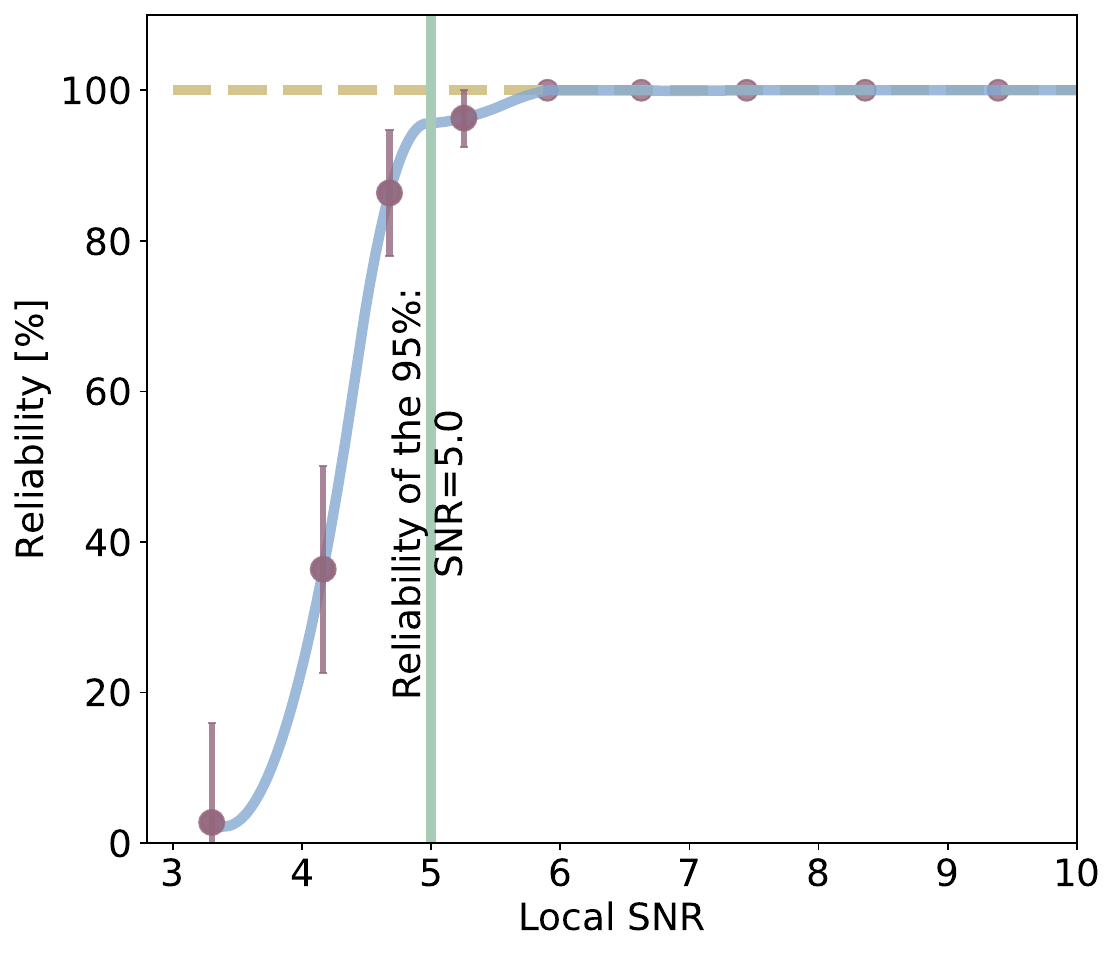}
    \caption{Reliability of the SHORES deep field (DEEP-1 on the top and DEEP-2 below) at 2.1GHz. The green solid line indicates $\rm{SNR}=5.08$ (DEEP-1) and  $\rm{SNR}=5.00$ (DEEP-2), for which the reliability is 95\%.}
    \label{fig:rel}
\end{figure}

The reliability saturates at 1 when the number of false detections is null at a given flux density level, corresponding to a given SNR. For both the SHORES deep fields, we obtain a reliability of 95\% at SNR$\sim$5.0 (Figure \ref{fig:rel}).

At SNR$\sim$4.5 the reliability is $\sim$ 85\% for both fields. 
We use this threshold combined with the visual inspection, as was already done for the shallower fields \citep{shores}, to identify the sources that can be included in our catalogue and used for scientific analysis, if not stated differently (e.g. Section \ref{sec:counts}). We also verify the presence of counterparts at other wavelengths to strengthen the hypothesis that sources with $4.5<\rm{SNR}<5.0$ are also true (therefore, the catalogue reliability is indeed higher than $85\%$). {The FDR quantifies the intrinsic statistical reliability, following the framework of \cite{shores}. Cross-identifications do not modify the formal FDR but provide external support for the adopted selection of 4.5–5$\sigma$detections, which are retained only after the quality checks described above.}

To calculate our completeness we perform a 10,000 simulation placing a single point source with known flux density in random positions across the negative maps using the \textsc{MIRIAD} task \textsc{IMGEN}, and calculate the percentage of input sources detected with \textsc{BLOBCAT} at a given flux density. A survey is 100$\%$ complete above a limiting flux $S_{lim}$, if it is possible to recover all the input sources with flux density $S>S_{lim}$.
For the SHORES deep fields, we obtain 95\% completeness at 182.3 $\mu$Jy (SNR$\sim$20) for DEEP-1 and at 198.1 $\mu$Jy (SNR$\sim$10.9) for DEEP-2.

\begin{figure}[ht]
    \centering
    \includegraphics[width=\linewidth]{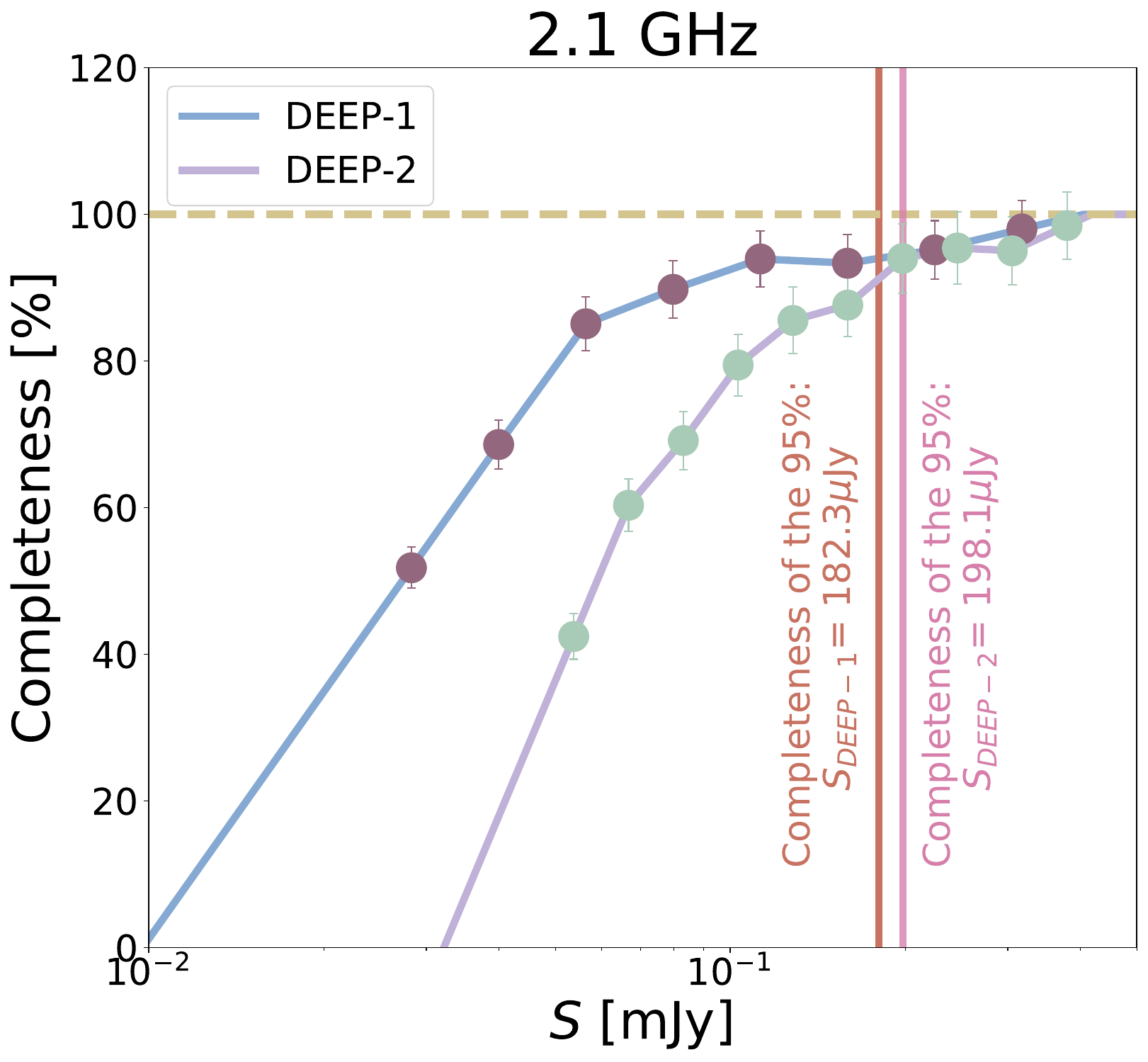}
    \caption{Completeness of the SHORES deep fields at 2.1GHz. The solid vertical lines indicate the flux $S$ at which the completeness is $95\%$.}
    \label{fig:enter-label}
\end{figure}

Furthermore, we performed source extraction with \textsc{BLOBCAT} also on the maps of the $2.1\,$GHz sub-bands, corrected with the SHORES-I polynomial expression at the respective frequencies.

\begin{figure*}[ht]
    \centering
\includegraphics[width=0.45\linewidth]{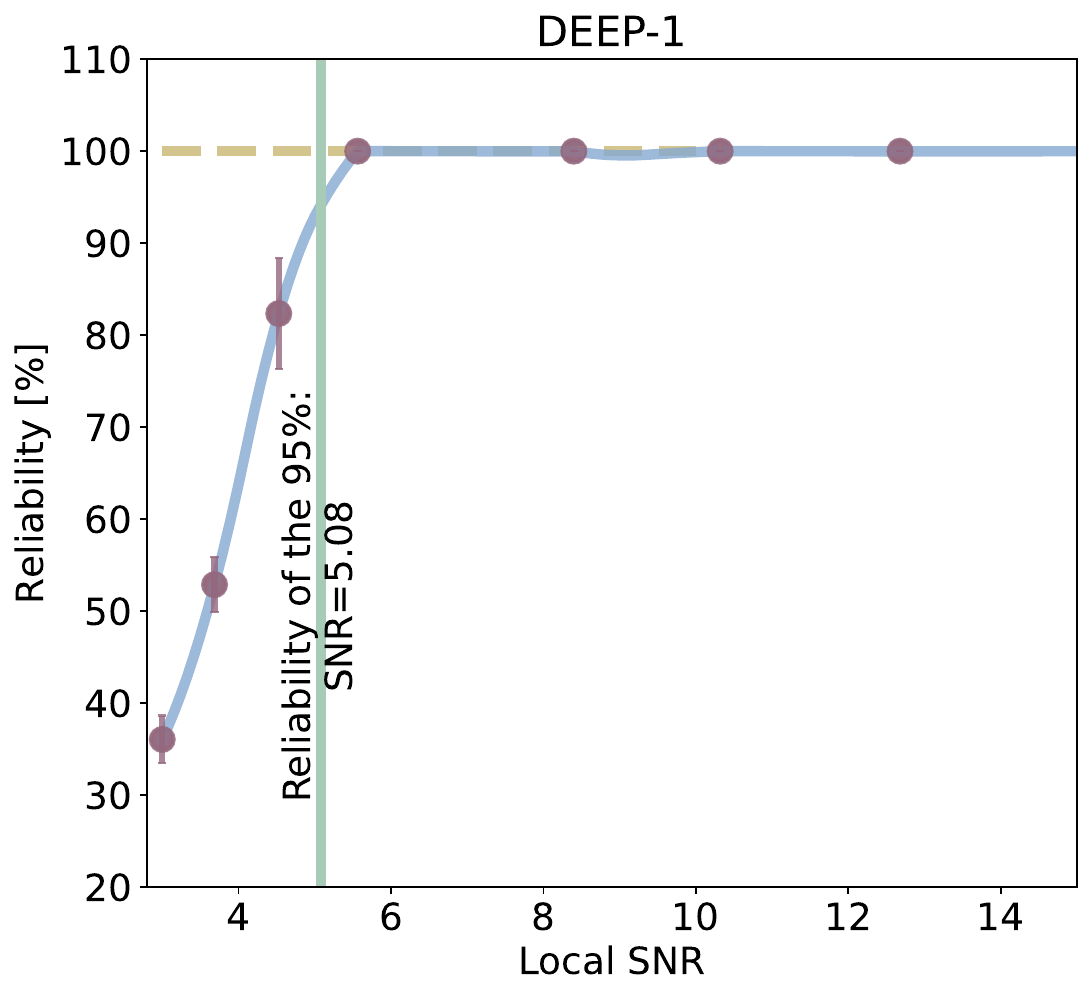}
\includegraphics[width=0.45\linewidth]{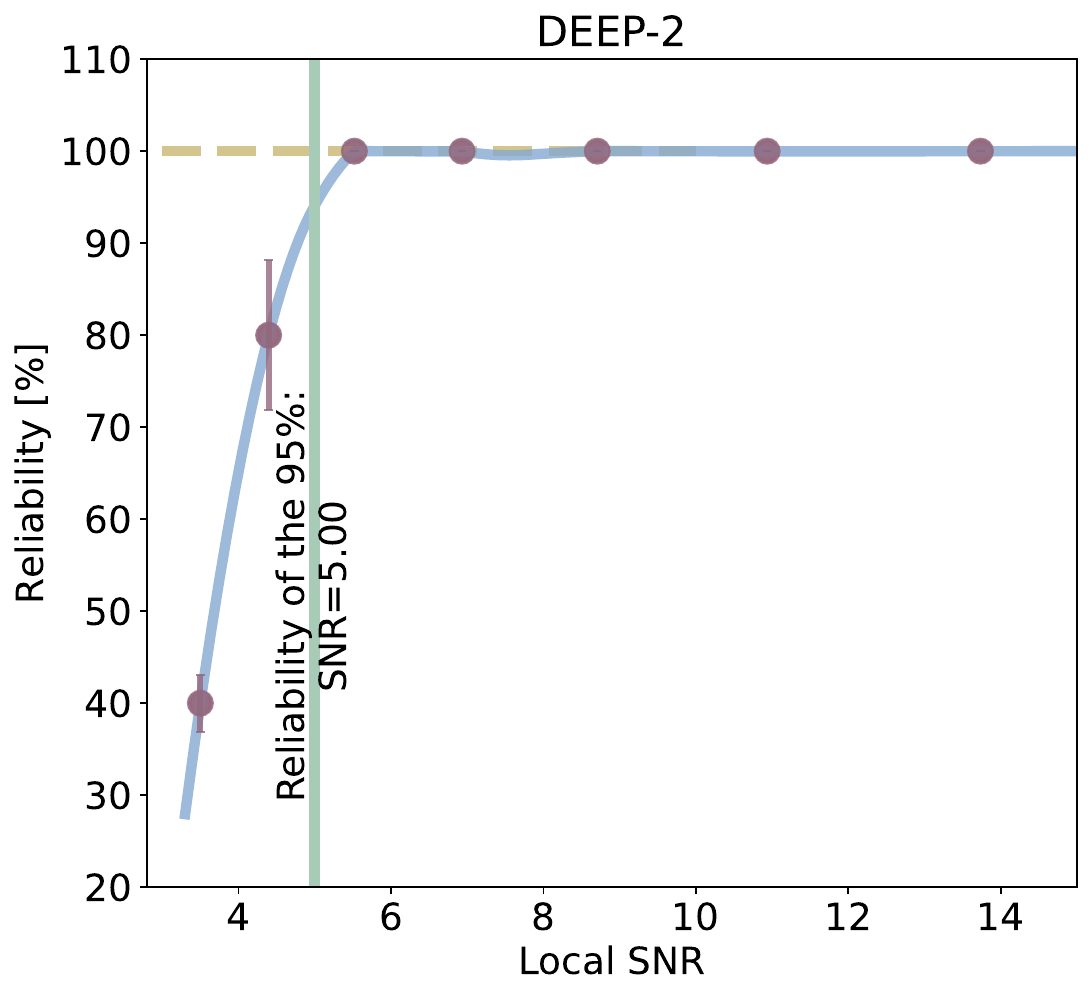}\quad
\includegraphics[width=0.45\linewidth]{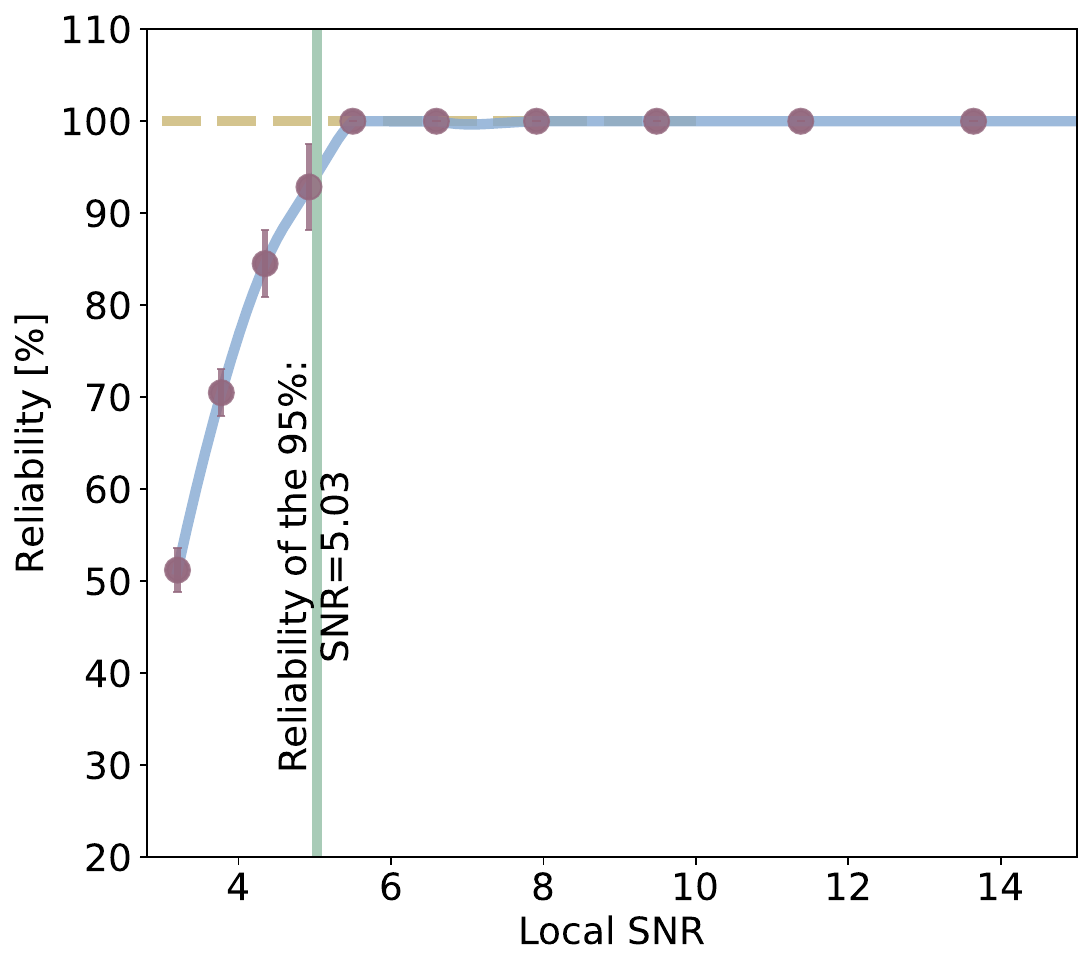}
\includegraphics[width=0.45\linewidth]{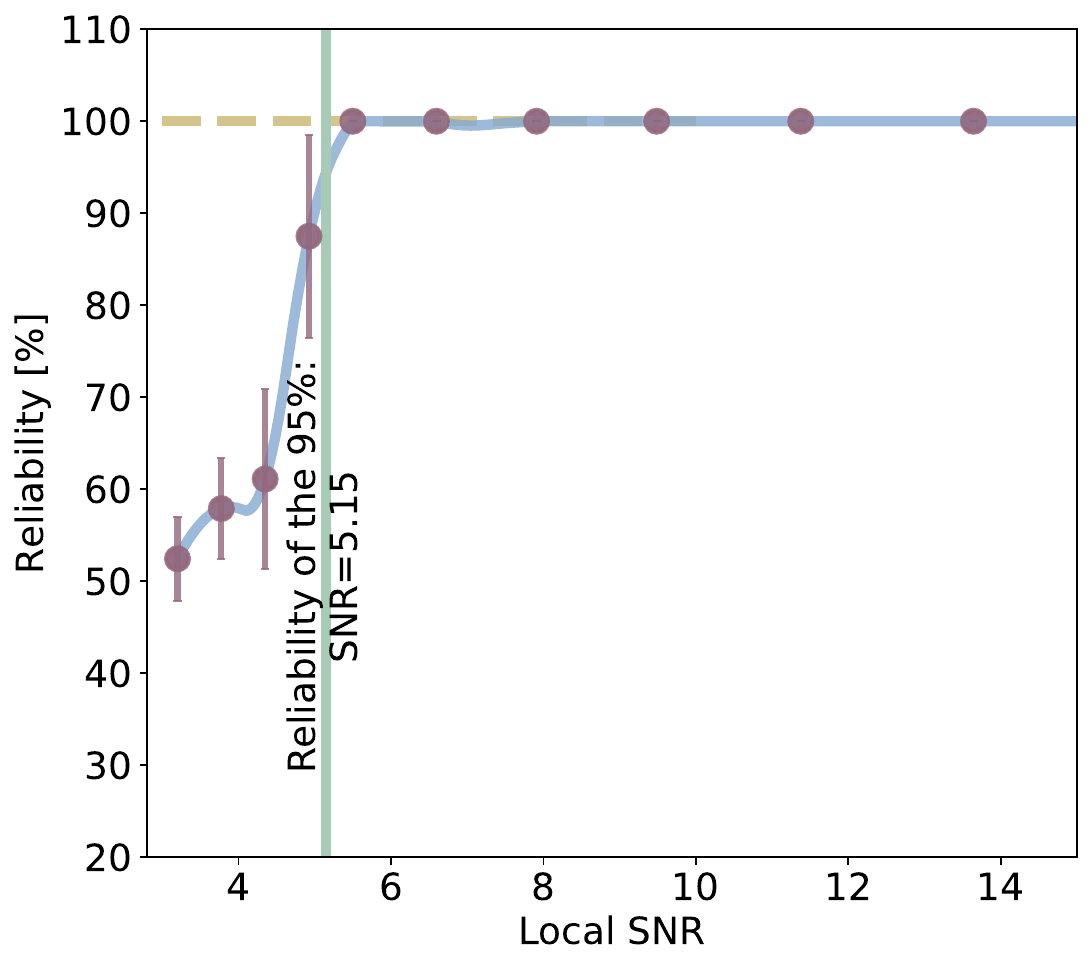}\quad \includegraphics[width=0.45\linewidth]{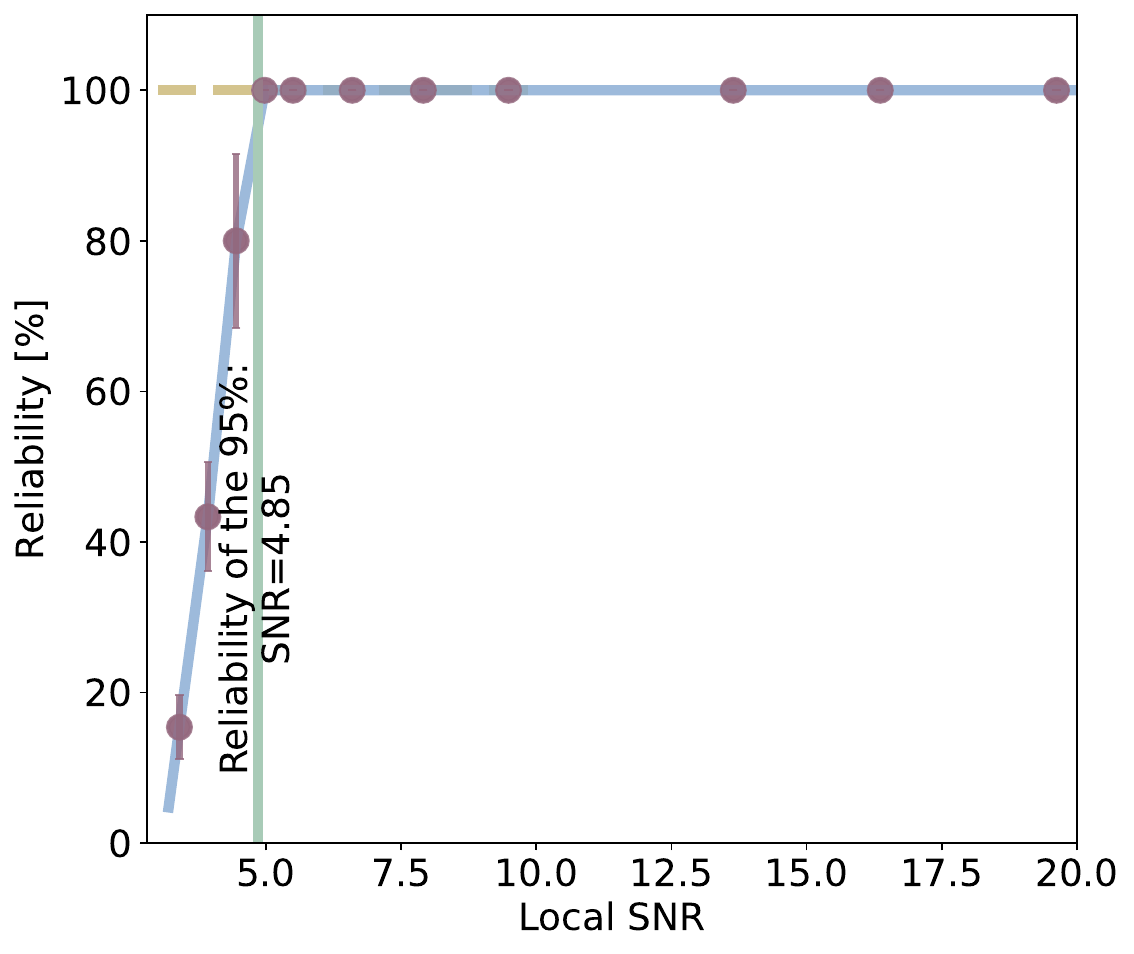}
\includegraphics[width=0.45\linewidth]{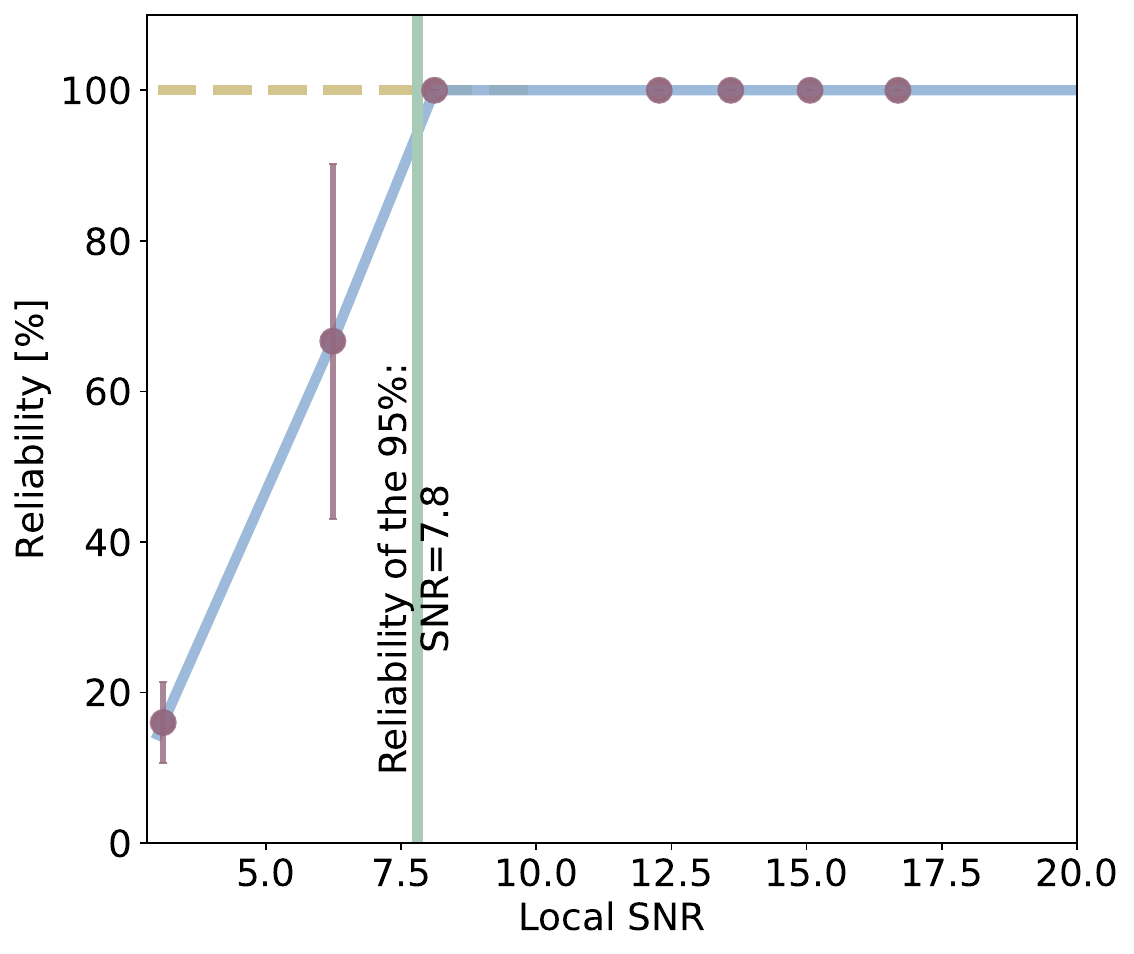}
    \caption{Reliability of the SHORES deep fields (DEEP-1 on the \textit{left} and DEEP-2 on the \textit{right}) at 5.5, 7.25 and 9.0 GHz. The green solid line indicates the SNR for which the reliability is $95\%$.}
    \label{fig:rel_5p5}
\end{figure*}

Similarly to {what was done} at 2.1 GHz, we blindly extracted sources from the 5.5 GHz and 9 GHz mosaics, and from the combined 7.25 GHz map with \textsc{BLOBCAT}.

We perform a cross-match between the mosaics and the $2.1\,$GHz catalogues using a radius of 3'', i.e. the lower limit of the resolution at 2.1 GHz, to identify counterparts. 

After visual inspection, we obtain:
\begin{itemize}
    \item At 5.5 GHz: 101 sources with SNR $> 4.5\,\sigma$, all of which have counterparts at $2.1\,$GHz. Among them, two are classified as composite sources.
    \item At 9.0 GHz: 33 sources with SNR $> 4.5\,\sigma$, all of which have counterparts at $2.1\,$GHz. Among them, one is classified as a composite source.
    \item{
In the composite 7.25\,GHz map we detect 85 sources with SNR $>4.5\sigma$, all matched at 2.1\,GHz; none is composite. Notably, 29 of these are recovered \emph{only} in the 7.25 GHz MFS image and not in the 5.5 and 9 GHz, i.e. at an even higher frequency, thanks to the improved sensitivity obtained by combining the two IFs (5.5 and 9.0\,GHz).}
\end{itemize}

We estimated the reliability and completeness of the mosaics, similarly to what is described in the previous sub-section (see Figure ~\ref{fig:rel_5p5}). 
The 95\% reliability threshold at 5.5 GHz is reached at an SNR of approximately 5.1 for DEEP-1 and 5.0 for DEEP-2. At 9 GHz, the same reliability level is achieved at $\rm{SNR}=4.8$ for DEEP-1 and $7.8$ for DEEP-2. The worse performance of DEEP-2 at 9\,GHz is driven by its less homogeneous $uv$-coverage (fewer integrations) compared to DEEP-1, which makes the mosaic less effective at mitigating pointing undersampling at this frequency (where the primary beam is smaller), resulting in a more structured noise pattern.

\section{Radio spectral behaviour} \label{sec:prop}
\begin{figure*}
  \centering
  \includegraphics[width=0.33\linewidth]{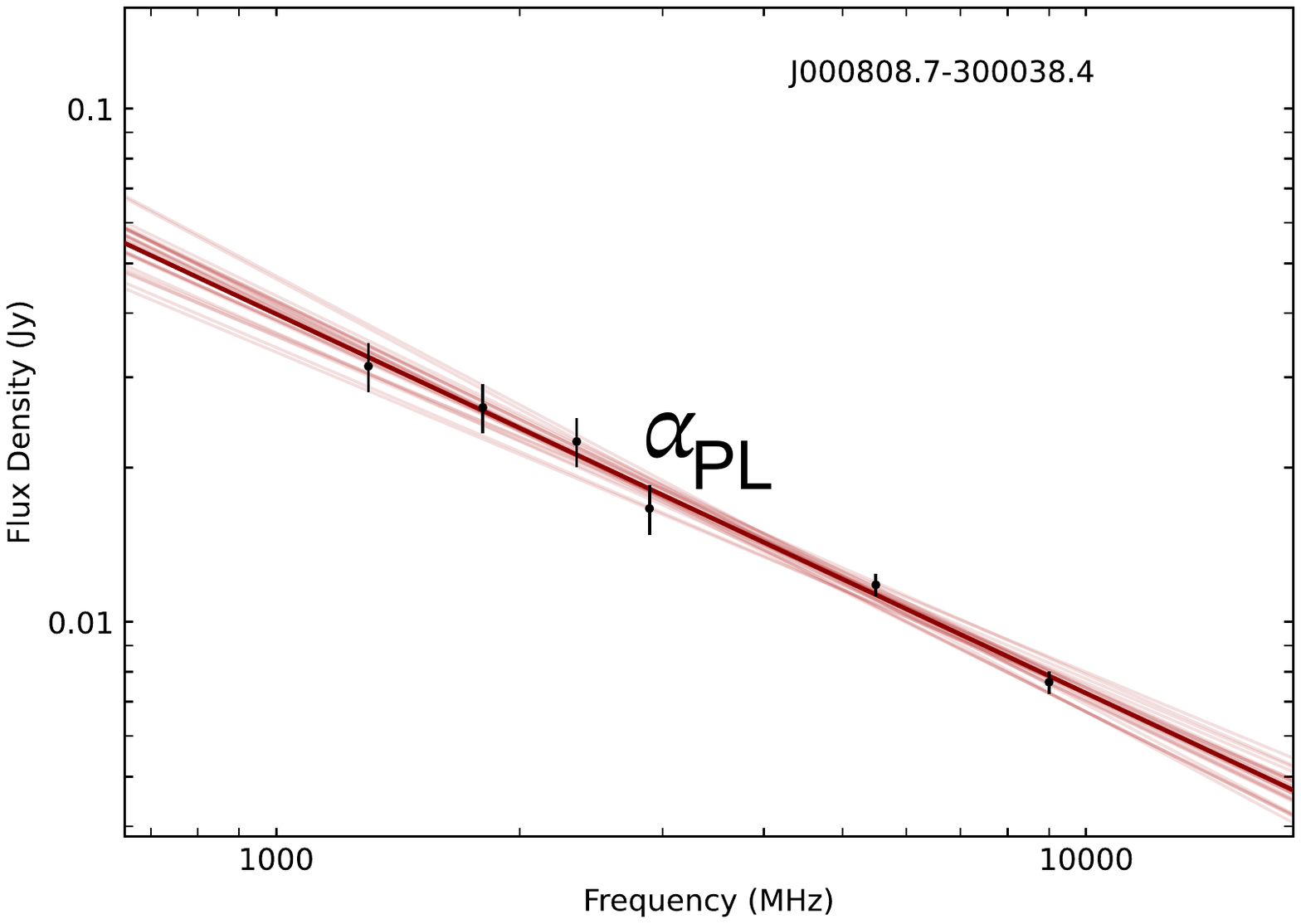}\hfill
  \includegraphics[width=0.33\linewidth]{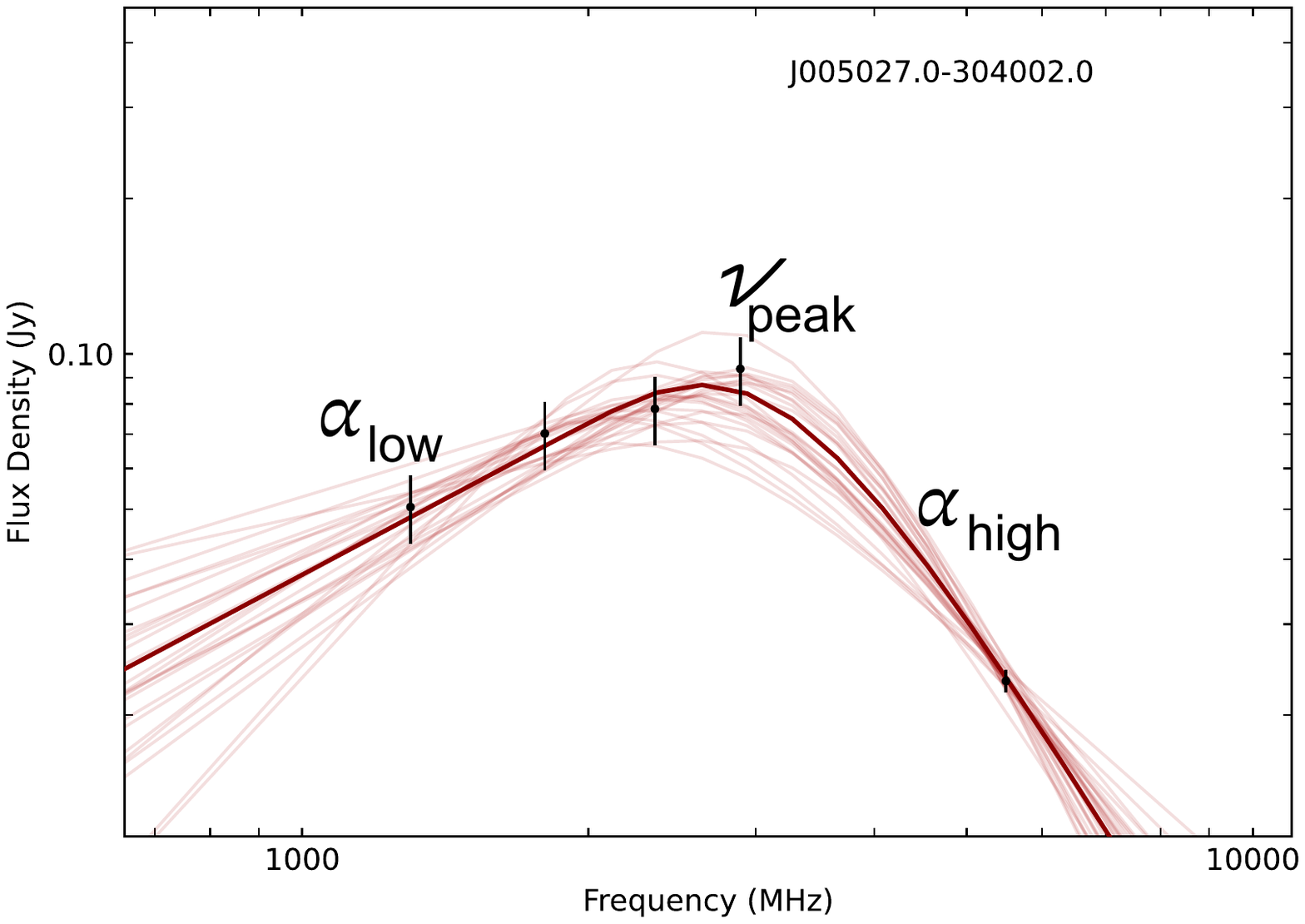}\hfill
  \includegraphics[width=0.33\linewidth]{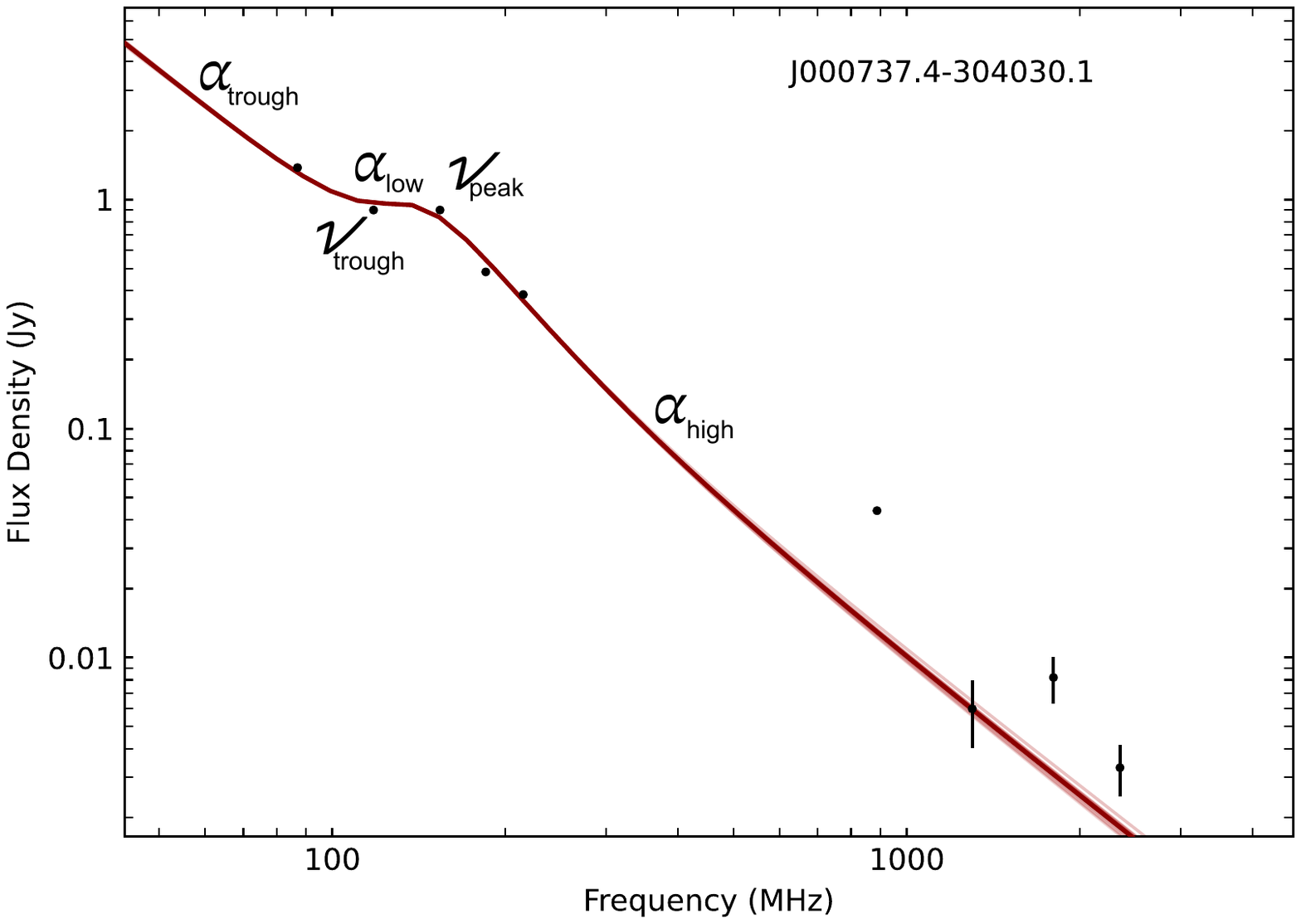}
  \caption{Example radio SEDs illustrating the fitted models and the \emph{trough/low/high} convention. 
  Points show the measured flux densities in the 0.1-10 GHz range.
  The best-fitting model is shown as a bright solid curve, with the different spectral index and frequency nomenclature highlighted by labels, accordingly. 
  Shaded bands represent the 68\% credible region from the \textsc{RadioSED} posterior.}
  \label{fig:sed-examples}
\end{figure*}
\begin{figure}[h]
    \centering
\includegraphics[width=\linewidth]{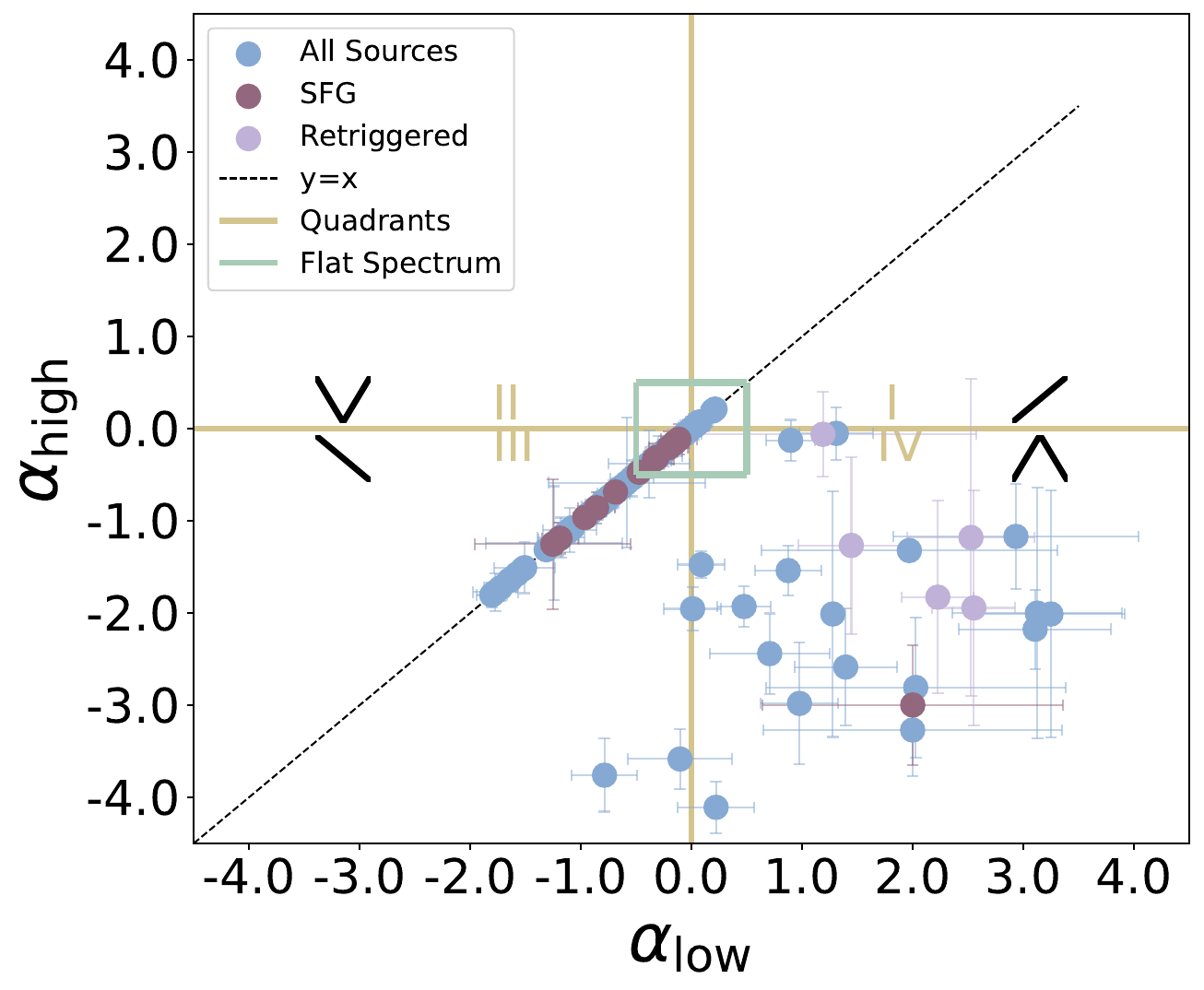}
    \caption{Colour–colour diagram of $\alpha_{\rm low}$ vs.\ $\alpha_{\rm high}$ from SHORES-only fits. 
SFGs (radio-normal) are identified via the FIR–radio correlation (Section \ref{sec:FIRRC}): sources with $q_{\rm FIR}\ge 1.69$ (the lower $1\sigma$ bound of the \citealt{ivison10} relation, in our convention) are highlighted as \textit{maroon circles}. 
Retriggered sources require an additional low-frequency index, $\alpha_{\rm trough}$, defined below the SED minimum $\nu_{\rm trough}$; in this diagram we show $(\alpha_{\rm low},\alpha_{\rm high})$ measured for the post-minimum branch ($\nu>\nu_{\rm trough}$). Axes follow the low/high convention defined in Section \ref{sec:prop_SHORES}.The green squared area indicates the locus of the flat-spectrum sources.} 
    \label{fig:ccplot-SHORES}
\end{figure}

\begin{figure}[h]
    \centering
\includegraphics[width=\linewidth]{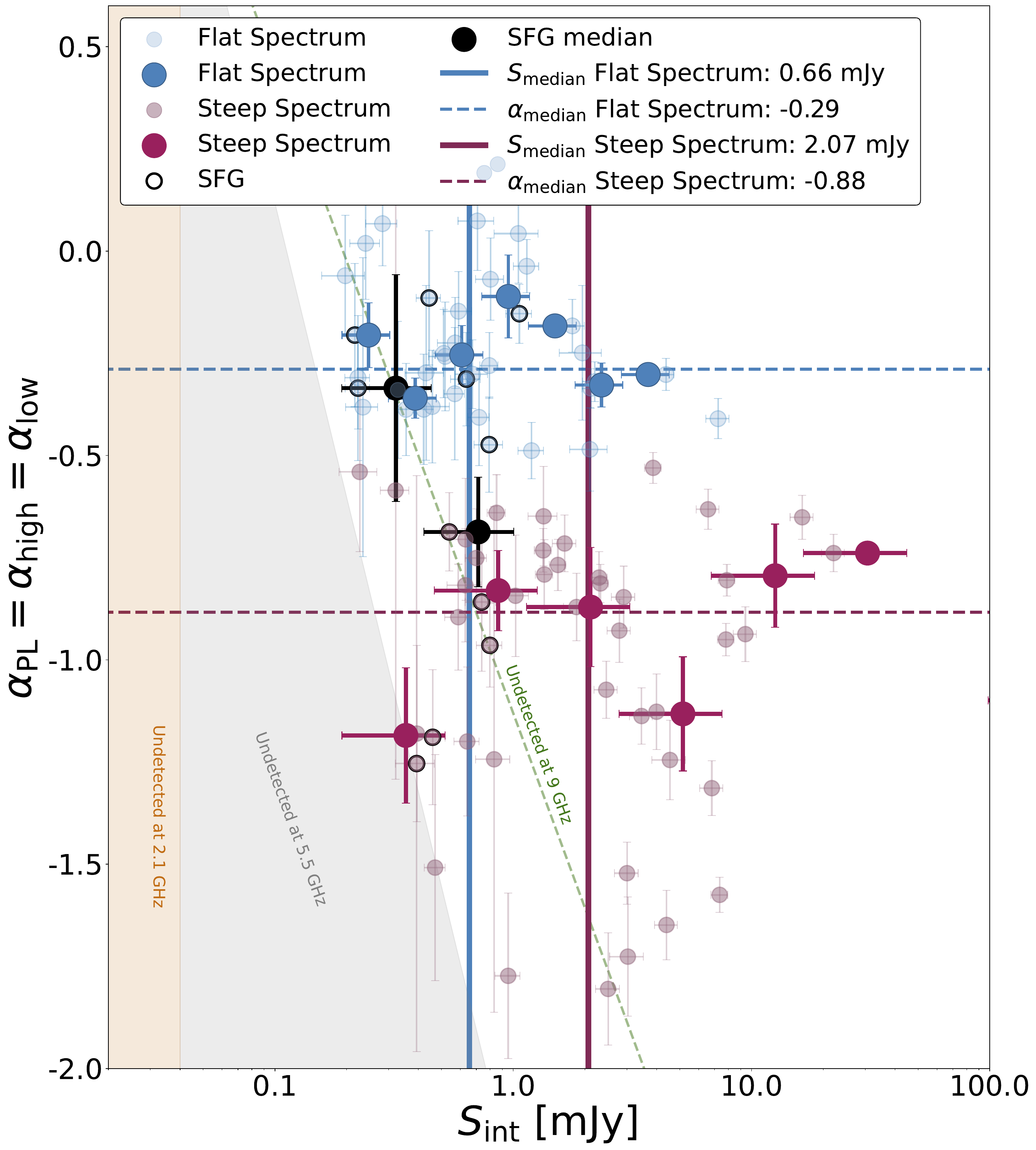}
    \caption{Distribution of the spectral indices of sources that show flat or steep spectral indices across the SHORES bands as a function of their 2.1 GHz flux density. {The shaded regions mark the parameter space inaccessible due to sensitivity limits, indicating where a source, despite being detected at 2.1 GHz, would fall below the detection limit at the comparison frequencies (5.5/9 GHz) due to its spectral shape, resulting in an undefined spectral index.}}
\label{fig:alpha_S_SHORES}
\end{figure}

\subsection{Spectral behaviour in the SHORES bands} \label{sec:prop_SHORES}

By matching the detections in our catalogues, we reconstructed the spectra of our sources across the 2.1–9.0\,GHz range. Where sensitivity permitted (see Section \ref{sec:catalogue}), the 2.1 GHz sub-bands provided up to six frequency points to define the spectra. In addition, we used the 7.25\,GHz composite MFS map to increase the likelihood of high-frequency counterparts for the faintest 2.1 GHz sources, as it reduces the noise of the 5.5 and 9.0 GHz maps by the expected factor of $\sqrt{2}$.

In total, we obtained 18 sources with only three spectral points at the main observed frequencies, and 96 sources with at least four points, of which 39 have six.

For all sources we used \textsc{RadioSED} \citep{radiosed}, a Bayesian radio SED-fitting tool that implements power-law, peaked, curved, and retriggered-AGN models to reconstruct the radio SED. We adopt the \emph{low/high} convention: $\alpha_{\rm low}$ denotes the slope in the low-frequency branch \emph{below} the spectral peak $\nu_{\rm peak}$ (typically attributed to the self-absorbed regime), while $\alpha_{\rm high}$ denotes the high-frequency slope \emph{above} $\nu_{\rm peak}$ (usually considered optically thin). 

The main SED parameters returned by the fitting ( Fig.~\ref{fig:sed-examples}) are:
\begin{itemize}
    \item $\nu_{\mathrm{trough}}$: the frequency of the SED minimum in the \textit{retriggered} model; if present, the spectrum rises for $\nu>\nu_{\rm trough}$.
    \item $\alpha_{\mathrm{trough}}$: the spectral index at $\nu<\nu_{\mathrm{trough}}$ (typically constrained by sub-GHz points in our sample).
    \item $\nu_{\mathrm{peak}}$: the frequency of the SED peak in \textit{peaked}/\textit{curved}/\textit{retriggered} solutions.
    \item $\alpha_{\mathrm{low}}$: the spectral index across the low-frequency branch, i.e. for $\nu_{\rm trough}<\nu<\nu_{\rm peak}$ (or, in the absence of a trough, immediately below $\nu_{\rm peak}$).
    \item $\alpha_{\mathrm{high}}$: the spectral index across the high-frequency branch, i.e. for $\nu>\nu_{\rm peak}$.
\end{itemize}

To illustrate our convention, Fig.~\ref{fig:sed-examples} shows three representative radio SEDs with their best-fitting models: a single power law, a peaked spectrum, and a retriggered spectrum. For clarity, we highlight in each model the trough-frequency (trough), the low-frequency (low) and the high-frequency (high) branches of the model, corresponding to $\alpha_{\rm low}$ (below $\nu_{\rm peak}$) and $\alpha_{\rm high}$ (above $\nu_{\rm peak}$), and mark $\nu_{\rm trough}$ and $\nu_{\rm peak}$ where present. Here and throughout, the term \emph{components} refers to the parametric branches of the fitted model, and should not be interpreted as necessarily distinct spatial emission regions.

As shown in the $\alpha_{\rm high}-\alpha_{\rm low}$ radio colour-colour diagram in Figure \ref{fig:ccplot-SHORES}, for the majority of sources (89 over 114)  the best-fit model is a single power law, i.e. $\alpha_{\rm high}=\alpha_{\rm low}\equiv\alpha_{\rm PL}$: 42.1\% are classified as steep ($\alpha_{\rm PL}\le -0.5$) and 33.3\% as flat ($|\alpha_{\rm PL}|\lesssim 0.5$), while no truly inverted spectra ($\alpha_{\rm PL}>0.5$) are found. Furthermore, the 18.4\% exhibit peaked spectra, where $\alpha_{\rm low}$ and $\alpha_{\rm high}$ have opposite signs, indicating a clear turnover in the synchrotron SED. Two sources (1.8\%) show curved but non-peaked spectra, suggestive of gentle curvature without a well-defined peak. Finally, 4.4\% (5 objects) are best fit by the \textit{retriggered} model, which describes spectra with a ‘valley’ followed by re-brightening: for these we use three indices—$\alpha_{\mathrm{trough}}$ (before the minimum), $\alpha_{\mathrm{low}}$ (between the minimum and $\nu_{\rm peak}$), and $\alpha_{\mathrm{high}}$ (above $\nu_{\rm peak}$).

Compared to the analysis at higher flux densities in the same frequency regime by \citet{galluzzi17} (their Table 2), the fraction of steep-spectrum sources at SHORES depths appears slightly higher but still comparable (30\% vs.\ our 34\%).

Sources peaking above 2 GHz (i.e. those in the fourth quadrant of Fig.~\ref{fig:ccplot-SHORES}) are likely AGN-dominated, with emission enhanced by jetted components. In contrast, flat- and steep-spectrum sources comprise a mix of boosted quasar-like cores, optically thin lobe emission, and star-formation–related radio emission. Disentangling these contributions with $\sim$GHz data alone is challenging.

By plotting the distribution of spectral indices for flat- and steep-spectrum sources as a function of their 2.1 GHz flux density (Figure~\ref{fig:alpha_S_SHORES}), we find that the two populations occupy distinct flux regimes. Steep-spectrum sources dominate at brighter flux densities, consistent with powerful AGN. This is partly due to selection effects: sources with rapidly declining spectra are harder to detect at high frequency, causing faint steep-spectrum sources to be under-represented.

Interestingly, the median spectral index flattens around $\sim$1\,mJy before steepening again at lower fluxes. If observational bias were the only driver, a monotonic flattening with decreasing flux would be expected. The observed turnover instead points to a population transition: at intermediate fluxes, the emergence of flat-spectrum cores (e.g. low-luminosity AGN) flattens the average spectral index, while at the faintest fluxes a steepening trend reappears, likely driven by starburst galaxies and possibly high-redshift AGN. These results align with similar behaviours reported in previous studies \citep[e.g.][]{prandoni06}.

We emphasise that all flux densities were cross-checked for consistency across bands, and no evidence of calibration offsets was found that could artificially affect the observed spectral trends (see \citealt{shores} and Section \ref{sec:prop_radio}).

Flat-spectrum sources, by contrast, cluster at the faint end and are more readily detected across our frequency range. Their distribution shows large scatter below $\sim$2 mJy but no clear trend with flux. No bright flat-spectrum AGN are found in the SHORES deep fields. Although this might indicate a real scarcity of such sources at these depths, we note that our spectral analysis relies on the availability of at least 3 photometric detections, which could introduce selection effects against flat-spectrum sources with low-SNR in some bands. At these depths, flat indices likely arise from compact AGN cores or radio-quiet quasi-stellar objects (QSOs), possibly mixed with steeper star-forming components. Such composite systems have been reported in sub-mJy samples \citep[e.g.][]{delvecchio17, smolcic2017, whittam22}.

In summary, the faint population shows a non-monotonic spectral trend, with a flattening at intermediate fluxes and a steepening at the faint end. This likely reflects a mix of flat-spectrum AGN and star-forming galaxies. Broader spectral sampling, especially at sub-GHz frequencies, will help disentangle these contributions more robustly.

\subsection{Spectral behaviour in the 0.1-10 GHz regime} \label{sec:prop_radio}

We reconstructed the SED across the radio domain for our sources by including the available radio ancillary data in the SED fit. In particular, we cross-matched our sources to identify counterparts in the Rapid ASKAP Continuum Survey (RACS, \citealt{racs1,racs2}) and in the Galactic and Extra-Galactic All-Sky MWA Extended Survey (GLEAM-X,\citealt{gleamx}).

RACS covers the whole sky at $-90<\delta<49$ at $\sim$887.5/943.5 (RACS-low,\citealt{racs1}), 1367.5 (RACS-mid, \citealt{racs2}) and 1655.5 MHz (RACS-high), reaching a sensitivity of $\sim$0.25 (RACS-low) and 0.2 mJy/beam (RACS-mid, high) with a resolution of $\sim$ 15$^{\prime\prime}$ (RACS-low), 10$^{\prime\prime}$ (RACS-mid) and 8$^{\prime\prime}$ (RACS-high). 
We find that 138 of our sources at SNR$>$4.5 a RACS-low counterpart within a radius of 15$^{\prime\prime}$, while 115 of the sources at SNR$>$4.5 have a RACS-mid counterpart and 143 in RACS-high

\begin{figure*}[ht]
    \centering
\includegraphics[width=\linewidth]{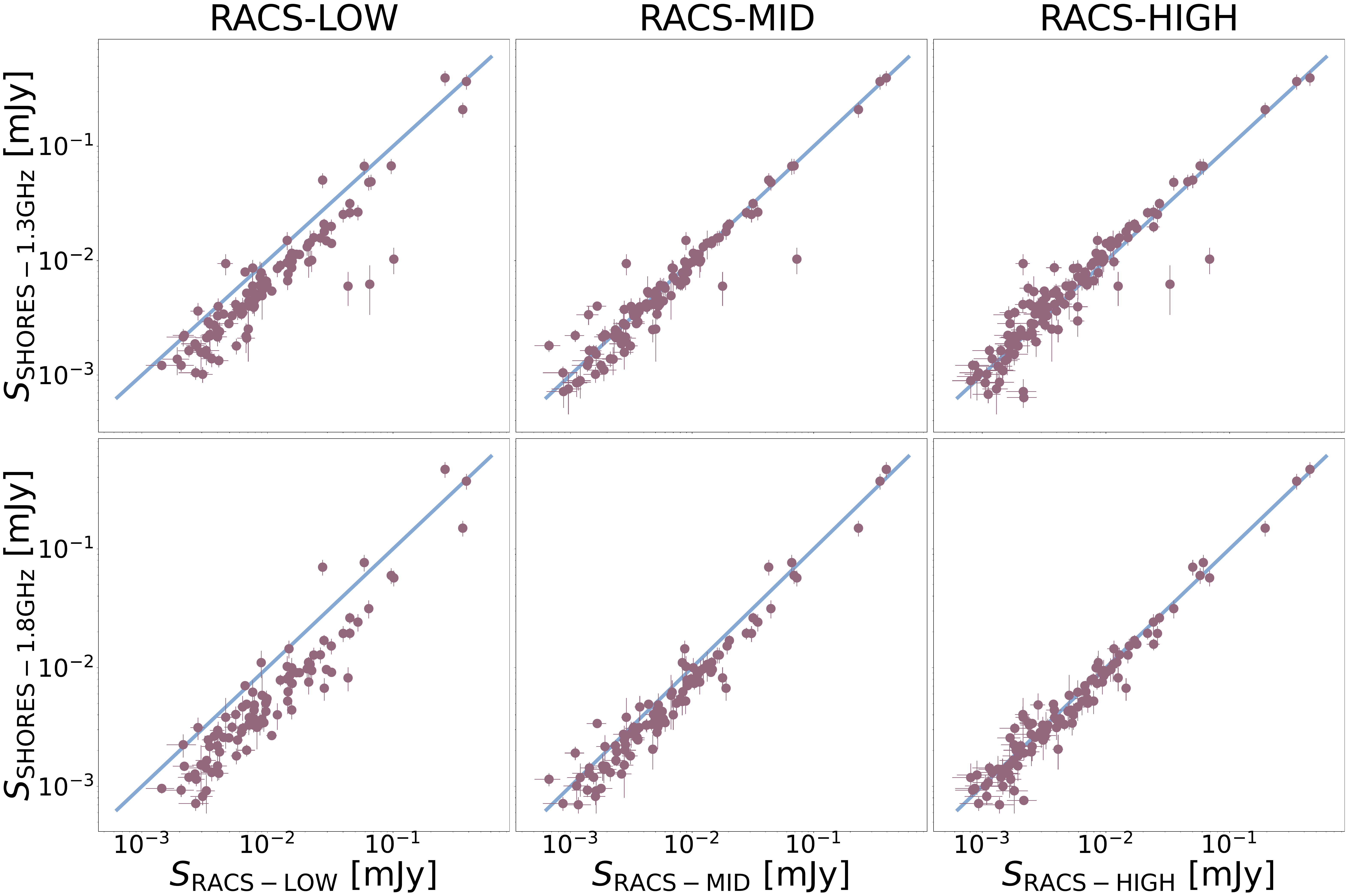}
    \caption{Comparison of SHORES 1.3 and 1.8 GHz sub-bands flux densities with RACS-mid, RACS-low and RACS-high ones. }
    
    \label{fig:flux-SHORESvsRACS}
\end{figure*}

We compared the RACS-mid and -high flux densities with those of our lowest sub-bands to confirm our findings (see Figure \ref{fig:flux-SHORESvsRACS}). 
Note that the systematic offset with RACS-low is expected.
{The flux density differences with RACS-low are fully consistent with the expected spectral behaviour between 900 MHz and our observing bands. When comparing RACS-low with our 1.3 GHz measurements, closer in frequency to 900 MHz, the discrepancy is significantly reduced. 
In contrast, the closer observing frequencies of RACS-mid (1367.5 MHz; $\approx$10$^{\prime\prime}$) and RACS-high (1655.5 MHz; $\approx$8$^{\prime\prime}$) yield a much tighter agreement with the SHORES flux scale.}

GLEAM-X covers an area of 1447 deg$^2$ around the South Galactic Pole region at $-32.7<\delta<-20.7$ with a median sensitivity $6.35$mJy/beam and a median resolution of $45^{\prime\prime}$ \citep{gleamx,mwa} with a broad spectral extension, spanning twenty frequency bands in the range 72 to 231 MHz, complementing the spectral coverage towards the MHz side of the radio SED, where self-absorption processes take place. To match the statistic load of the GLEAM-X points in the fit with that of the higher frequencies, we average them in three broad band points. We cross-match our catalogue with GLEAM-X using a $30^{\prime\prime}$ radius and find that 123 of our sources at SNR$>$4.5 have a GLEAM-X counterpart. {Because our synthesized beam ($\sim$4'') is much smaller than the GLEAM-X beam ($\sim$45''), the main cross-matching concern is blending (multiple SHORES sources falling within a single GLEAM beam) rather than spurious matches. Using the confirmed sources per field (DEEP-1: 315; DEEP-2: 174) and assuming a circular GLEAM beam with r $=$ FWHM/2 $=$ 22.5'', we estimate $\approx$ 5 GLEAM beams in DEEP-1 and $\approx$ 1–2 in DEEP-2 to potentially contain $\gtrsim$ 2 SHORES sources. However, after a careful inspection during the cross-matching procedure, no such ambiguous cases were found.}

\begin{figure}[ht]
\centering
    \includegraphics[width=\linewidth]{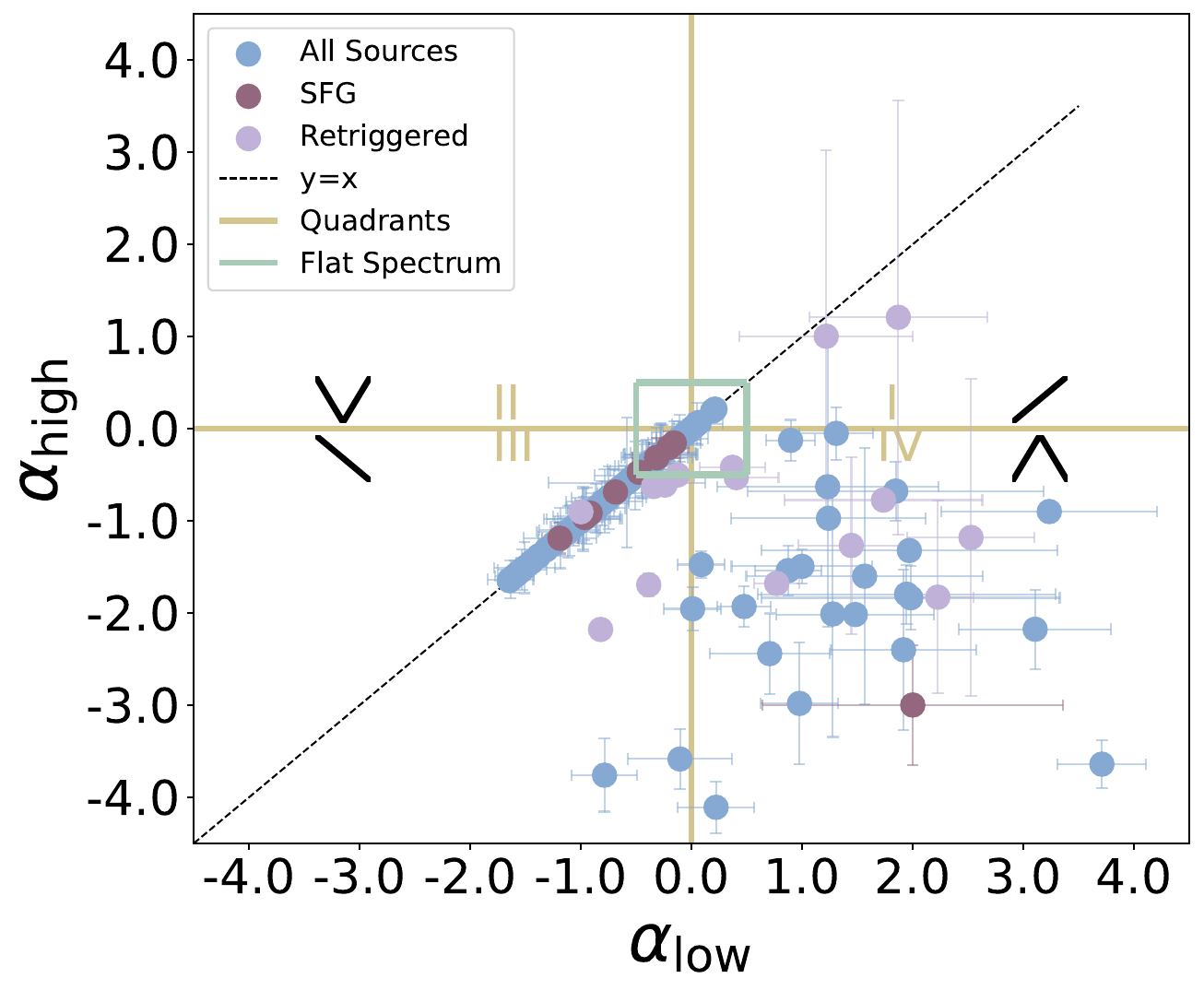}
    \caption{Radio colour-colour diagram of the sources in our sample, based on the spectral indices retrieved from the SED fitting across the 0.1-10 GHz radio regime.}
    \label{fig:radio-cc}
\end{figure}

\begin{figure}[ht]
\centering
    \includegraphics[width=\linewidth]{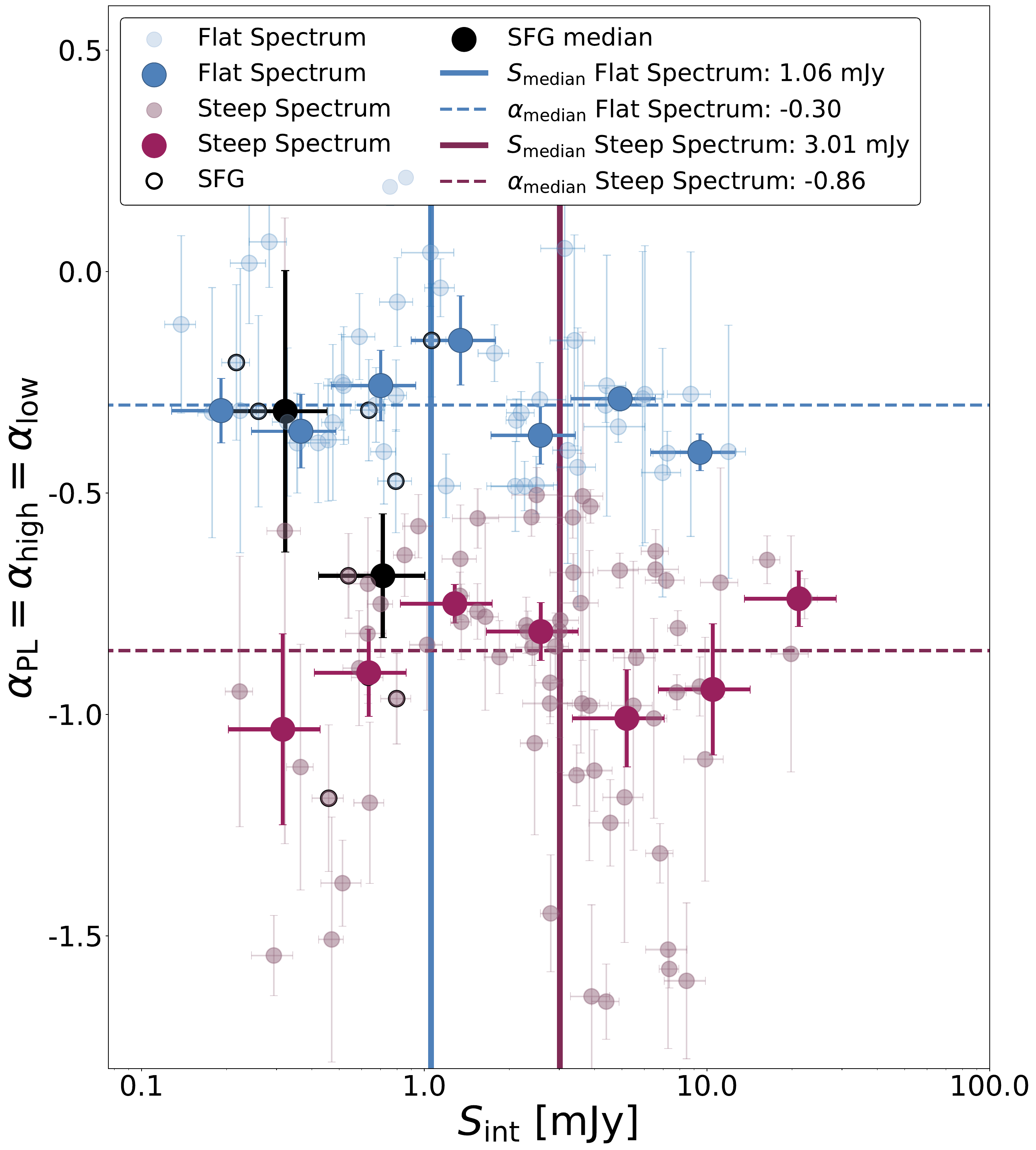}

\caption{Distribution of the spectral indices of sources that show a flat or steep spectrum across the 0.1-10GHz radio regime as a function of the 2.1 GHz. We highlight with black empty circles the candidate SFGs based on the $q_{FIR}$.}
    \label{fig:radio-hist}
\end{figure}

As a result, 159 of our sources are detected in a minimum of 4 different radio bands at SNR $>$4.5, that we used for the analysis described in the following sub-sections. Their radio spectral behaviour fitting resulted in the colour-colour plot presented in Figure \ref{fig:radio-cc} with the spectral indices distributions shown in Figure \ref{fig:radio-hist}.

In Figure \ref{fig:radio-cc}, we show the radio colour-colour diagram. Most sources follow a simple power-law spectrum: 44.2\% are steep and 29.4\% are flat, with no genuinely inverted also in this case. About 15.3\% show pure peaked spectra and 1.2\% (two sources) have curved but non-peaked spectra. Finally, 8.6\% (14 sources) are fit by the retriggered model.

The resulting overall behaviour of the spectral index as a function of flux is the same as the one observed for the SHORES-only SED-fitting. In particular, there are 2.1 GHz-bright flat sources missed by the fitting based only on SHORES data, all but one because they are out of the mosaic area. The one in the mosaic might have a spectral change at higher frequencies that makes it invisible at 5.5-9 GHz.

The steep spectral indices distribution as a function of flux density has a turn-off (Figure \ref{fig:radio-hist}) in this frequency range too, indicating the possibility of distinguishing emission due to nuclear activity or star formation. To help in this effort the combination with FIR data is necessary, as will be discussed in the next session.

\section{Radio to FIR relations.}
\label{sec:FIRRC}

The FIRRC is a well-established empirical relationship observed in star-forming galaxies, linking their FIR and radio emissions \citep{helou85,Delhaize+92,yun01,ivison10,jarvis10,smith14,Giulietti+22}. 
The SHORES fields are selected in the H-ATLAS \citep{eales10} survey to combine radio and FIR data. 

We cross-matched the catalogues using a search radius of 15$^{\prime\prime}$, comparable with the H-ATLAS resolution.
Of our sources with a SNR$>4.5$, 93 have an H-ATLAS counterpart with a detection in at least 3 different FIR bands: these are the sources considered in the following analysis of the FIRRC. In fact, this criterion ensures that we have reliable FIR measurements to estimate the FIR luminosity and to establish a robust correlation with radio data.

To quantify the FIRRC, we use the parameter $q_{\text{FIR}} $, which is defined as \citep{yun01,Magnelli+15}:
\begin{equation}
q_{\text{FIR}} = \log_{10} \left( \frac{L_{\text{FIR}}}{3.75 \times 10^{12} \text{W}} \right) - \log_{10} \left( \frac{L_{\text{1.4 \text{GHz}}}}{\text{W Hz}^{-1}} \right)
\end{equation}
where $L_{\text{FIR}}$ is the total infrared luminosity (integrated over 8–1000 $\mu$m) and $L_{\text{1.4 GHz}}$ is the radio luminosity at 1.4 GHz.
To calculate the luminosities based on the fluxes, we use the SPIRE redshift estimates reported in the H-ATLAS catalogue. 
In particular, the $L_{\text{FIR}}$ is defined as

\begin{equation}
    L_{\text{FIR}} = \frac{4\pi D_L^2}{(1 + z)} \int_{8\mu m}^{1000\mu m} S_{\nu, e} d\nu,  
\end{equation}
where $D_L$ is the luminosity distance, $z$ is the SPIRE redshift and $S_{\nu,e}$ is the monochromatic flux density at a given infrared wavelength.

The rest-frame $L_{\text{1.4 GHz}} $is computed as
\begin{equation}
    L_{\nu, e} = \frac{4\pi D_L^2(z)}{(1 + z)^{1+\alpha}} \left( \frac{\nu_e}{\nu_o} \right)^{\alpha} S_{\nu, o} \text{,}
\end{equation}
where $\alpha$ is the spectral index and the monochromatic radio flux density is $S_{\nu}=\nu^\alpha$.

Our median $q_{\text{FIR}}$ is $\sim$ 2.36, in full agreement with \cite{ivison10}: $\sim$ 81 \% of our analysed sources have a $q_{\text{FIR}}$ within 1$\sigma$ from the value found by \cite{ivison10} for SFGs. \cite{condon02} suggest $q_{\text{FIR}}\sim1.8$ as a threshold to distinguish between SFGs and AGN: only $\sim$ 18.2 \% of the considered sources are below this limit and should, thus, be considered AGN with FIR counterparts.
\begin{figure}
   \centering
    \includegraphics[width=\linewidth]{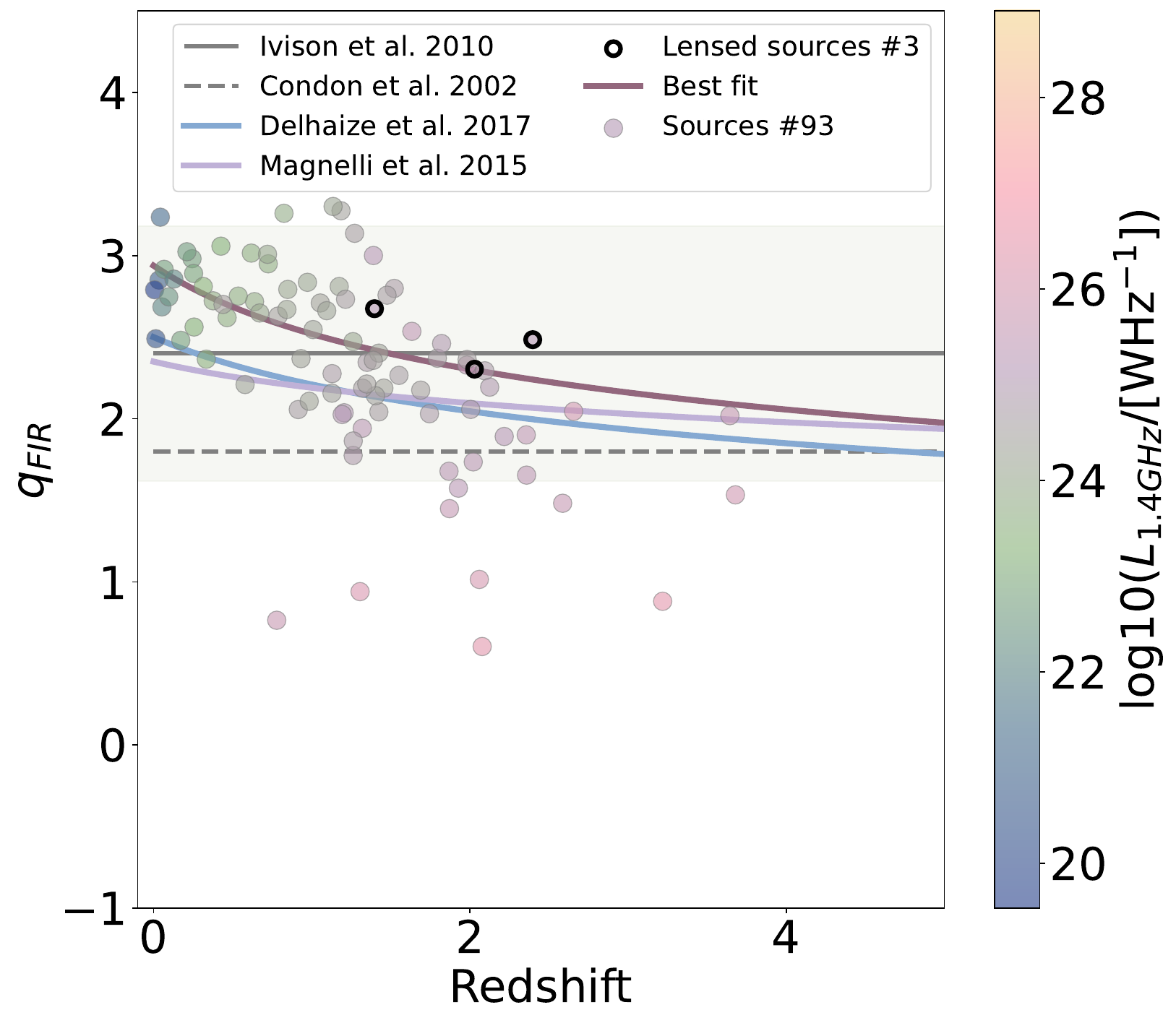}

     \caption{FIRRC $q_{\rm FIR}$ is shown as a function of redshift for 93 sources in our catalog, specifically those with at least three FIR photometric points, including 3 candidate lensed galaxies (black circles).The green shaded area represents the 1$\sigma$ dispersion around the value indicated by \cite{ivison10} for SFGs. The colour bar represent the 1.4 GHz luminosity and the 500 $\mu$m flux, respectively. The fractional uncertainty on the SPIRE photometric redshift amounts to $\Delta z/(1+z)\approx 28\%$.}
    \label{fig:qfir}
\end{figure}
In Figure \ref{fig:qfir} we show the relationship between $q_{\text{FIR}}$ and redshift and radio luminosity. The existence of a trend of $q_{\text{FIR}}$ with redshift remains a subject of considerable debate. In this work, we find that 
\begin{equation}
    q_{\text{FIR}}\sim 2.94(1+z)^{-0.22},
\end{equation}
i.e. a mild dependency on the redshift. We compare our results with \cite{condon02}, \cite{ivison10}, \cite{Magnelli+15} and \cite{delhaize17}. \cite{Magnelli+15} analysed a mass-selected sample of star-forming galaxies up to a redshift of approximately 2, and observed a marginal evolution characterized by the relation $q_{\text{FIR}}(z) = (2.35 \pm 0.08)(1 + z)^{-0.12 \pm 0.04}$. Comparable findings were achieved by \cite{delhaize17} using a radio-selected sample of star-forming galaxies extending up to a redshift of approximately 6, with the relation $q_{\text{FIR}}(z) = (2.52 \pm 0.03)(1 + z)^{-0.21 \pm 0.01}$.

In Figure \ref{fig:radio-hist}, we show the spectral index distribution of the 19 sources that have both H-ATLAS and radio ancillary counterparts, to compare FIR and radio spectral behaviour. Of these, 10 have $q_{\mathrm{FIR}}\ge 1.69$, i.e. the 1$\sigma$ lower bound of the \citet{ivison10} FIR–radio relation for star-forming galaxies.
In these objects, the emission is expected to be dominated by SF, the median spectral index is $-0.6\pm0.1$, in agreement with expectations for SFGs, as predicted by models of synchrotron emission from supernovae remnants (e.g. \citealt{Condon+92,Murphy+09}) and confirmed in statistical studies of FIR-selected SFGs (e.g. \citealt{delhaize17, Magnelli+15}).

However, we note that some of the power-law radio sources without FIR counterparts may in fact be SFGs, undetected in FIR. The H-ATLAS survey is limited by instrumental confusion noise, particularly at 250–500 $\mu$m, where the typical confusion limit is $\sim$6–7 mJy (e.g. \citealt{nguyen10, valiante16}). As a result, faint FIR emission from high-redshift SFGs may go undetected even when their radio counterparts are observable in deep maps. The spectral index distribution of these radio-only sources, peaking around $\alpha \sim -0.6$, is consistent with typical values expected for radio jets, but also of SFGs. This finding suggests that these sources could indeed be jets without any FIR counterpart, but yet a fraction of them could be composed by dust-obscured star-forming galaxies at $z \gtrsim 1.5$–2, falling below the Herschel detection limit. 

{Unfortunately, since our redshift estimates come solely from the H-ATLAS catalogue, we cannot meaningfully assess whether the \textit{radio-detected but FIR-undetected} sources could exhibit a $q_{\mathrm{FIR}}$ value consistent with star-forming emission: without FIR photometry, there is no redshift estimate, and $q_{\mathrm{FIR}}$ cannot be evaluated. }

\begin{figure}
    \centering
    \includegraphics[width=\linewidth]{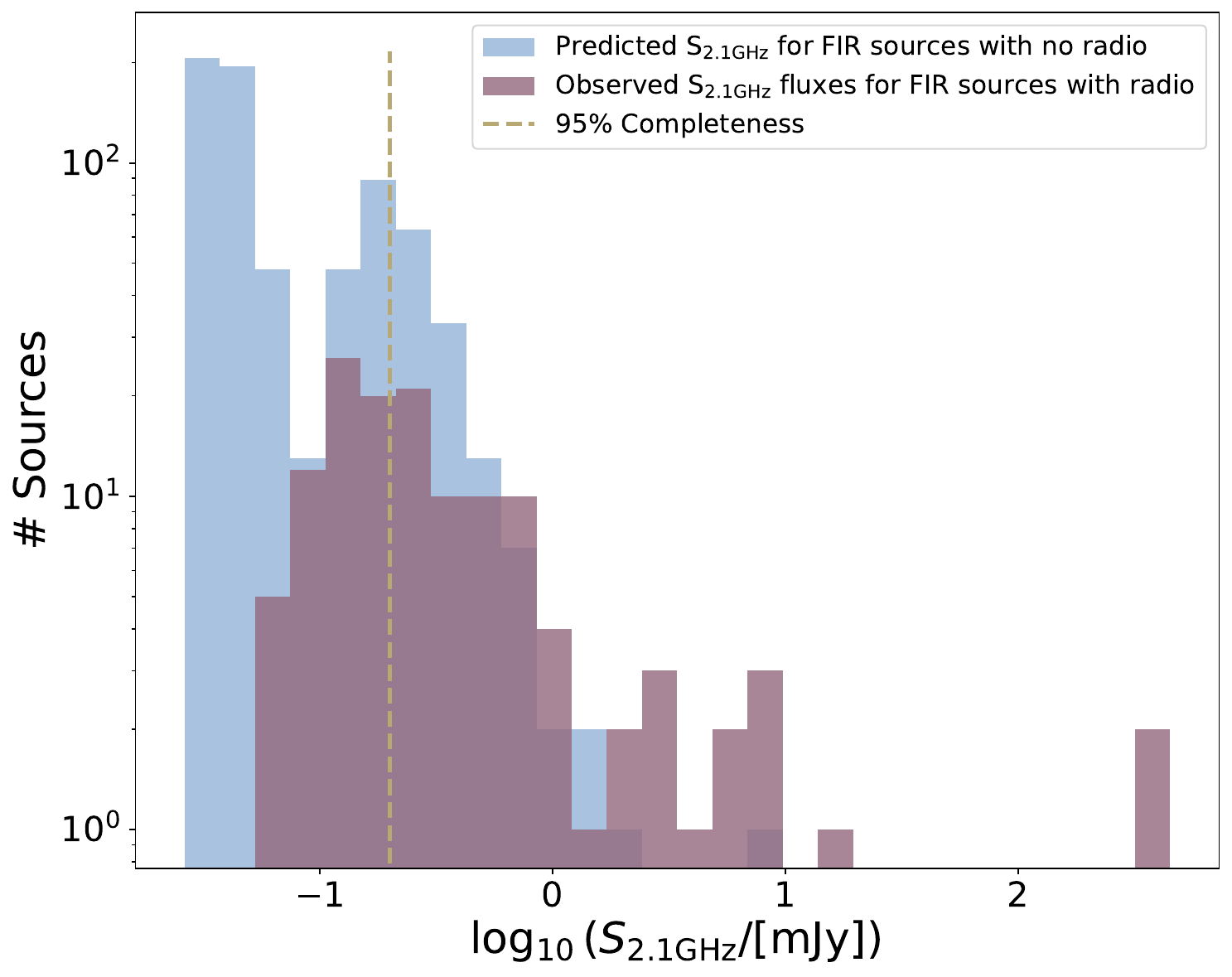}
    \caption{Observed 2.1 GHz flux densities for FIR-detected radio sources (purple) 
compared to the predicted 2.1 GHz fluxes of H-ATLAS sources with FIR 
detections but no radio counterparts (blue), computed from their FIR 
luminosities and photometric redshifts using the adopted 
$q_{\mathrm{FIR}}(z)$ relation. 
The vertical dashed line shows the 95$\%$ completeness limit. 
Most predicted fluxes lie below this threshold, naturally explaining their absence from the radio catalogue.}
    \label{fig:fluxes_noradio}
\end{figure}

{However, we can perform a complementary and more robust test on the opposite class of objects: \textit{H-ATLAS sources with FIR detections (and therefore photometric redshifts) but without radio counterparts}. For these galaxies, we used their FIR luminosities together with the adopted $q_{\mathrm{FIR}}(z)$ relation to predict the corresponding radio luminosities and, from these, the expected 2.1-GHz flux densities under the assumption that they are star-forming.}
{
This yields the predicted radio-flux distribution of the FIR-selected galaxies that SHORES should have detected if the survey had been deeper. As shown in Figure \ref{fig:fluxes_noradio}, the vast majority of these predicted fluxes lie below the 95\% completeness limit of our radio observations and are significantly fainter than the fluxes of the FIR sources included in our catalogue.
}

This highlights the importance of combining deep radio and FIR observations when identifying and characterising faint SFGs. Nonetheless, caution is required when interpreting the spectral indices, as the observed radio emission may also be contaminated by non-stellar processes. These include galactic outflows or compact, low-luminosity AGN with no prominent FIR emission. 

In conclusion, to disentangle these contributions and confirm the nature of these ambiguous sources, deeper observations in complementary wavebands, as millimetre (e.g. with ALMA) and X-ray data to build panchromatic SED fitting, are essential. Such multi-wavelength approaches can help break degeneracies between AGN and SFG emission mechanisms and improve the accuracy of FIRRC-based classifications, especially in the low-luminosity and high-redshift regimes.

\section{Radio number counts}
\label{sec:counts}

\begin{figure*}[ht]
   \centering
    \includegraphics[width=0.8\linewidth]{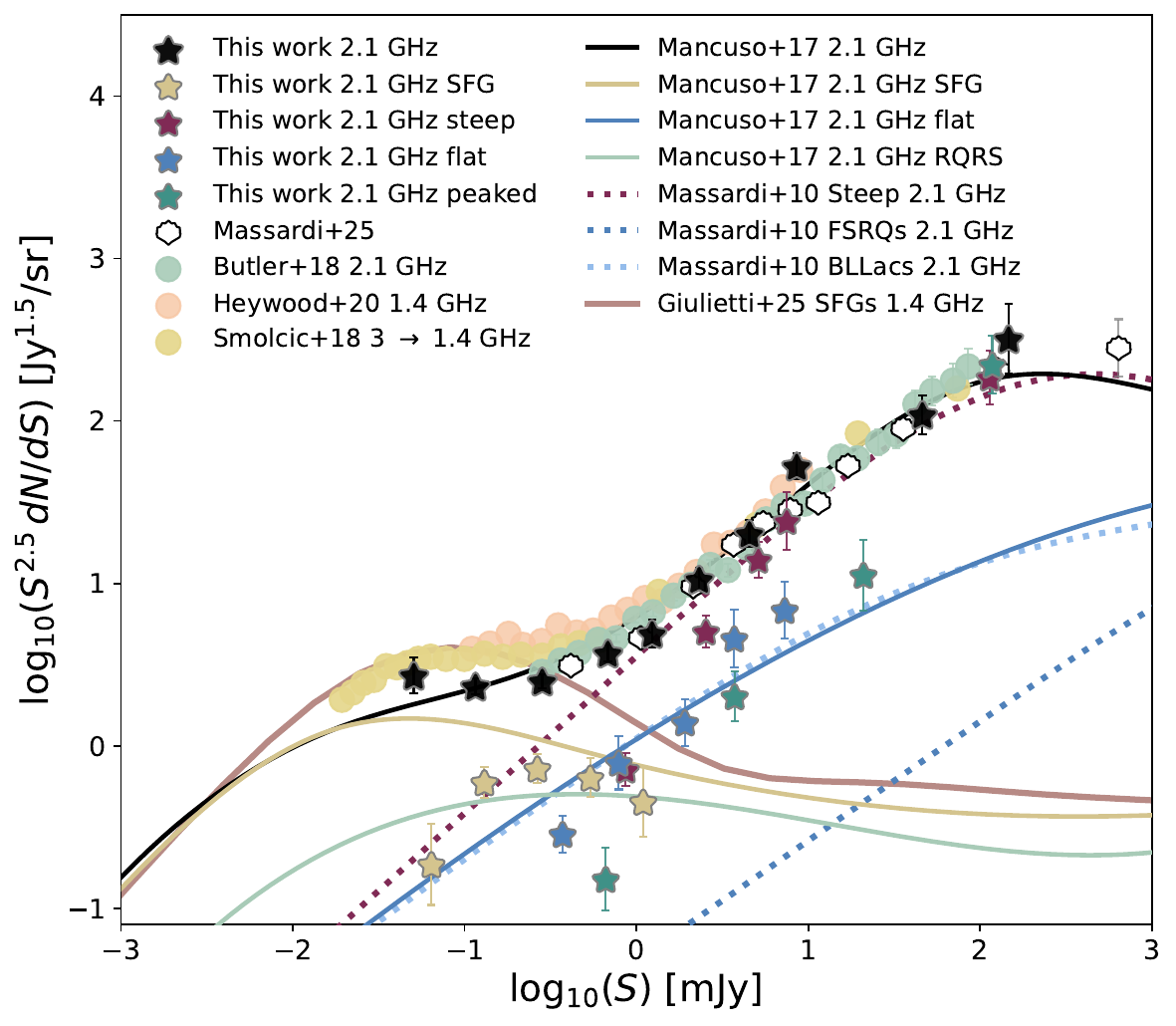}
    \caption{Euclidean differential source counts at 2.1 GHz of the two deep fields. For comparison we added recent estimates by SHORES-I, \cite{butler18}, \cite{heywood20} and \cite{smolcic2017} and the models by \cite{massardi10}, \cite{Mancuso+17} and \cite{giulietti25}. We highlight the differential number counts for the fiducial candidate star-forming galaxies with golden stars and the sources classified as steep (burgundy), flat (light blue) and peaked (green).}
    \label{fig:number_counts}
\end{figure*}

We finally estimated the differential radio number counts for the SHORES deep fields (Figure \ref{fig:number_counts} and Table \ref{tab:counts}-\ref{tab:counts_sfg}) at 2.1 and 5.5 GHz.

Combining the source counts of the current paper with those of SHORES-I, we show how the SHORES survey, thanks to the combination of wide-area shallow fields and small-area deep fields,  successfully recovers the bright and faint 2.1 GHz radio number counts down to flux densities of a few tens of $\mu$Jy, with comparable significance. The faintest bins are dominated by SFGs and possibly radio-quiet AGN, in line with what is expected from both empirical studies (e.g. \citealt{smolcic2017}) and theoretical predictions (e.g. \citealt{Mancuso+17}).\begin{table}
    \centering
\begin{tabular}{l|l|l|l}

\hline
Intervals & Bin Centers & $N$ & $S_{\rm 2.1GHz}^{2.5}dN/dS$\\
mJy & mJy & & Jy$^{1.5}$sr$^{-1}$  \\
\hline

0.026 - 0.076 & 0.051 & 15 & 2.7 $\pm$ 0.7 \\
0.076 - 0.156& 0.116 & 81 & 2.3 $\pm$ 0.3 \\
0.156 - 0.410 & 0.283 & 109 & 2.5 $\pm$ 0.2 \\
0.410 - 0.953 & 0.681 & 68 & 3.7 $\pm$ 0.5 \\
0.953 - 1.512 & 1.232 & 25 & 5.0 $\pm$ 1.9 \\
1.511 - 3.118 & 2.315 & 38 & 10.6 $\pm$ 1.7 \\
3.118 - 5.958& 4.538 & 28 & 20.2 $\pm$ 3.8 \\
5.958 - 11.160& 8.559 & 30 & 52.7 $\pm$ 9.6 \\
11.160 - 81.160& 46.160 & 13 & 109.4 $\pm$ 30.4 \\
81.160 - 211.600 & 146.381& 4 & 320.7 $\pm$ 160.4 \\
\hline
\end{tabular}

\caption{Differential Euclidean source counts at 2.1 GHz.}\label{tab:counts}

\end{table}
\begin{table}
    
\centering
\begin{tabular}{l|l|l|l}

\hline
Intervals & Bin Centers & $N$ & $S_{\rm 2.1GHz}^{2.5}dN/dS$\\
mJy & mJy & & Jy$^{1.5}$sr$^{-1}$  \\
\hline
0.042 - 0.086 & 0.064 & 3 & 0.2 $\pm$ 0.1 \\
0.086 - 0.175 & 0.130 & 21 & 0.6 $\pm$  0.1 \\
0.174 - 0.355 & 0.265 & 25 & 0.7 $\pm$  0.1 \\
0.355 - 0.721 & 0.538 & 13 & 0.6 $\pm$  0.2 \\
0.721 - 1.466 & 1.093 & 4 & 0.5 $\pm$ 0.2 \\
\hline
\end{tabular}

\caption{Differential Euclidean source counts 2.1 GHz for the fiducial SFGs.}\label{tab:counts_sfg}
\end{table} {It is worth noting that the selection of SFGs is based on the $q_{\rm{FIR}}$ parameter, which could be estimated only for sources with at least three FIR detections. As a result, additional sources potentially contributing to the SFG number counts in the faint-end may be missed due to the sensitivity limits of H-ATLAS. Our observed number counts for fiducial SFGs qualitatively appear to follow the trend predicted for radio-quiet AGN in the evolutionary model by \citet{Mancuso+17}. Considering the limited availability of FIR counterparts due to \textit{Herschel}’s resolution and depth, along with incomplete radio photometric sampling for the faintest sources, we cautiously note that a fraction of sub-mJy sources classified as SFGs may host radio-quiet AGN.} The observed turnover in the differential counts below $\sim$1 mJy matches the well-known transition in source populations (see e.g. \citealt{dezotti10, massardi10}), and our measurements help fill the observational gap in the range $0.05–0.2\,$mJy where previous surveys often suffer from incompleteness.

We also compared our number counts with other deep surveys such as VLA-COSMOS (e.g. \citealt{smolcic2017}) and MIGHTEE \citep{heywood22}, finding consistent trends.
By leveraging matches with multi-wavelength ancillary data, we attempted a statistical separation of SFGs and AGNs with different spectral properties at low flux densities—bearing in mind the limitations due to incomplete FIR coverage and multi-frequency detection thresholds.
While this spectral and multi-band decomposition does not allow us to disantangle the different evolutionary models, the resulting distributions appear qualitatively consistent with the expected behaviour of SFGs and radio-quiet AGN on the one hand (e.g. \citealt{Mancuso+17}), and of radio-loud AGN on the other (e.g. \citealt{bonato21}). Within the limits of the current completeness, this broad consistency supports the tentative use of such classifications to statistically explore the nature of the faint radio population.

\section{Summary and Perspectives} \label{sec:conclusions}

We presented the building and analysis of the catalogue of the SHORES survey deep fields at $2.1\,$GHz, and of their follow-up at $5.5$ and $9.0\,$GHz. The catalogue at $2.1\,$GHz is based on source extraction from two  $\sim 1.2\,$deg$^2$ mosaics (Deep\,1 and Deep\,2), selected for their  rich ancillary coverage and lack of bright contaminating sources.  The Deep\,1 mosaic reaches an rms of $\sigma \simeq 9\,\mu$Jy/beam, while Deep\,2 reaches $\sigma \simeq 18\,\mu$Jy/beam over the same area. Thus, the SHORES deep survey covers 2.4\,deg$^2$ with a mean rms sensitivity of $\sim$13.5\,$\mu$Jy/beam. The extracted catalogue is $95\%$ complete above $183\,\mu$Jy ($4.5\,\sigma$ in the shallower field) and $85\%$ reliable above the adopted $4.5\,\sigma$ threshold. The follow-up in mosaic mode of the inner $0.5\times0.5$ deg$^2$ of each field provided a catalogue of $101$ counterparts at $5.5\,$GHz and $33$ at $9.0\,$GHz.

The analysis of the spectral behaviour in the observed frequency range pointed out that to such sensitivity levels the flat-steep spectra sources dominates the population, while peaked spectra remain a 15.7\% fraction, indicating the relevant role that SF plays over jetted AGN components. The combination with lower frequency radio and FIR ancillary data stressed the importance of a multi-wavelength analysis for $<1\mu$Jy radio sources classification. About 18.7\% of our sources have a counterpart in H-ATLAS, mostly with $\rm{FIRRC}>1.7$, stressing the relevance of SF over nuclear activity in such sources. We identified for them a trend of FIRRC with redshift and an average spectral index over the whole radio domain $\alpha\sim-0.6\pm0.1$. 

We finally presented the source counts at 2.1 for our sample, including in the former the separation between SFG and AGN-dominated sources as made possible by our attempted classification. Our findings are in good agreement with previous datasets and with in-situ theoretical models for galaxy formation \citep{Mancuso+17}. Our findings highlight the importance of deep, multi-frequency radio surveys over well-characterised extragalactic fields and demonstrate the value of combining radio and FIR data for disentangling source populations. In this context, SHORES represents a key preparatory step for the SKA era, bridging the observational gap between current large-area shallow surveys and pencil-beam ultra-deep fields.

\begin{acknowledgments}
We acknowledge Q. D'Amato for his contributions in the early stages of the SHORES survey. M. Behiri acknowledges the support given by L. Capuano, A. Traina and L. Leuzzi and the 06:39 Frecciarossa train from Trieste Centrale to Bologna Centrale. This work was partially funded from the projects: INAF GO-GTO Normal 2023 funding scheme with the project ``Serendipitous H-ATLAS-fields Observations of Radio Extragalactic Sources (SHORES)''; INAF Large Grant 2022 project ``MeerKAT and LOFAR Team up: a Unique Radio Window on Galaxy/AGN co-Evolution''; INAF Large GO 2024 project ``MeerKAT and Euclid Team up: Exploring the galaxy-halo connection at cosmic noon''; ``Data Science methods for MultiMessenger Astrophysics $\&$ Multi-Survey Cosmology'', funded by the Italian Ministry of University and Research, Programmazione triennale 2021/2023 (DM n.2503 dd. 9 December 2019), Programma Congiunto Scuole; Italian Research Center on High Performance Computing Big Data and Quantum Computing (ICSC), project funded by European Union - NextGenerationEU - and National Recovery and Resilience Plan (NRRP) - Mission 4 Component 2 within the activities of Spoke 3 (Astrophysics and Cosmos Observations); European Union - NextGenerationEU under the PRIN MUR 2022 project n.20224JR28W ``Charting unexplored avenues in Dark Matter''. 
\end{acknowledgments}

\vspace{5mm}
\facilities{ATCA}
\software{    \textsc{RADIOSED} \citep{radiosed}
          \textsc{BLOBCAT} \citep{BLOBCAT}
            \textsc{Miriad} \citep{miriad}
            \textsc{PySE} \citep{pyse}
            \textsc{WSCLEAN} \citep{offringa14}}

\bibliography{bibliography}{}

@ARTICLE{galluzzi17,
       author = {{Galluzzi}, V. and {Massardi}, M. and {Bonaldi}, A. and {Casasola}, V. and {Gregorini}, L. and {Trombetti}, T. and {Burigana}, C. and {De Zotti}, G. and {Ricci}, R. and {Stevens}, J. and {Ekers}, R.~D. and {Bonavera}, L. and {di Serego Alighieri}, S. and {Liuzzo}, E. and {L{\'o}pez-Caniego}, M. and {Mignano}, A. and {Paladino}, R. and {Toffolatti}, L. and {Tucci}, M.},
        title = "{Multifrequency polarimetry of a complete sample of PACO radio sources}",
      journal = {\mnras},
         year = 2017,
        month = mar,
       volume = {465},
       number = {4},
        pages = {4085-4098},
          doi = {10.1093/mnras/stw3017},
archivePrefix = {arXiv},
       eprint = {1611.07746},
 primaryClass = {astro-ph.GA},
       adsurl = {https://ui.adsabs.harvard.edu/abs/2017MNRAS.465.4085G}
}

@ARTICLE{sargent10,
       author = {{Sargent}, M.~T. and {Schinnerer}, E. and {Murphy}, E. and {Aussel}, H. and {Le Floc'h}, E. and {Frayer}, D.~T. and {Mart{\'\i}nez-Sansigre}, A. and {Oesch}, P. and {Salvato}, M. and {Smol{\v{c}}i{\'c}}, V. and {Zamorani}, G. and {Brusa}, M. and {Cappelluti}, N. and {Carilli}, C.~L. and {Carollo}, C.~M. and {Ilbert}, O. and {Kartaltepe}, J. and {Koekemoer}, A.~M. and {Lilly}, S.~J. and {Sanders}, D.~B. and {Scoville}, N.~Z.},
        title = "{The VLA-COSMOS Perspective on the Infrared-Radio Relation. I. New Constraints on Selection Biases and the Non-Evolution of the Infrared/Radio Properties of Star-Forming and Active Galactic Nucleus Galaxies at Intermediate and High Redshift}",
      journal = {\apjs},
         year = 2010,
        month = feb,
       volume = {186},
       number = {2},
        pages = {341-377},
          doi = {10.1088/0067-0049/186/2/341},
archivePrefix = {arXiv},
       eprint = {1001.1354},
 primaryClass = {astro-ph.CO},
       adsurl = {https://ui.adsabs.harvard.edu/abs/2010ApJS..186..341S}
}

@ARTICLE{ivison10b,
       author = {{Ivison}, R.~J. and {Swinbank}, A.~M. and {Swinyard}, B. and {Smail}, I. and {Pearson}, C.~P. and {Rigopoulou}, D. and {Polehampton}, E. and {Baluteau}, J. -P. and {Barlow}, M.~J. and {Blain}, A.~W. and {Bock}, J. and {Clements}, D.~L. and {Coppin}, K. and {Cooray}, A. and {Danielson}, A. and {Dwek}, E. and {Edge}, A.~C. and {Franceschini}, A. and {Fulton}, T. and {Glenn}, J. and {Griffin}, M. and {Isaak}, K. and {Leeks}, S. and {Lim}, T. and {Naylor}, D. and {Oliver}, S.~J. and {Page}, M.~J. and {P{\'e}rez Fournon}, I. and {Rowan-Robinson}, M. and {Savini}, G. and {Scott}, D. and {Spencer}, L. and {Valtchanov}, I. and {Vigroux}, L. and {Wright}, G.~S.},
        title = "{Herschel and SCUBA-2 imaging and spectroscopy of a bright, lensed submillimetre galaxy at z = 2.3}",
      journal = {\aap},
         year = 2010,
        month = jul,
       volume = {518},
          eid = {L35},
        pages = {L35},
          doi = {10.1051/0004-6361/201014548},
archivePrefix = {arXiv},
       eprint = {1005.1071},
 primaryClass = {astro-ph.CO},
       adsurl = {https://ui.adsabs.harvard.edu/abs/2010A&A...518L..35I}
}

@ARTICLE{bell03,
       author = {{Bell}, Eric F.},
        title = "{Estimating Star Formation Rates from Infrared and Radio Luminosities: The Origin of the Radio-Infrared Correlation}",
      journal = {\apj},
         year = 2003,
        month = apr,
       volume = {586},
       number = {2},
        pages = {794-813},
          doi = {10.1086/367829},
archivePrefix = {arXiv},
       eprint = {astro-ph/0212121},
 primaryClass = {astro-ph},
       adsurl = {https://ui.adsabs.harvard.edu/abs/2003ApJ...586..794B}
}

@ARTICLE{pyse,
       author = {{Carbone}, D. and {Garsden}, H. and {Spreeuw}, H. and {Swinbank}, J.~D. and {van der Horst}, A.~J. and {Rowlinson}, A. and {Broderick}, J.~W. and {Rol}, E. and {Law}, C. and {Molenaar}, G. and {Wijers}, R.~A.~M.~J.},
        title = "{PySE: Software for extracting sources from radio images}",
      journal = {Astronomy and Computing},
         year = 2018,
        month = apr,
       volume = {23},
          eid = {92},
        pages = {92},
          doi = {10.1016/j.ascom.2018.02.003},
archivePrefix = {arXiv},
       eprint = {1802.09604},
 primaryClass = {astro-ph.IM},
       adsurl = {https://ui.adsabs.harvard.edu/abs/2018A&C....23...92C}
}

@article{heywood20,
	title        = {{VLA imaging of the XMM-LSS/VIDEO deep field at 1-2 GHz}},
	author       = {{Heywood}, I. and {Hale}, C.~L. and {Jarvis}, M.~J. and {Makhathini}, S. and {Peters}, J.~A. and {Sebokolodi}, M.~L.~L. and {Smirnov}, O.~M.},
	year         = 2020,
	month        = aug,
	journal      = {\mnras},
	volume       = 496,
	number       = 3,
	pages        = {3469--3481},
	doi          = {10.1093/mnras/staa1770},
	archiveprefix = {arXiv},
	eprint       = {2006.08551},
	primaryclass = {astro-ph.GA},
	adsurl       = {https://ui.adsabs.harvard.edu/abs/2020MNRAS.496.3469H}
}

@article{butler18,
	title        = {{The XXL Survey. XVIII. ATCA 2.1 GHz radio source catalogue and source counts for the XXL-South field}},
	author       = {{Butler}, Andrew and {Huynh}, Minh and {Delhaize}, Jacinta and {Smol{\v{c}}i{\'c}}, Vernesa and {Kapi{\'n}ska}, Anna and {Milakovi{\'c}}, Dinko and {Novak}, Mladen and {Baran}, Nikola and {O'Brien}, Andrew and {Chiappetti}, Lucio and {Desai}, Shantanu and {Fotopoulou}, Sotiria and {Horellou}, Cathy and {Lidman}, Chris and {Pierre}, Marguerite},
	year         = 2018,
	month        = nov,
	journal      = {\aap},
	volume       = 620,
	pages        = {A3},
	doi          = {10.1051/0004-6361/201630129},
	eid          = {A3},
	archiveprefix = {arXiv},
	eprint       = {1703.10296},
	primaryclass = {astro-ph.GA},
	adsurl       = {https://ui.adsabs.harvard.edu/abs/2018A&A...620A...3B}
}

@article{eales10,
	title        = {{The Herschel ATLAS}},
	author       = {{Eales}, S. and {Dunne}, L. and {Clements}, D. and {Cooray}, A. and {De Zotti}, G. and {Dye}, S. and {Ivison}, R. and {Jarvis}, M. and {Lagache}, G. and {Maddox}, S. and {Negrello}, M. and {Serjeant}, S. and {Thompson}, M.~A. and {Van Kampen}, E. and {Amblard}, A. and {Andreani}, P. and {Baes}, M. and {Beelen}, A. and {Bendo}, G.~J. and {Benford}, D. and {Bertoldi}, F. and {Bock}, J. and {Bonfield}, D. and {Boselli}, A. and {Bridge}, C. and {Buat}, V. and {Burgarella}, D. and {Carlberg}, R. and {Cava}, A. and {Chanial}, P. and {Charlot}, S. and {Christopher}, N. and {Coles}, P. and {Cortese}, L. and {Dariush}, A. and {da Cunha}, E. and {Dalton}, G. and {Danese}, L. and {Dannerbauer}, H. and {Driver}, S. and {Dunlop}, J. and {Fan}, L. and {Farrah}, D. and {Frayer}, D. and {Frenk}, C. and {Geach}, J. and {Gardner}, J. and {Gomez}, H. and {Gonz{\'a}lez-Nuevo}, J. and {Gonz{\'a}lez-Solares}, E. and {Griffin}, M. and {Hardcastle}, M. and {Hatziminaoglou}, E. and {Herranz}, D. and {Hughes}, D. and {Ibar}, E. and {Jeong}, Woong-Seob and {Lacey}, C. and {Lapi}, A. and {Lawrence}, A. and {Lee}, M. and {Leeuw}, L. and {Liske}, J. and {L{\'o}pez-Caniego}, M. and {M{\"u}ller}, T. and {Nandra}, K. and {Panuzzo}, P. and {Papageorgiou}, A. and {Patanchon}, G. and {Peacock}, J. and {Pearson}, C. and {Phillipps}, S. and {Pohlen}, M. and {Popescu}, C. and {Rawlings}, S. and {Rigby}, E. and {Rigopoulou}, M. and {Robotham}, A. and {Rodighiero}, G. and {Sansom}, A. and {Schulz}, B. and {Scott}, D. and {Smith}, D.~J.~B. and {Sibthorpe}, B. and {Smail}, I. and {Stevens}, J. and {Sutherland}, W. and {Takeuchi}, T. and {Tedds}, J. and {Temi}, P. and {Tuffs}, R. and {Trichas}, M. and {Vaccari}, M. and {Valtchanov}, I. and {van der Werf}, P. and {Verma}, A. and {Vieria}, J. and {Vlahakis}, C. and {White}, Glenn J.},
	year         = 2010,
	month        = may,
	journal      = {\pasp},
	volume       = 122,
	number       = 891,
	pages        = 499,
	doi          = {10.1086/653086},
	archiveprefix = {arXiv},
	eprint       = {0910.4279},
	primaryclass = {astro-ph.CO},
	adsurl       = {https://ui.adsabs.harvard.edu/abs/2010PASP..122..499E}
}

@ARTICLE{nguyen10,
       author = {{Nguyen}, H.~T. and {Schulz}, B. and {Levenson}, L. and {Amblard}, A. and {Arumugam}, V. and {Aussel}, H. and {Babbedge}, T. and {Blain}, A. and {Bock}, J. and {Boselli}, A. and {Buat}, V. and {Castro-Rodriguez}, N. and {Cava}, A. and {Chanial}, P. and {Chapin}, E. and {Clements}, D.~L. and {Conley}, A. and {Conversi}, L. and {Cooray}, A. and {Dowell}, C.~D. and {Dwek}, E. and {Eales}, S. and {Elbaz}, D. and {Fox}, M. and {Franceschini}, A. and {Gear}, W. and {Glenn}, J. and {Griffin}, M. and {Halpern}, M. and {Hatziminaoglou}, E. and {Ibar}, E. and {Isaak}, K. and {Ivison}, R.~J. and {Lagache}, G. and {Lu}, N. and {Madden}, S. and {Maffei}, B. and {Mainetti}, G. and {Marchetti}, L. and {Marsden}, G. and {Marshall}, J. and {O'Halloran}, B. and {Oliver}, S.~J. and {Omont}, A. and {Page}, M.~J. and {Panuzzo}, P. and {Papageorgiou}, A. and {Pearson}, C.~P. and {Perez Fournon}, I. and {Pohlen}, M. and {Rangwala}, N. and {Rigopoulou}, D. and {Rizzo}, D. and {Roseboom}, I.~G. and {Rowan-Robinson}, M. and {Scott}, D. and {Seymour}, N. and {Shupe}, D.~L. and {Smith}, A.~J. and {Stevens}, J.~A. and {Symeonidis}, M. and {Trichas}, M. and {Tugwell}, K.~E. and {Vaccari}, M. and {Valtchanov}, I. and {Vigroux}, L. and {Wang}, L. and {Ward}, R. and {Wiebe}, D. and {Wright}, G. and {Xu}, C.~K. and {Zemcov}, M.},
        title = "{HerMES: The SPIRE confusion limit}",
      journal = {\aap},
         year = 2010,
        month = jul,
       volume = {518},
          eid = {L5},
        pages = {L5},
          doi = {10.1051/0004-6361/201014680},
archivePrefix = {arXiv},
       eprint = {1005.2207},
 primaryClass = {astro-ph.CO},
       adsurl = {https://ui.adsabs.harvard.edu/abs/2010A&A...518L...5N}
}

@article{radiosed,
	title        = {{RADIOSED - I. Bayesian inference of radio SEDs from inhomogeneous surveys}},
	author       = {{Kerrison}, Emily F. and {Allison}, James R. and {Moss}, Vanessa A. and {Sadler}, Elaine M. and {Rees}, Glen A.},
	year         = 2024,
	month        = oct,
	journal      = {\mnras},
	volume       = 533,
	number       = 4,
	pages        = {4248--4267},
	doi          = {10.1093/mnras/stae1796},
	archiveprefix = {arXiv},
	eprint       = {2407.16201},
	primaryclass = {astro-ph.GA},
	adsurl       = {https://ui.adsabs.harvard.edu/abs/2024MNRAS.533.4248K}
}

@article{dezotti10,
	title        = {{Radio and millimeter continuum surveys and their astrophysical implications}},
	author       = {{De Zotti}, Gianfranco and {Massardi}, Marcella and {Negrello}, Mattia and {Wall}, Jasper},
	year         = 2010,
	month        = feb,
	journal      = {\aapr},
	volume       = 18,
	number       = {1-2},
	pages        = {1--65},
	doi          = {10.1007/s00159-009-0026-0},
	archiveprefix = {arXiv},
	eprint       = {0908.1896},
	primaryclass = {astro-ph.CO},
	adsurl       = {https://ui.adsabs.harvard.edu/abs/2010A&ARv..18....1D}
}

@ARTICLE{talia21,
       author = {{Talia}, Margherita and {Cimatti}, Andrea and {Giulietti}, Marika and {Zamorani}, Gianni and {Bethermin}, Matthieu and {Faisst}, Andreas and {Le F{\`e}vre}, Olivier and {Smol{\c{c}}i{\'c}}, Vernesa},
        title = "{Illuminating the Dark Side of Cosmic Star Formation Two Billion Years after the Big Bang}",
      journal = {\apj},
         year = 2021,
        month = mar,
       volume = {909},
       number = {1},
          eid = {23},
        pages = {23},
          doi = {10.3847/1538-4357/abd6e3},
archivePrefix = {arXiv},
       eprint = {2011.03051},
 primaryClass = {astro-ph.CO},
       adsurl = {https://ui.adsabs.harvard.edu/abs/2021ApJ...909...23T}
}

@article{Blanford,
    author = {Blandford, R. D. and Znajek, R. L.},
    title = {Electromagnetic extraction of energy from Kerr black holes},
    journal = {Monthly Notices of the Royal Astronomical Society},
    volume = {179},
    number = {3},
    pages = {433-456},
    year = {1977},
    month = {07},
    issn = {0035-8711},
    doi = {10.1093/mnras/179.3.433},
    url = {https://doi.org/10.1093/mnras/179.3.433},
    eprint = {https://academic.oup.com/mnras/article-pdf/179/3/433/9333653/mnras179-0433.pdf},
}

@INPROCEEDINGS{jarvis16,
       author = {{Jarvis}, M. and {Taylor}, R. and {Agudo}, I. and {Allison}, J.~R. and {Deane}, R.~P. and {Frank}, B. and {Gupta}, N. and {Heywood}, I. and {Maddox}, N. and {McAlpine}, K. and {Santos}, M. and {Scaife}, A.~M.~M. and {Vaccari}, M. and {Zwart}, J.~T.~L. and {Adams}, E. and {Bacon}, D.~J. and {Baker}, A.~J. and {Bassett}, B.~A. and {Best}, P.~N. and {Beswick}, R. and {Blyth}, S. and {Brown}, M.~L. and {Bruggen}, M. and {Cluver}, M. and {Colafrancesco}, S. and {Cotter}, G. and {Cress}, C. and {Dav{\'e}}, R. and {Ferrari}, C. and {Hardcastle}, M.~J. and {Hale}, C.~L. and {Harrison}, I. and {Hatfield}, P.~W. and {Klockner}, H.~R. and {Kolwa}, S. and {Malefahlo}, E. and {Marubini}, T. and {Mauch}, T. and {Moodley}, K. and {Morganti}, R. and {Norris}, R.~P. and {Peters}, J.~A. and {Prandoni}, I. and {Prescott}, M. and {Oliver}, S. and {Oozeer}, N. and {Rottgering}, H.~J.~A. and {Seymour}, N. and {Simpson}, C. and {Smirnov}, O. and {Smith}, D.~J.~B.},
        title = "{The MeerKAT International GHz Tiered Extragalactic Exploration (MIGHTEE) Survey}",
    booktitle = {MeerKAT Science: On the Pathway to the SKA},
         year = 2016,
        month = jan,
          eid = {6},
        pages = {6},
          doi = {10.22323/1.277.0006},
archivePrefix = {arXiv},
       eprint = {1709.01901},
 primaryClass = {astro-ph.GA},
       adsurl = {https://ui.adsabs.harvard.edu/abs/2016mks..confE...6J}
}

@article{negrello2017,
	title        = {{The Herschel-ATLAS: a sample of 500 {\ensuremath{\mu}}m-selected lensed galaxies over 600 deg$^{2}$}},
	author       = {{Negrello}, M. and {Amber}, S. and {Amvrosiadis}, A. and {Cai}, Z. -Y. and {Lapi}, A. and {Gonzalez-Nuevo}, J. and {De Zotti}, G. and {Furlanetto}, C. and {Maddox}, S.~J. and {Allen}, M. and {Bakx}, T. and {Bussmann}, R.~S. and {Cooray}, A. and {Covone}, G. and {Danese}, L. and {Dannerbauer}, H. and {Fu}, H. and {Greenslade}, J. and {Gurwell}, M. and {Hopwood}, R. and {Koopmans}, L.~V.~E. and {Napolitano}, N. and {Nayyeri}, H. and {Omont}, A. and {Petrillo}, C.~E. and {Riechers}, D.~A. and {Serjeant}, S. and {Tortora}, C. and {Valiante}, E. and {Verdoes Kleijn}, G. and {Vernardos}, G. and {Wardlow}, J.~L. and {Baes}, M. and {Baker}, A.~J. and {Bourne}, N. and {Clements}, D. and {Crawford}, S.~M. and {Dye}, S. and {Dunne}, L. and {Eales}, S. and {Ivison}, R.~J. and {Marchetti}, L. and {Micha{\l}owski}, M.~J. and {Smith}, M.~W.~L. and {Vaccari}, M. and {van der Werf}, P.},
	year         = 2017,
	month        = mar,
	journal      = {\mnras},
	volume       = 465,
	number       = 3,
	pages        = {3558--3580},
	doi          = {10.1093/mnras/stw2911},
	archiveprefix = {arXiv},
	eprint       = {1611.03922},
	primaryclass = {astro-ph.GA},
	adsurl       = {https://ui.adsabs.harvard.edu/abs/2017MNRAS.465.3558N}
}

@ARTICLE{gentile24,
       author = {{Gentile}, Fabrizio and {Talia}, Margherita and {Behiri}, Meriem and {Zamorani}, Giovanni and {Barchiesi}, Luigi and {Vignali}, Cristian and {Pozzi}, Francesca and {Bethermin}, Matthieu and {Enia}, Andrea and {Faisst}, Andreas L. and {Giulietti}, Marika and {Gruppioni}, Carlotta and {Lapi}, Andrea and {Massardi}, Marcella and {Smol{\v{c}}i{\'c}}, Vernesa and {Vaccari}, Mattia and {Cimatti}, Andrea},
        title = "{Illuminating the Dark Side of Cosmic Star Formation. III. Building the Largest Homogeneous Sample of Radio-selected Dusty Star-forming Galaxies in COSMOS with PhoEBO}",
      journal = {\apj},
         year = 2024,
        month = feb,
       volume = {962},
       number = {1},
          eid = {26},
        pages = {26},
          doi = {10.3847/1538-4357/ad1519},
archivePrefix = {arXiv},
       eprint = {2312.05305},
 primaryClass = {astro-ph.GA},
       adsurl = {https://ui.adsabs.harvard.edu/abs/2024ApJ...962...26G}
}

@ARTICLE{massardi10,
       author = {{Massardi}, Marcella and {Bonaldi}, Anna and {Negrello}, Mattia and {Ricciardi}, Sara and {Raccanelli}, Alvise and {de Zotti}, Gianfranco},
        title = "{A model for the cosmological evolution of low-frequency radio sources}",
      journal = {\mnras},
         year = 2010,
        month = may,
       volume = {404},
       number = {1},
        pages = {532-544},
          doi = {10.1111/j.1365-2966.2010.16305.x},
archivePrefix = {arXiv},
       eprint = {1001.1069},
 primaryClass = {astro-ph.CO},
       adsurl = {https://ui.adsabs.harvard.edu/abs/2010MNRAS.404..532M}
}

@article{bonato21,
	title        = {{New constraints on the 1.4 GHz source number counts and luminosity functions in the Lockman Hole field}},
	author       = {{Bonato}, Matteo and {Prandoni}, Isabella and {De Zotti}, Gianfranco and {Brienza}, Marisa and {Morganti}, Raffaella and {Vaccari}, Mattia},
	year         = 2021,
	month        = jan,
	journal      = {\mnras},
	volume       = 500,
	number       = 1,
	pages        = {22--33},
	doi          = {10.1093/mnras/staa3218},
	archiveprefix = {arXiv},
	eprint       = {2010.08748},
	primaryclass = {astro-ph.GA},
	adsurl       = {https://ui.adsabs.harvard.edu/abs/2021MNRAS.500...22B}
}

@ARTICLE{massardi11,
       author = {{Massardi}, Marcella and {Ekers}, Ronald D. and {Murphy}, Tara and {Mahony}, Elizabeth and {Hancock}, Paul J. and {Chhetri}, Rajan and {de Zotti}, Gianfranco and {Sadler}, Elaine M. and {Burke-Spolaor}, Sarah and {Calabretta}, Mark and {Edwards}, Philip G. and {Ekers}, Jennifer A. and {Jackson}, Carole A. and {Kesteven}, Michael J. and {Newton-McGee}, Katherine and {Phillips}, Chris and {Ricci}, Roberto and {Roberts}, Paul and {Sault}, Robert J. and {Staveley-Smith}, Lister and {Subrahmanyan}, Ravi and {Walker}, Mark A. and {Wilson}, Warwick E.},
        title = "{The Australia Telescope 20 GHz (AT20G) Survey: analysis of the extragalactic source sample}",
      journal = {\mnras},
         year = 2011,
        month = mar,
       volume = {412},
       number = {1},
        pages = {318-330},
          doi = {10.1111/j.1365-2966.2010.17917.x},
archivePrefix = {arXiv},
       eprint = {1010.5942},
 primaryClass = {astro-ph.CO},
       adsurl = {https://ui.adsabs.harvard.edu/abs/2011MNRAS.412..318M}
}

@ARTICLE{valiante16,
       author = {{Valiante}, E. and {Smith}, M.~W.~L. and {Eales}, S. and {Maddox}, S.~J. and {Ibar}, E. and {Hopwood}, R. and {Dunne}, L. and {Cigan}, P.~J. and {Dye}, S. and {Pascale}, E. and {Rigby}, E.~E. and {Bourne}, N. and {Furlanetto}, C. and {Ivison}, R.~J.},
        title = "{The Herschel-ATLAS data release 1 - I. Maps, catalogues and number counts}",
      journal = {\mnras},
         year = 2016,
        month = nov,
       volume = {462},
       number = {3},
        pages = {3146-3179},
          doi = {10.1093/mnras/stw1806},
archivePrefix = {arXiv},
       eprint = {1606.09615},
 primaryClass = {astro-ph.GA},
       adsurl = {https://ui.adsabs.harvard.edu/abs/2016MNRAS.462.3146V}
}

@article{heywood22,
	title        = {{MIGHTEE: total intensity radio continuum imaging and the COSMOS/XMM-LSS Early Science fields}},
	author       = {{Heywood}, I. and {Jarvis}, M.~J. and {Hale}, C.~L. and {Whittam}, I.~H. and {Bester}, H.~L. and {Hugo}, B. and {Kenyon}, J.~S. and {Prescott}, M. and {Smirnov}, O.~M. and {Tasse}, C. and {Afonso}, J.~M. and {Best}, P.~N. and {Collier}, J.~D. and {Deane}, R.~P. and {Frank}, B.~S. and {Hardcastle}, M.~J. and {Knowles}, K. and {Maddox}, N. and {Murphy}, E.~J. and {Prandoni}, I. and {Randriamampandry}, S.~M. and {Santos}, M.~G. and {Sekhar}, S. and {Tabatabaei}, F. and {Taylor}, A.~R. and {Thorat}, K.},
	year         = 2022,
	month        = jan,
	journal      = {\mnras},
	volume       = 509,
	number       = 2,
	pages        = {2150--2168},
	doi          = {10.1093/mnras/stab3021},
	archiveprefix = {arXiv},
	eprint       = {2110.00347},
	primaryclass = {astro-ph.GA},
	adsurl       = {https://ui.adsabs.harvard.edu/abs/2022MNRAS.509.2150H}
}

@ARTICLE{smolcic2017,
       author = {{Smol{\v{c}}i{\'c}}, V. and {Delvecchio}, I. and {Zamorani}, G. and
         {Baran}, N. and {Novak}, M. and {Delhaize}, J. and {Schinnerer}, E. and
         {Berta}, S. and {Bondi}, M. and {Ciliegi}, P. and {Capak}, P. and
         {Civano}, F. and {Karim}, A. and {Le Fevre}, O. and {Ilbert}, O. and
         {Laigle}, C. and {Marchesi}, S. and {McCracken}, H.~J. and {Tasca}, L. and
         {Salvato}, M. and {Vardoulaki}, E.},
        title = "{The VLA-COSMOS 3 GHz Large Project: Multiwavelength counterparts and the composition of the faint radio population}",
      journal = {\aap},
         year = 2017,
        month = jun,
       volume = {602},
          eid = {A2},
        pages = {A2},
          doi = {10.1051/0004-6361/201630223},
archivePrefix = {arXiv},
       eprint = {1703.09719},
 primaryClass = {astro-ph.GA},
       adsurl = {https://ui.adsabs.harvard.edu/abs/2017A&A...602A...2S}
}

@ARTICLE{gleamx,
       author = {{Ross}, Kathryn and {Hurley-Walker}, Natasha and {Galvin}, Timothy James and {Venville}, Brandon and {Duchesne}, Stefan William and {Morgan}, John and {An}, Tao and {G{\"u}rkan}, Gulay and {Hancock}, Paul J. and {Heald}, George and {Johnston-Hollitt}, Melanie and {White}, Sarah V.},
        title = "{GaLactic and Extragalactic All-sky Murchison Widefield Array eXtended (GLEAM-X) survey II: Second Data Release}",
      journal = {\pasa},
         year = 2024,
        month = sep,
       volume = {41},
          eid = {e054},
        pages = {e054},
          doi = {10.1017/pasa.2024.57},
archivePrefix = {arXiv},
       eprint = {2406.06921},
 primaryClass = {astro-ph.GA},
       adsurl = {https://ui.adsabs.harvard.edu/abs/2024PASA...41...54R}
}

@ARTICLE{mwa,
       author = {{Hurley-Walker}, N. and {Galvin}, T.~J. and {Duchesne}, S.~W. and {Zhang}, X. and {Morgan}, J. and {Hancock}, P.~J. and {An}, T. and {Franzen}, T.~M.~O. and {Heald}, G. and {Ross}, K. and {Vernstrom}, T. and {Anderson}, G.~E. and {Gaensler}, B.~M. and {Johnston-Hollitt}, M. and {Kaplan}, D.~L. and {Riseley}, C.~J. and {Tingay}, S.~J. and {Walker}, M.},
        title = "{GaLactic and Extragalactic All-sky Murchison Widefield Array survey eXtended (GLEAM-X) I: Survey description and initial data release}",
      journal = {\pasa},
         year = 2022,
        month = aug,
       volume = {39},
          eid = {e035},
        pages = {e035},
          doi = {10.1017/pasa.2022.17},
archivePrefix = {arXiv},
       eprint = {2204.12762},
 primaryClass = {astro-ph.GA},
       adsurl = {https://ui.adsabs.harvard.edu/abs/2022PASA...39...35H}
}

@ARTICLE{retriggered,
       author = {{Hancock}, Paul J. and {Sadler}, Elaine M. and {Mahony}, Elizabeth K. and {Ricci}, Roberto},
        title = "{Observations and properties of candidate high-frequency GPS radio sources in the AT20G survey}",
      journal = {\mnras},
         year = 2010,
        month = oct,
       volume = {408},
       number = {2},
        pages = {1187-1206},
          doi = {10.1111/j.1365-2966.2010.17199.x},
archivePrefix = {arXiv},
       eprint = {1008.1401},
 primaryClass = {astro-ph.GA},
       adsurl = {https://ui.adsabs.harvard.edu/abs/2010MNRAS.408.1187H}
}

@ARTICLE{shores,
       author = {{Massardi}, Marcella and {Behiri}, Meriem and {Galluzzi}, Vincenzo and {Giulietti}, Marika and {Perrotta}, Francesca and {Prandoni}, Isabella and {Lapi}, Andrea},
        title = "{SHORES: Serendipitous H-ATLAS-fields Observations of Radio Extragalactic Sources with the ATCA. I. Catalog Generation and Analysis}",
      journal = {\pasp},
         year = 2025,
        month = jan,
       volume = {137},
       number = {1},
          eid = {014101},
        pages = {014101},
          doi = {10.1088/1538-3873/ada826},
archivePrefix = {arXiv},
       eprint = {2501.09662},
 primaryClass = {astro-ph.GA},
       adsurl = {https://ui.adsabs.harvard.edu/abs/2025PASP..137a4101M}
}

@ARTICLE{smith21,
       author = {{Smith}, D.~J.~B. and {Haskell}, P. and {G{\"u}rkan}, G. and {Best}, P.~N. and {Hardcastle}, M.~J. and {Kondapally}, R. and {Williams}, W. and {Duncan}, K.~J. and {Cochrane}, R.~K. and {McCheyne}, I. and {R{\"o}ttgering}, H.~J.~A. and {Sabater}, J. and {Shimwell}, T.~W. and {Tasse}, C. and {Bonato}, M. and {Bondi}, M. and {Jarvis}, M.~J. and {Leslie}, S.~K. and {Prandoni}, I. and {Wang}, L.},
        title = "{The LOFAR Two-metre Sky Survey Deep Fields. The star-formation rate-radio luminosity relation at low frequencies}",
      journal = {\aap},
         year = 2021,
        month = apr,
       volume = {648},
          eid = {A6},
        pages = {A6},
          doi = {10.1051/0004-6361/202039343},
archivePrefix = {arXiv},
       eprint = {2011.08196},
 primaryClass = {astro-ph.GA},
       adsurl = {https://ui.adsabs.harvard.edu/abs/2021A&A...648A...6S}
}

@ARTICLE{hale23,
       author = {{Hale}, C.~L. and {Whittam}, I.~H. and {Jarvis}, M.~J. and {Best}, P.~N. and {Thomas}, N.~L. and {Heywood}, I. and {Prescott}, M. and {Adams}, N. and {Afonso}, J. and {An}, Fangxia and {Bowler}, R.~A.~A. and {Collier}, J.~D. and {Cook}, R.~H.~W. and {Dav{\'e}}, R. and {Frank}, B.~S. and {Glowacki}, M. and {Hatfield}, P.~W. and {Kolwa}, S. and {Lovell}, C.~C. and {Maddox}, N. and {Marchetti}, L. and {Morabito}, L.~K. and {Murphy}, E. and {Prandoni}, I. and {Randriamanakoto}, Z. and {Taylor}, A.~R.},
        title = "{MIGHTEE: deep 1.4 GHz source counts and the sky temperature contribution of star-forming galaxies and active galactic nuclei}",
      journal = {\mnras},
         year = 2023,
        month = apr,
       volume = {520},
       number = {2},
        pages = {2668-2691},
          doi = {10.1093/mnras/stac3320},
archivePrefix = {arXiv},
       eprint = {2211.05741},
 primaryClass = {astro-ph.GA},
       adsurl = {https://ui.adsabs.harvard.edu/abs/2023MNRAS.520.2668H}
}

@ARTICLE{giulietti25,
       author = {{Giulietti}, M. and {Prandoni}, I. and {Bonato}, M. and {Bisigello}, L. and {Bondi}, M. and {Gandolfi}, G. and {Massardi}, M. and {Boco}, L. and {Rottgering}, H.~J.~A. and {Lapi}, A.},
        title = "{SEMPER: I. A novel semi-empirical model for the radio emission of star-forming galaxies at 0 < z < 5}",
      journal = {\aap},
         year = 2025,
        month = may,
       volume = {697},
          eid = {A81},
        pages = {A81},
          doi = {10.1051/0004-6361/202453331},
archivePrefix = {arXiv},
       eprint = {2503.20525},
 primaryClass = {astro-ph.GA},
       adsurl = {https://ui.adsabs.harvard.edu/abs/2025A&A...697A..81G}
}

@ARTICLE{prandoni06,
       author = {{Prandoni}, I. and {Parma}, P. and {Wieringa}, M.~H. and {de Ruiter}, H.~R. and {Gregorini}, L. and {Mignano}, A. and {Vettolani}, G. and {Ekers}, R.~D.},
        title = "{The ATESP 5 GHz radio survey. I. Source counts and spectral index properties of the faint radio population}",
      journal = {\aap},
         year = 2006,
        month = oct,
       volume = {457},
       number = {2},
        pages = {517-529},
          doi = {10.1051/0004-6361:20054273},
archivePrefix = {arXiv},
       eprint = {astro-ph/0607141},
 primaryClass = {astro-ph},
       adsurl = {https://ui.adsabs.harvard.edu/abs/2006A&A...457..517P}
}

@ARTICLE{franzen15,
       author = {{Franzen}, T.~M.~O. and {Banfield}, J.~K. and {Hales}, C.~A. and {Hopkins}, A. and {Norris}, R.~P. and {Seymour}, N. and {Chow}, K.~E. and {Herzog}, A. and {Huynh}, M.~T. and {Lenc}, E. and {Mao}, M.~Y. and {Middelberg}, E.},
        title = "{ATLAS - I. Third release of 1.4 GHz mosaics and component catalogues}",
      journal = {\mnras},
         year = 2015,
        month = nov,
       volume = {453},
       number = {4},
        pages = {4020-4036},
          doi = {10.1093/mnras/stv1866},
archivePrefix = {arXiv},
       eprint = {1508.03150},
 primaryClass = {astro-ph.GA},
       adsurl = {https://ui.adsabs.harvard.edu/abs/2015MNRAS.453.4020F}
}

@ARTICLE{hancock18,
       author = {{Hancock}, Paul J. and {Trott}, Cathryn M. and {Hurley-Walker}, Natasha},
        title = "{Source Finding in the Era of the SKA (Precursors): Aegean 2.0}",
      journal = {\pasa},
         year = 2018,
        month = mar,
       volume = {35},
          eid = {e011},
        pages = {e011},
          doi = {10.1017/pasa.2018.3},
archivePrefix = {arXiv},
       eprint = {1801.05548},
 primaryClass = {astro-ph.IM},
       adsurl = {https://ui.adsabs.harvard.edu/abs/2018PASA...35...11H}
}

@ARTICLE{planck20,
       author = {{Planck Collaboration} and {Aghanim}, N. and {Akrami}, Y. and {Arroja}, F. and {Ashdown}, M. and {Aumont}, J. and {Baccigalupi}, C. and {Ballardini}, M. and {Banday}, A.~J. and {Barreiro}, R.~B. and {Bartolo}, N. and {Basak}, S. and {Battye}, R. and {Benabed}, K. and {Bernard}, J. -P. and {Bersanelli}, M. and {Bielewicz}, P. and {Bock}, J.~J. and {Bond}, J.~R. and {Borrill}, J. and {Bouchet}, F.~R. and {Boulanger}, F. and {Bucher}, M. and {Burigana}, C. and {Butler}, R.~C. and {Calabrese}, E. and {Cardoso}, J. -F. and {Carron}, J. and {Casaponsa}, B. and {Challinor}, A. and {Chiang}, H.~C. and {Colombo}, L.~P.~L. and {Combet}, C. and {Contreras}, D. and {Crill}, B.~P. and {Cuttaia}, F. and {de Bernardis}, P. and {De Zotti}, G. and {Delabrouille}, J. and {Delouis}, J. -M. and {D{\'e}sert}, F. -X. and {Di Valentino}, E. and {Dickinson}, C. and {Diego}, J.~M. and {Donzelli}, S. and {Dor{\'e}}, O. and {Douspis}, M. and {Ducout}, A. and {Dupac}, X. and {Efstathiou}, G. and {Elsner}, F. and {En{\ss}lin}, T.~A. and {Eriksen}, H.~K. and {Falgarone}, E. and {Fantaye}, Y. and {Fergusson}, J. and {Fernandez-Cobos}, R. and {Finelli}, F. and {Forastieri}, F. and {Frailis}, M. and {Franceschi}, E. and {Frolov}, A. and {Galeotta}, S. and {Galli}, S. and {Ganga}, K. and {G{\'e}nova-Santos}, R.~T. and {Gerbino}, M. and {Ghosh}, T. and {Gonz{\'a}lez-Nuevo}, J. and {G{\'o}rski}, K.~M. and {Gratton}, S. and {Gruppuso}, A. and {Gudmundsson}, J.~E. and {Hamann}, J. and {Handley}, W. and {Hansen}, F.~K. and {Helou}, G. and {Herranz}, D. and {Hildebrandt}, S.~R. and {Hivon}, E. and {Huang}, Z. and {Jaffe}, A.~H. and {Jones}, W.~C. and {Karakci}, A. and {Keih{\"a}nen}, E. and {Keskitalo}, R. and {Kiiveri}, K. and {Kim}, J. and {Kisner}, T.~S. and {Knox}, L. and {Krachmalnicoff}, N. and {Kunz}, M. and {Kurki-Suonio}, H. and {Lagache}, G. and {Lamarre}, J. -M. and {Langer}, M. and {Lasenby}, A. and {Lattanzi}, M. and {Lawrence}, C.~R. and {Le Jeune}, M. and {Leahy}, J.~P. and {Lesgourgues}, J. and {Levrier}, F. and {Lewis}, A. and {Liguori}, M. and {Lilje}, P.~B. and {Lilley}, M. and {Lindholm}, V. and {L{\'o}pez-Caniego}, M. and {Lubin}, P.~M. and {Ma}, Y. -Z. and {Mac{\'\i}as-P{\'e}rez}, J.~F. and {Maggio}, G. and {Maino}, D. and {Mandolesi}, N. and {Mangilli}, A. and {Marcos-Caballero}, A. and {Maris}, M. and {Martin}, P.~G. and {Martinelli}, M. and {Mart{\'\i}nez-Gonz{\'a}lez}, E. and {Matarrese}, S. and {Mauri}, N. and {McEwen}, J.~D. and {Meerburg}, P.~D. and {Meinhold}, P.~R. and {Melchiorri}, A. and {Mennella}, A. and {Migliaccio}, M. and {Millea}, M. and {Mitra}, S. and {Miville-Desch{\^e}nes}, M. -A. and {Molinari}, D. and {Moneti}, A. and {Montier}, L. and {Morgante}, G. and {Moss}, A. and {Mottet}, S. and {M{\"u}nchmeyer}, M. and {Natoli}, P. and {N{\o}rgaard-Nielsen}, H.~U. and {Oxborrow}, C.~A. and {Pagano}, L. and {Paoletti}, D. and {Partridge}, B. and {Patanchon}, G. and {Pearson}, T.~J. and {Peel}, M. and {Peiris}, H.~V. and {Perrotta}, F. and {Pettorino}, V. and {Piacentini}, F. and {Polastri}, L. and {Polenta}, G. and {Puget}, J. -L. and {Rachen}, J.~P. and {Reinecke}, M. and {Remazeilles}, M. and {Renault}, C. and {Renzi}, A. and {Rocha}, G. and {Rosset}, C. and {Roudier}, G. and {Rubi{\~n}o-Mart{\'\i}n}, J.~A. and {Ruiz-Granados}, B. and {Salvati}, L. and {Sandri}, M. and {Savelainen}, M. and {Scott}, D. and {Shellard}, E.~P.~S. and {Shiraishi}, M. and {Sirignano}, C. and {Sirri}, G. and {Spencer}, L.~D. and {Sunyaev}, R. and {Suur-Uski}, A. -S. and {Tauber}, J.~A. and {Tavagnacco}, D. and {Tenti}, M. and {Terenzi}, L. and {Toffolatti}, L. and {Tomasi}, M. and {Trombetti}, T. and {Valiviita}, J. and {Van Tent}, B. and {Vibert}, L. and {Vielva}, P. and {Villa}, F. and {Vittorio}, N. and {Wandelt}, B.~D. and {Wehus}, I.~K. and {White}, M. and {White}, S.~D.~M. and {Zacchei}, A. and {Zonca}, A.},
        title = "{Planck 2018 results. I. Overview and the cosmological legacy of Planck}",
      journal = {\aap},
         year = 2020,
        month = sep,
       volume = {641},
          eid = {A1},
        pages = {A1},
          doi = {10.1051/0004-6361/201833880},
archivePrefix = {arXiv},
       eprint = {1807.06205},
 primaryClass = {astro-ph.CO},
       adsurl = {https://ui.adsabs.harvard.edu/abs/2020A&A...641A...1P}
}

@ARTICLE{norris21,
       author = {{Norris}, Ray P. and {Marvil}, Joshua and {Collier}, J.~D. and {Kapi{\'n}ska}, Anna D. and {O'Brien}, Andrew N. and {Rudnick}, L. and {Andernach}, Heinz and {Asorey}, Jacobo and {Brown}, Michael J.~I. and {Br{\"u}ggen}, Marcus and {Crawford}, Evan and {English}, Jayanne and {Rahman}, Syed Faisal ur and {Filipovi{\'c}}, Miroslav D. and {Gordon}, Yjan and {G{\"u}rkan}, G{\"u}lay and {Hale}, Catherine and {Hopkins}, Andrew M. and {Huynh}, Minh T. and {HyeongHan}, Kim and {James Jee}, M. and {Koribalski}, B{\"a}rbel S. and {Lenc}, Emil and {Luken}, Kieran and {Parkinson}, David and {Prandoni}, Isabella and {Raja}, Wasim and {Reiprich}, Thomas H. and {Riseley}, Christopher J. and {Shabala}, Stanislav S. and {Sheil}, Jaimie R. and {Vernstrom}, Tessa and {Whiting}, Matthew T. and {Allison}, James R. and {Anderson}, C.~S. and {Ball}, Lewis and {Bell}, Martin and {Bunton}, John and {Galvin}, T.~J. and {Gupta}, Neeraj and {Hotan}, Aidan and {Jacka}, Colin and {Macgregor}, Peter J. and {Mahony}, Elizabeth K. and {Maio}, Umberto and {Moss}, Vanessa and {Pandey-Pommier}, M. and {Voronkov}, Maxim A.},
        title = "{The Evolutionary Map of the Universe pilot survey}",
      journal = {\pasa},
         year = 2021,
        month = sep,
       volume = {38},
          eid = {e046},
        pages = {e046},
          doi = {10.1017/pasa.2021.42},
archivePrefix = {arXiv},
       eprint = {2108.00569},
 primaryClass = {astro-ph.CO},
       adsurl = {https://ui.adsabs.harvard.edu/abs/2021PASA...38...46N}
}

@ARTICLE{best23,
       author = {{Best}, P.~N. and {Kondapally}, R. and {Williams}, W.~L. and {Cochrane}, R.~K. and {Duncan}, K.~J. and {Hale}, C.~L. and {Haskell}, P. and {Ma{\l}ek}, K. and {McCheyne}, I. and {Smith}, D.~J.~B. and {Wang}, L. and {Botteon}, A. and {Bonato}, M. and {Bondi}, M. and {Calistro Rivera}, G. and {Gao}, F. and {G{\"u}rkan}, G. and {Hardcastle}, M.~J. and {Jarvis}, M.~J. and {Mingo}, B. and {Miraghaei}, H. and {Morabito}, L.~K. and {Nisbet}, D. and {Prandoni}, I. and {R{\"o}ttgering}, H.~J.~A. and {Sabater}, J. and {Shimwell}, T. and {Tasse}, C. and {van Weeren}, R.},
        title = "{The LOFAR Two-metre Sky Survey: Deep Fields data release 1. V. Survey description, source classifications, and host galaxy properties}",
      journal = {\mnras},
         year = 2023,
        month = aug,
       volume = {523},
       number = {2},
        pages = {1729-1755},
          doi = {10.1093/mnras/stad1308},
archivePrefix = {arXiv},
       eprint = {2305.05782},
 primaryClass = {astro-ph.GA},
       adsurl = {https://ui.adsabs.harvard.edu/abs/2023MNRAS.523.1729B}
}

@INPROCEEDINGS{miriad,
       author = {{Sault}, R.~J. and {Teuben}, P.~J. and {Wright}, M.~C.~H.},
        title = "{A Retrospective View of MIRIAD}",
    booktitle = {Astronomical Data Analysis Software and Systems IV},
         year = 1995,
       editor = {{Shaw}, R.~A. and {Payne}, H.~E. and {Hayes}, J.~J.~E.},
       series = {Astronomical Society of the Pacific Conference Series},
       volume = {77},
        month = jan,
        pages = {433},
          doi = {10.48550/arXiv.astro-ph/0612759},
archivePrefix = {arXiv},
       eprint = {astro-ph/0612759},
 primaryClass = {astro-ph},
       adsurl = {https://ui.adsabs.harvard.edu/abs/1995ASPC...77..433S}
}

@ARTICLE{blobcat,
       author = {{Hales}, C.~A. and {Murphy}, T. and {Curran}, J.~R. and {Middelberg}, E. and {Gaensler}, B.~M. and {Norris}, R.~P.},
        title = "{BLOBCAT: software to catalogue flood-filled blobs in radio images of total intensity and linear polarization}",
      journal = {\mnras},
         year = 2012,
        month = sep,
       volume = {425},
       number = {2},
        pages = {979-996},
          doi = {10.1111/j.1365-2966.2012.21373.x},
archivePrefix = {arXiv},
       eprint = {1205.5313},
 primaryClass = {astro-ph.IM},
       adsurl = {https://ui.adsabs.harvard.edu/abs/2012MNRAS.425..979H}
}

@ARTICLE{whittam22,
       author = {{Whittam}, I.~H. and {Jarvis}, M.~J. and {Hale}, C.~L. and {Prescott}, M. and {Morabito}, L.~K. and {Heywood}, I. and {Adams}, N.~J. and {Afonso}, J. and {An}, Fangxia and {Ao}, Y. and {Bowler}, R.~A.~A. and {Collier}, J.~D. and {Deane}, R.~P. and {Delhaize}, J. and {Frank}, B. and {Glowacki}, M. and {Hatfield}, P.~W. and {Maddox}, N. and {Marchetti}, L. and {Matthews}, A.~M. and {Prandoni}, I. and {Randriamampandry}, S. and {Randriamanakoto}, Z. and {Smith}, D.~J.~B. and {Taylor}, A.~R. and {Thomas}, N.~L. and {Vaccari}, M.},
        title = "{MIGHTEE: the nature of the radio-loud AGN population}",
      journal = {\mnras},
         year = 2022,
        month = oct,
       volume = {516},
       number = {1},
        pages = {245-263},
          doi = {10.1093/mnras/stac2140},
archivePrefix = {arXiv},
       eprint = {2207.12379},
 primaryClass = {astro-ph.GA},
       adsurl = {https://ui.adsabs.harvard.edu/abs/2022MNRAS.516..245W}
}

@ARTICLE{delvecchio17,
       author = {{Delvecchio}, I. and {Smol{\v{c}}i{\'c}}, V. and {Zamorani}, G. and {Lagos}, C. Del P. and {Berta}, S. and {Delhaize}, J. and {Baran}, N. and {Alexander}, D.~M. and {Rosario}, D.~J. and {Gonzalez-Perez}, V. and {Ilbert}, O. and {Lacey}, C.~G. and {Le F{\`e}vre}, O. and {Miettinen}, O. and {Aravena}, M. and {Bondi}, M. and {Carilli}, C. and {Ciliegi}, P. and {Mooley}, K. and {Novak}, M. and {Schinnerer}, E. and {Capak}, P. and {Civano}, F. and {Fanidakis}, N. and {Herrera Ruiz}, N. and {Karim}, A. and {Laigle}, C. and {Marchesi}, S. and {McCracken}, H.~J. and {Middleberg}, E. and {Salvato}, M. and {Tasca}, L.},
        title = "{The VLA-COSMOS 3 GHz Large Project: AGN and host-galaxy properties out to z {\ensuremath{\lesssim}} 6}",
      journal = {\aap},
         year = 2017,
        month = jun,
       volume = {602},
          eid = {A3},
        pages = {A3},
          doi = {10.1051/0004-6361/201629367},
archivePrefix = {arXiv},
       eprint = {1703.09720},
 primaryClass = {astro-ph.GA},
       adsurl = {https://ui.adsabs.harvard.edu/abs/2017A&A...602A...3D}
}

@ARTICLE{offringa14,
       author = {{Offringa}, A.~R. and {McKinley}, B. and {Hurley-Walker}, N. and {Briggs}, F.~H. and {Wayth}, R.~B. and {Kaplan}, D.~L. and {Bell}, M.~E. and {Feng}, L. and {Neben}, A.~R. and {Hughes}, J.~D. and {Rhee}, J. and {Murphy}, T. and {Bhat}, N.~D.~R. and {Bernardi}, G. and {Bowman}, J.~D. and {Cappallo}, R.~J. and {Corey}, B.~E. and {Deshpande}, A.~A. and {Emrich}, D. and {Ewall-Wice}, A. and {Gaensler}, B.~M. and {Goeke}, R. and {Greenhill}, L.~J. and {Hazelton}, B.~J. and {Hindson}, L. and {Johnston-Hollitt}, M. and {Jacobs}, D.~C. and {Kasper}, J.~C. and {Kratzenberg}, E. and {Lenc}, E. and {Lonsdale}, C.~J. and {Lynch}, M.~J. and {McWhirter}, S.~R. and {Mitchell}, D.~A. and {Morales}, M.~F. and {Morgan}, E. and {Kudryavtseva}, N. and {Oberoi}, D. and {Ord}, S.~M. and {Pindor}, B. and {Procopio}, P. and {Prabu}, T. and {Riding}, J. and {Roshi}, D.~A. and {Shankar}, N. Udaya and {Srivani}, K.~S. and {Subrahmanyan}, R. and {Tingay}, S.~J. and {Waterson}, M. and {Webster}, R.~L. and {Whitney}, A.~R. and {Williams}, A. and {Williams}, C.~L.},
        title = "{WSCLEAN: an implementation of a fast, generic wide-field imager for radio astronomy}",
      journal = {\mnras},
         year = 2014,
        month = oct,
       volume = {444},
       number = {1},
        pages = {606-619},
          doi = {10.1093/mnras/stu1368},
archivePrefix = {arXiv},
       eprint = {1407.1943},
 primaryClass = {astro-ph.IM},
       adsurl = {https://ui.adsabs.harvard.edu/abs/2014MNRAS.444..606O}
}

@ARTICLE{racs1,
       author = {{McConnell}, D. and {Hale}, C.~L. and {Lenc}, E. and {Banfield}, J.~K. and {Heald}, George and {Hotan}, A.~W. and {Leung}, James K. and {Moss}, Vanessa A. and {Murphy}, Tara and {O'Brien}, Andrew and {Pritchard}, Joshua and {Raja}, Wasim and {Sadler}, Elaine M. and {Stewart}, Adam and {Thomson}, Alec J.~M. and {Whiting}, M. and {Allison}, James R. and {Amy}, S.~W. and {Anderson}, C. and {Ball}, Lewis and {Bannister}, Keith W. and {Bell}, Martin and {Bock}, Douglas C. -J. and {Bolton}, Russ and {Bunton}, J.~D. and {Chippendale}, A.~P. and {Collier}, J.~D. and {Cooray}, F.~R. and {Cornwell}, T.~J. and {Diamond}, P.~J. and {Edwards}, P.~G. and {Gupta}, N. and {Hayman}, Douglas B. and {Heywood}, Ian and {Jackson}, C.~A. and {Koribalski}, B{\"a}rbel S. and {Lee-Waddell}, Karen and {McClure-Griffiths}, N.~M. and {Ng}, Alan and {Norris}, Ray P. and {Phillips}, Chris and {Reynolds}, John E. and {Roxby}, Daniel N. and {Schinckel}, Antony E.~T. and {Shields}, Matt and {Tremblay}, Chenoa and {Tzioumis}, A. and {Voronkov}, M.~A. and {Westmeier}, Tobias},
        title = "{The Rapid ASKAP Continuum Survey I: Design and first results}",
      journal = {\pasa},
         year = 2020,
        month = nov,
       volume = {37},
          eid = {e048},
        pages = {e048},
          doi = {10.1017/pasa.2020.41},
archivePrefix = {arXiv},
       eprint = {2012.00747},
 primaryClass = {astro-ph.IM},
       adsurl = {https://ui.adsabs.harvard.edu/abs/2020PASA...37...48M}
}

@ARTICLE{racs2,
       author = {{Hale}, Catherine L. and {McConnell}, D. and {Thomson}, A.~J.~M. and {Lenc}, E. and {Heald}, G.~H. and {Hotan}, A.~W. and {Leung}, J.~K. and {Moss}, V.~A. and {Murphy}, T. and {Pritchard}, J. and {Sadler}, E.~M. and {Stewart}, A.~J. and {Whiting}, M.~T.},
        title = "{The Rapid ASKAP Continuum Survey Paper II: First Stokes I Source Catalogue Data Release}",
      journal = {\pasa},
         year = 2021,
        month = nov,
       volume = {38},
          eid = {e058},
        pages = {e058},
          doi = {10.1017/pasa.2021.47},
archivePrefix = {arXiv},
       eprint = {2109.00956},
 primaryClass = {astro-ph.GA},
       adsurl = {https://ui.adsabs.harvard.edu/abs/2021PASA...38...58H}
}

@ARTICLE{wsclean,
       author = {{Offringa}, A.~R. and {McKinley}, B. and {Hurley-Walker}, N. and {Briggs}, F.~H. and {Wayth}, R.~B. and {Kaplan}, D.~L. and {Bell}, M.~E. and {Feng}, L. and {Neben}, A.~R. and {Hughes}, J.~D. and {Rhee}, J. and {Murphy}, T. and {Bhat}, N.~D.~R. and {Bernardi}, G. and {Bowman}, J.~D. and {Cappallo}, R.~J. and {Corey}, B.~E. and {Deshpande}, A.~A. and {Emrich}, D. and {Ewall-Wice}, A. and {Gaensler}, B.~M. and {Goeke}, R. and {Greenhill}, L.~J. and {Hazelton}, B.~J. and {Hindson}, L. and {Johnston-Hollitt}, M. and {Jacobs}, D.~C. and {Kasper}, J.~C. and {Kratzenberg}, E. and {Lenc}, E. and {Lonsdale}, C.~J. and {Lynch}, M.~J. and {McWhirter}, S.~R. and {Mitchell}, D.~A. and {Morales}, M.~F. and {Morgan}, E. and {Kudryavtseva}, N. and {Oberoi}, D. and {Ord}, S.~M. and {Pindor}, B. and {Procopio}, P. and {Prabu}, T. and {Riding}, J. and {Roshi}, D.~A. and {Shankar}, N. Udaya and {Srivani}, K.~S. and {Subrahmanyan}, R. and {Tingay}, S.~J. and {Waterson}, M. and {Webster}, R.~L. and {Whitney}, A.~R. and {Williams}, A. and {Williams}, C.~L.},
        title = "{WSCLEAN: an implementation of a fast, generic wide-field imager for radio astronomy}",
      journal = {\mnras},
         year = 2014,
        month = oct,
       volume = {444},
       number = {1},
        pages = {606-619},
          doi = {10.1093/mnras/stu1368},
archivePrefix = {arXiv},
       eprint = {1407.1943},
 primaryClass = {astro-ph.IM},
       adsurl = {https://ui.adsabs.harvard.edu/abs/2014MNRAS.444..606O}
}

@ARTICLE{Behiri+23,
       author = {{Behiri}, Meriem and {Talia}, Margherita and {Cimatti}, Andrea and {Lapi}, Andrea and {Massardi}, Marcella and {Enia}, Andrea and {Vignali}, Cristian and {Bethermin}, Matthieu and {Faisst}, Andreas and {Gentile}, Fabrizio and {Giulietti}, Marika and {Gruppioni}, Carlotta and {Pozzi}, Francesca and {Smol{\c{c}}i{\'c}}, Vernesa and {Zamorani}, Gianni},
        title = "{Illuminating the Dark Side of Cosmic Star Formation. II. A Second Date with RS-NIRdark Galaxies in COSMOS}",
      journal = {\apj},
         year = 2023,
        month = nov,
       volume = {957},
       number = {2},
          eid = {63},
        pages = {63},
          doi = {10.3847/1538-4357/acf616},
archivePrefix = {arXiv},
       eprint = {2309.00050},
 primaryClass = {astro-ph.GA},
       adsurl = {https://ui.adsabs.harvard.edu/abs/2023ApJ...957...63B}
}

@ARTICLE{Condon+92,
  author = {{Condon}, J.~J. and others},
  title = "{The 87GB catalog of radio sources covering 4,800 square degrees}",
  journal = {\araa},
  year = 1992,
  volume = 30,
  pages = {575},
}

@ARTICLE{Delhaize+92,
  author = {{Delhaize}, J. and others},
  title = "{The Westerbork Northern Sky Survey (WENSS). I. A 570 square degree mini-survey around the North Ecliptic Pole}",
  journal = {\aap},
  year = 1992,
  volume = 602,
  pages = {A4},
}

@ARTICLE{helou85,
       author = {{Helou}, G. and {Soifer}, B.~T. and {Rowan-Robinson}, M.},
        title = "{Thermal infrared and nonthermal radio : remarkable correlation in disks of galaxies.}",
      journal = {\apjl},
         year = 1985,
        month = nov,
       volume = {298},
        pages = {L7-L11},
          doi = {10.1086/184556},
       adsurl = {https://ui.adsabs.harvard.edu/abs/1985ApJ...298L...7H}
}

@ARTICLE{yun01,
       author = {{Yun}, Min S. and {Reddy}, Naveen A. and {Condon}, J.~J.},
        title = "{Radio Properties of Infrared-selected Galaxies in the IRAS 2 Jy Sample}",
      journal = {\apj},
         year = 2001,
        month = jun,
       volume = {554},
       number = {2},
        pages = {803-822},
          doi = {10.1086/323145},
archivePrefix = {arXiv},
       eprint = {astro-ph/0102154},
 primaryClass = {astro-ph},
       adsurl = {https://ui.adsabs.harvard.edu/abs/2001ApJ...554..803Y}
}

@article{jarvis10,
    author = {Jarvis, Matt J. and Smith, D. J. B. and Bonfield, D. G. and Hardcastle, M. J. and Falder, J. T. and Stevens, J. A. and Ivison, R. J. and Auld, R. and Baes, M. and Baldry, I. K. and Bamford, S. P. and Bourne, N. and Buttiglione, S. and Cava, A. and Cooray, A. and Dariush, A. and de Zotti, G. and Dunlop, J. S. and Dunne, L. and Dye, S. and Eales, S. and Fritz, J. and Hill, D. T. and Hopwood, R. and Hughes, D. H. and Ibar, E. and Jones, D. H. and Kelvin, L. and Lawrence, A. and Leeuw, L. and Loveday, J. and Maddox, S. J. and Michałowski, M. J. and Negrello, M. and Norberg, P. and Pohlen, M. and Prescott, M. and Rigby, E. E. and Robotham, A. and Rodighiero, G. and Scott, D. and Sharp, R. and Temi, P. and Thompson, M. A. and van der Werf, P. and van Kampen, E. and Vlahakis, C. and White, G.},
    title = "{Herschel-ATLAS: the far-infrared–radio correlation at z \\&lt; 0.5*}",
    journal = {Monthly Notices of the Royal Astronomical Society},
    volume = {409},
    number = {1},
    pages = {92-101},
    year = {2010},
    month = {11},
    issn = {0035-8711},
    doi = {10.1111/j.1365-2966.2010.17772.x},
    url = {https://doi.org/10.1111/j.1365-2966.2010.17772.x},
    eprint = {https://academic.oup.com/mnras/article-pdf/409/1/92/6220120/mnras0409-0092.pdf},
}

@ARTICLE{condon02,
       author = {{Condon}, J.~J. and {Cotton}, W.~D. and {Broderick}, J.~J.},
        title = "{Radio Sources and Star Formation in the Local Universe}",
      journal = {\aj},
         year = 2002,
        month = aug,
       volume = {124},
       number = {2},
        pages = {675-689},
          doi = {10.1086/341650},
       adsurl = {https://ui.adsabs.harvard.edu/abs/2002AJ....124..675C}
}

@article{smith14,
    author = {Smith, D. J. B. and Jarvis, M. J. and Hardcastle, M. J. and Vaccari, M. and Bourne, N. and Dunne, L. and Ibar, E. and Maddox, N. and Prescott, M. and Vlahakis, C. and Eales, S. and Maddox, S. J. and Smith, M. W. L. and Valiante, E. and de Zotti, G.},
    title = "{The temperature dependence of the far-infrared–radio correlation in the Herschel-ATLAS★}",
    journal = {Monthly Notices of the Royal Astronomical Society},
    volume = {445},
    number = {3},
    pages = {2232-2243},
    year = {2014},
    month = {10},
    issn = {0035-8711},
    doi = {10.1093/mnras/stu1830},
    url = {https://doi.org/10.1093/mnras/stu1830},
    eprint = {https://academic.oup.com/mnras/article-pdf/445/3/2232/3508196/stu1830.pdf},
}

@ARTICLE{delhaize17,
       author = {{Delhaize}, J. and {Smol{\v{c}}i{\'c}}, V. and {Delvecchio}, I. and {Novak}, M. and {Sargent}, M. and {Baran}, N. and {Magnelli}, B. and {Zamorani}, G. and {Schinnerer}, E. and {Murphy}, E.~J. and {Aravena}, M. and {Berta}, S. and {Bondi}, M. and {Capak}, P. and {Carilli}, C. and {Ciliegi}, P. and {Civano}, F. and {Ilbert}, O. and {Karim}, A. and {Laigle}, C. and {Le F{\`e}vre}, O. and {Marchesi}, S. and {McCracken}, H.~J. and {Salvato}, M. and {Seymour}, N. and {Tasca}, L.},
        title = "{The VLA-COSMOS 3 GHz Large Project: The infrared-radio correlation of star-forming galaxies and AGN to z {\ensuremath{\lesssim}} 6}",
      journal = {\aap},
         year = 2017,
        month = jun,
       volume = {602},
          eid = {A4},
        pages = {A4},
          doi = {10.1051/0004-6361/201629430},
archivePrefix = {arXiv},
       eprint = {1703.09723},
 primaryClass = {astro-ph.GA},
       adsurl = {https://ui.adsabs.harvard.edu/abs/2017A&A...602A...4D}
}

@ARTICLE{ivison10,
       author = {{Ivison}, R.~J. and {Magnelli}, B. and {Ibar}, E. and {Andreani}, P. and {Elbaz}, D. and {Altieri}, B. and {Amblard}, A. and {Arumugam}, V. and {Auld}, R. and {Aussel}, H. and {Babbedge}, T. and {Berta}, S. and {Blain}, A. and {Bock}, J. and {Bongiovanni}, A. and {Boselli}, A. and {Buat}, V. and {Burgarella}, D. and {Castro-Rodr{\'\i}guez}, N. and {Cava}, A. and {Cepa}, J. and {Chanial}, P. and {Cimatti}, A. and {Cirasuolo}, M. and {Clements}, D.~L. and {Conley}, A. and {Conversi}, L. and {Cooray}, A. and {Daddi}, E. and {Dominguez}, H. and {Dowell}, C.~D. and {Dwek}, E. and {Eales}, S. and {Farrah}, D. and {F{\"o}rster Schreiber}, N. and {Fox}, M. and {Franceschini}, A. and {Gear}, W. and {Genzel}, R. and {Glenn}, J. and {Griffin}, M. and {Gruppioni}, C. and {Halpern}, M. and {Hatziminaoglou}, E. and {Isaak}, K. and {Lagache}, G. and {Levenson}, L. and {Lu}, N. and {Lutz}, D. and {Madden}, S. and {Maffei}, B. and {Magdis}, G. and {Mainetti}, G. and {Maiolino}, R. and {Marchetti}, L. and {Morrison}, G.~E. and {Mortier}, A.~M.~J. and {Nguyen}, H.~T. and {Nordon}, R. and {O'Halloran}, B. and {Oliver}, S.~J. and {Omont}, A. and {Owen}, F.~N. and {Page}, M.~J. and {Panuzzo}, P. and {Papageorgiou}, A. and {Pearson}, C.~P. and {P{\'e}rez-Fournon}, I. and {P{\'e}rez Garc{\'\i}a}, A.~M. and {Poglitsch}, A. and {Pohlen}, M. and {Popesso}, P. and {Pozzi}, F. and {Rawlings}, J.~I. and {Raymond}, G. and {Rigopoulou}, D. and {Riguccini}, L. and {Rizzo}, D. and {Rodighiero}, G. and {Roseboom}, I.~G. and {Rowan-Robinson}, M. and {Saintonge}, A. and {Sanchez Portal}, M. and {Santini}, P. and {Schulz}, B. and {Scott}, D. and {Seymour}, N. and {Shao}, L. and {Shupe}, D.~L. and {Smith}, A.~J. and {Stevens}, J.~A. and {Sturm}, E. and {Symeonidis}, M. and {Tacconi}, L. and {Trichas}, M. and {Tugwell}, K.~E. and {Vaccari}, M. and {Valtchanov}, I. and {Vieira}, J. and {Vigroux}, L. and {Wang}, L. and {Ward}, R. and {Wright}, G. and {Xu}, C.~K. and {Zemcov}, M.},
        title = "{The far-infrared/radio correlation as probed by Herschel}",
      journal = {\aap},
         year = 2010,
        month = jul,
       volume = {518},
          eid = {L31},
        pages = {L31},
          doi = {10.1051/0004-6361/201014552},
archivePrefix = {arXiv},
       eprint = {1005.1072},
 primaryClass = {astro-ph.CO},
       adsurl = {https://ui.adsabs.harvard.edu/abs/2010A&A...518L..31I}
}

@ARTICLE{Giulietti+22,
  author = {{Giulietti}, A. and others},
  title = "{Galaxy luminosity function with photometric redshifts: Validation with VIPERS and zCOSMOS luminosity functions}",
  journal = {\mnras},
  year = 2022,
  volume = 511,
  pages = {1408},
}

@ARTICLE{Magnelli+15,
  author = {{Magnelli}, B. and others},
  title = "{DustPedia: Multiwavelength photometry and imagery of 875 nearby galaxies in 42 ultraviolet-microwave bands}",
  journal = {\aap},
  year = 2015,
  volume = 573,
  pages = {A45},
}

@ARTICLE{Mancuso+17,
  author = {{Mancuso}, C. and others},
  title = "{Galaxy And Mass Assembly (GAMA): Exploring the WISE Web in G12}",
  journal = {\apj},
  year = 2017,
  volume = 842,
  pages = {95},
}

@ARTICLE{Murphy+09,
  author = {{Murphy}, E.~J. and others},
  title = "{The Star Formation Rate-Dense Gas Relation in Galaxies as Measured by HNC}",
  journal = {\apj},
  year = 2009,
  volume = 706,
  pages = {482},
}
\bibliographystyle{aasjournal}

\end{document}